\documentclass[lettersize,journal]{IEEEtran}
\usepackage{amsmath,amsfonts}
\usepackage{algorithmic}
\usepackage{array}
\usepackage[caption=false,font=normalsize,labelfont=sf,textfont=sf]{subfig}
\usepackage{textcomp}
\usepackage{stfloats}
\usepackage{url}
\usepackage{verbatim}
\usepackage{graphicx}
\usepackage{threeparttable}
\usepackage{multirow}
\usepackage{bbm}
\usepackage{xcolor}
\def\BibTeX{{\rm B\kern-.05em{\sc i\kern-.025em b}\kern-.08em
    T\kern-.1667em\lower.7ex\hbox{E}\kern-.125emX}}
\usepackage{balance}
\begin{document}
\title{Process, Bias and Temperature Scalable CMOS Analog Computing Circuits for Machine Learning}
\author{\IEEEauthorblockN{Pratik Kumar$^\dagger$, Ankita Nandi$^\dagger$, Shantanu Chakrabartty$^{\ast}$, Chetan Singh Thakur$^\dagger$}\thanks{$^\star $This work has been accepted in IEEE for publication and is available with journal DOI: {10.1109/TCSI.2022.3216287}. Copyright may be transferred without notice, after which this version may no longer be accessible.}
\\
\IEEEauthorblockA{\{pratikkumar, ankitanandi, csthakur\}@iisc.ac.in, \{shantanu\}@wustl.edu \\
$^\dagger$Department of Electronic Systems Engineering, Indian Institute of Science, Bangalore, India, 560012\\
$^{\ast}$Department of Electrical and Systems Engineering, Washington University in St. Louis,USA, 63130}
}

\maketitle
\begin{abstract}
Analog computing is attractive compared to digital computing due to its potential for achieving higher computational density and higher energy efficiency. However, unlike digital circuits, conventional analog computing circuits cannot be easily mapped across different process nodes due to differences in transistor biasing regimes, temperature variations and limited dynamic range. In this work, we generalize the previously reported margin-propagation-based analog computing framework for designing novel \textit{shape-based analog computing} (S-AC) circuits that can be easily cross-mapped across different process nodes. Similar to digital designs S-AC designs can also be scaled for precision, speed, and power. As a proof-of-concept, we show several examples of S-AC circuits implementing mathematical functions that are commonly used in machine learning architectures. Using circuit simulations we demonstrate that the circuit input/output characteristics remain robust when mapped from a planar CMOS 180nm process to a FinFET 7nm process. Also, using benchmark datasets we demonstrate that the classification accuracy of a S-AC based neural network remains robust when mapped across the two processes and to changes in temperature. 
\end{abstract}
\begin{IEEEkeywords}
Machine Learning, Process Scalability, Approximate Computing, Margin Propagation, Shape-based Computing.
\end{IEEEkeywords}

\section{Introduction}

\IEEEPARstart{A}{nalog} computing techniques are attractive for implementing machine learning (ML) architecture because of the potential to achieve high computational density and high energy-efficiency when compared to an equivalent digital implementation. ML training also allows for offline and online calibration which can compensate for analog artifacts due to device mismatch and non-linearity~\cite{Cau99}. Examples of previous analog ML implementations include~\cite{chakrabartty2007sub, kang2009chip, yamasaki2003analog, kumar2022hybrid}. However, one of the key advantages of digital implementation is its \textit{process scalability} where a digital circuit module designed in one process node (typically with a larger feature size) can be mapped to a more advanced process node (with a smaller feature size) with minimal to no circuit modification~\cite{huang2021machine,vigoda_thesis}. Due to \textit{process scalability}, digital implementations (both ML and non-ML architectures) can benefit from sub-10nm technology scaling in terms of improved speed, power and compute density. On the other hand, scaling analog computing circuits across process nodes has been difficult due to several reasons~\cite{finfet,mismatch1} and can be highlighted using Fig.~\ref{gmid}. The figure compares transistor performance using the product of transconductance efficiency and speed as a figure-of-merit (FOM)~\cite{murmann} for a 180nm planar CMOS process and for a 7nm FinFET process~\cite{ptm22,asap7}. The FOM is shown in Fig.~\ref{gmid} for three different biasing regimes, weak-inversion (WI), moderate-inversion (MI), and strong-inversion (SI). Note that in Fig.~\ref{gmid}, biasing regimes can be differentiated from each other by their respective transconductance efficiencies ($g_m/I_d$). For ${V_{gs}} - {V_{th}} < 0$, transistors are operating in WI while the transition slope denotes MI regime. It can also be seen from Fig.~\ref{gmid} that for 7nm FinFET process, the highest dynamic range and the best FOM is achieved in the moderate inversion regime. As the feature size is scaled to a more advanced process node (7nm feature size), the moderate inversion region dominates the transistor dynamic range. Whereas, most analog computing circuits that exploit the large-signal transistor I-V characteristics operate either in the weak-inversion regime~\cite{translinear_weak} or in the strong-inversion regime~\cite{translinear_strong}. These circuits cannot be directly scaled to the sub-10nm process nodes without significant performance degradation. 
\begin{figure}[t]
\centering
\includegraphics[width=0.89\linewidth]{Images/ft_gmid.png}
\caption{Plot of transconductance efficiency~($g_m/I_d$) as a function of $V_{gs}-V_{th}$ at different process nodes~\cite{ptm22}. The plot also shows the product of $g_m/I_d$ and speed~($f_T$), denoting the efficiency peak obtained in moderate inversion. Plots are shown for n-type planar CMOS and FinFET at different process nodes. Here the maximum supply voltages of each process node can be noted as $1.8V$, $0.8V$ and $0.7V$ for 180nm, 22nm and 7nm, respectively.}
\label{gmid}
\end{figure}

In this paper, we propose analog computing circuits that are process scalable and hence similar to digital designs, can be used as synthesizable analog standard cells. At the core of the proposed circuits is a generalization of the previously reported bias scalable analog computing framework called margin propagation~\cite{synth_bias_scale}. In this work, we show that generalized margin propagation leads to a novel shape-based analog computing (S-AC) framework where circuit accuracy can be traded off with speed and power, like digital designs. Furthermore, S-AC circuits can be scaled across processes and can work across different biasing regimes and across different operating temperatures. As a proof-of-concept, this paper presents the design of several key analog computing circuits commonly used in ML architectures.        
The key contributions of this work in relation to previous approaches are as follows:
\begin{itemize}
    \item {Generalization of Margin Propagation (MP) design framework and introducing a multi-spline approach that allows trading off computational accuracy with speed and power.} Using this multi-spline approach, we design a basic prototype function and show that its characteristic is robust with respect to biasing, process nodes, and temperature variations. 
    \item Using the basic prototype function, we synthesize S-AC circuits that approximate different functions commonly used in ML architecture. 
    \item Using the basic S-AC circuits, we present a complete design of a 3-layer neural network that is process and bias scalable with respect to classification accuracy on benchmark datasets. 
\end{itemize}
The rest of the sections are organized as follows. Section~\ref{sac_math} presents the mathematical framework for generalized margin propagation (GMP) and the S-AC framework. Section~\ref{sac_ckt_analy} shows the MOS circuit implementation of the basic S-AC circuit and demonstrates its scalability at different biases and process nodes. In Section~\ref{designs} we {show} the design of widely used machine learning functions implemented using a basic S-AC unit and its performance trade-off analysis. Section~\ref{analog_system} presents a case study of S-AC based 3-layer neural network. Section~\ref{conc} concludes the papers with discussions and final remarks.

\begin{figure*}[t]
	\centering
	\subfloat[]{\includegraphics[width=0.24\linewidth]{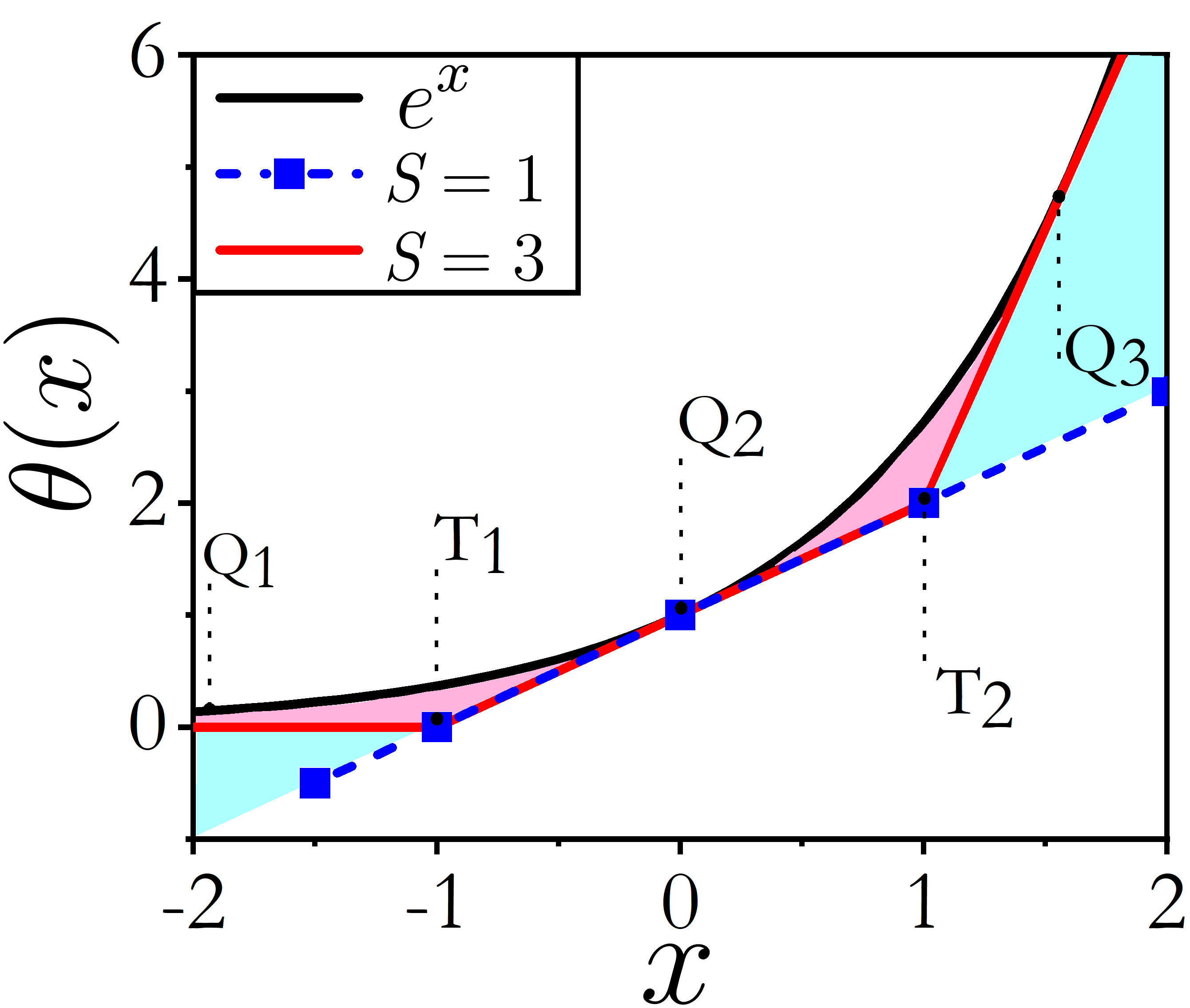}
		\label{exp}}
	\subfloat[]{\includegraphics[width=0.38\linewidth]{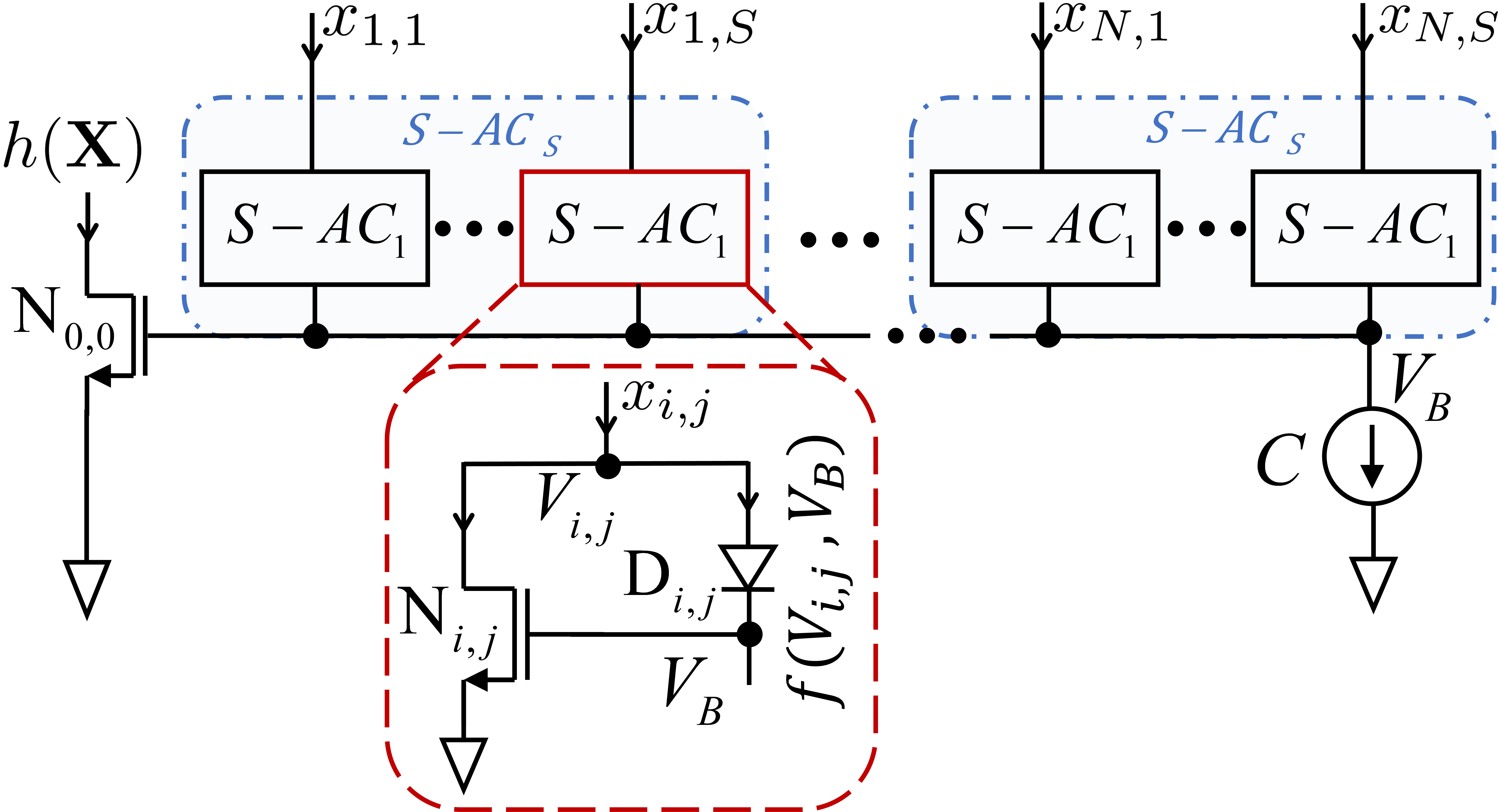}
		\label{n_sac_ckt}}
	\subfloat[]{\includegraphics[width=0.38\linewidth]{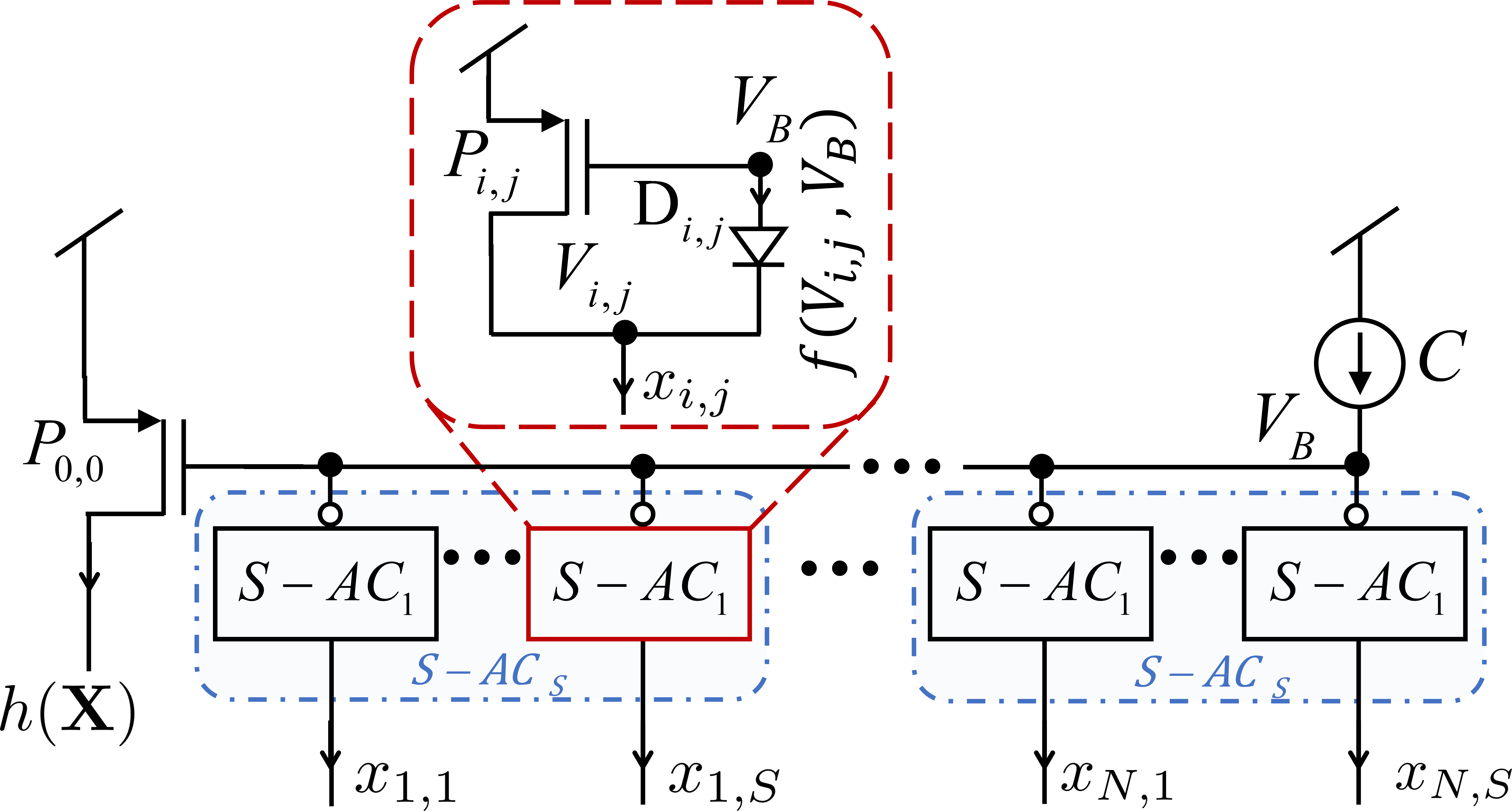}
		\label{p_sac_ckt}}

	\caption{\protect\subref{exp}~{Plot showing the approximation of a non-linear function $\theta(x) \simeq {e^x}$ using linear splines~($S$). Here, the approximations are shown for different spline ($S$) counts i.e, $S = 1,3$, where $Q_1,Q_2,Q_3$ are the tangential points and $T_1,T_2,T_3$ are the tuning points}; \protect\subref{n_sac_ckt}~Implementation of N-type S-AC circuit for $N$ inputs and $S$ splines, the inset shows the circuit implementation of a single S-AC unit using n-type FET and a diode; \protect\subref{p_sac_ckt}~Implementation of P-type S-AC circuit for $N$ inputs and $S$ splines, the inset shows the circuit implementation of a single S-AC unit using p-type FET and a diode. The circle on the P-type S-AC unit is used to differentiate between an N-type S-AC unit and a P-type S-AC unit.}
	\label{sac_ckt_shape}
\end{figure*}

\section{Generalization of Margin Propagation and Shape-based Analog Computing}\label{sac_math} 
In this section, we extend our previous work in the area of bias-scalable analog computing circuits~\cite{synth_bias_scale} using a multi-spline approach which is then generalized to shape-based computing. These shapes or prototype functions are then shown to be implemented using physical operating principles of MOSFETs and diodes. 

\subsection{Multi-Spline Approximation of log-sum-exp function}
Similar to the previous Margin Propagation framework~\cite{synth_bias_scale}, the starting point of our framework is an approximation of the log-sum-exp function~\cite{synth_bias_scale} $h_{\log}:\mathbb{R}^N \rightarrow \mathbb{R}$ given by
\begin{align}
h_{\log}(\textbf{x}) = C \cdot \log \left( {\sum\limits_{i = 1}^N {{e^{\frac{{{x_i}}}{C}}}} } \right)
\label{log-sum-exp}
\end{align}
where $C$ is a hyper-parameter and $\textbf{x} \in \mathbb{R}^N$ is a vector with elements $x_i \in \mathbb{R}$. 
Equation \eqref{log-sum-exp} can be written as
\begin{align}
\sum\limits_{i = 1}^N {{e^{\frac{{{x_i} - h_{\log}}}{C}}}}  = 1
\label{log-sum-exp_rewritten}
\end{align}
which is an equivalent non-linear constraint satisfaction problem. 
In our previous Margin Propagation related work~\cite{synth_bias_scale} we had approximated the $e^{(.)}$ using a single spline as  
\begin{align}
e^{x} \cong [x]_+
\label{mpapprox}
\end{align}
where $[.]_+$ denotes a rectifying linear unit (ReLU) function. {However, the rectification function can be substituted based upon some key properties to formulate the GMP formulation explained in detail in Section~\ref{gmp}.} 

The single-spline approximation ($S=1$) of the
exponential function is highlighted in Fig.~\ref{exp}.
Using a single spline, \eqref{log-sum-exp_rewritten} can be expressed as a piece-wise approximation $h \approx h_{\log}$ where $h$ is computed as a solution to the non-linear equation 
\begin{align}
	\sum\limits_{i = 1}^N \left[x_i - h \right]_+  = C
	\label{mp}
\end{align}
Fig.~\ref{exp} also provides the insight that the exponential function can be better approximated using multiple splines
($S > 1$) defined by different parameters $Q_j,T_j$, where $j = 1,..,S$. {It also shows that the single spline approximation has a wider margin due to one intersecting solution, while the three spline approximation shows a narrower margin. Thus introducing multiple splines in the approximation provides control over the precision and accuracy trade-off}. In this work, we, therefore, generalize the MP framework in~\eqref{mp} by approximating $e^{(.)}$ using multiple splines as
\begin{align}
e^x \cong \sum\limits_{j = 1}^S {\left( {{e^{{Q_j}}} - \sum\limits_{k = 1}^{j - 1} {{e^{{Q_k}}}} } \right)} {\left[ {x - {T_j}} \right]_ + }
\label{generic_msmp}
\end{align}
The rationale for choosing the values of $Q_j,T_j$, where $j = 1,..,S$ in~\eqref{generic_msmp} is provided in Appendix~\ref{Appendix:A}. Appendix~\ref{Appendix:A} also shows that, for specific values of $Q_j,T_j$, where $j = 1,..,S$, \eqref{generic_msmp} leads to a simplified multi-spline approximation like~\eqref{mp} and given by
\begin{align}
\sum\limits_{i = 1}^N {\sum\limits_{j = 1}^S {\left[ {{x_{i,j}} - h} \right]_{+}} }  &= C
\label{equation18wc_DUALsUM}
\end{align}
Thus, $h(.)$ is a function of a matrix whose elements are $x_{i,j}, i = 1,..,N; j = 1,..,S$. Equation~\eqref{equation18wc_DUALsUM} therefore serves as a generalization of margin-propagation using multiple splines.

\begin{figure*}[t]
	\centering
	\subfloat[]{\includegraphics[width=0.25\linewidth]{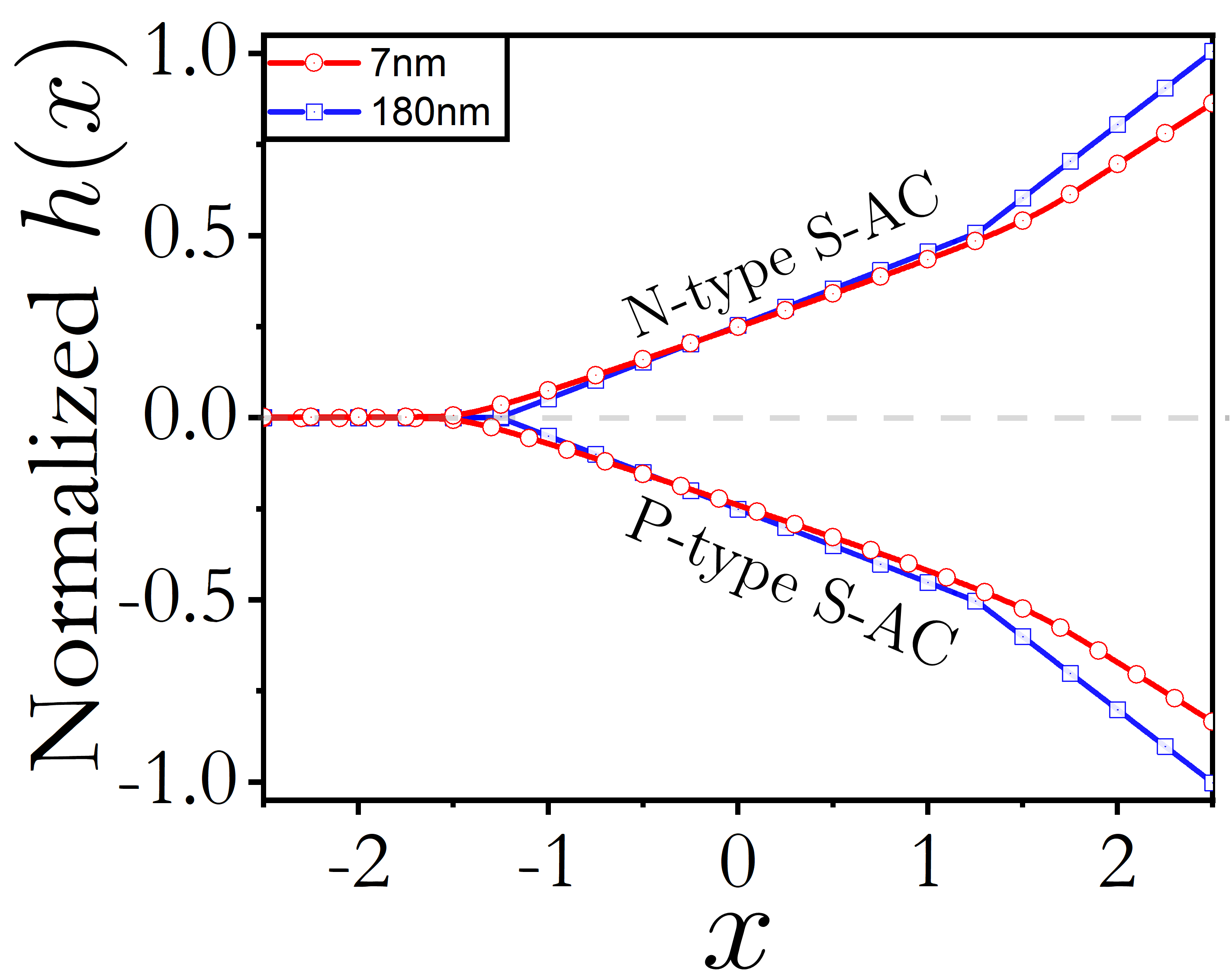}
		\label{protoS1}}
	\subfloat[]{\includegraphics[width=0.25\linewidth]{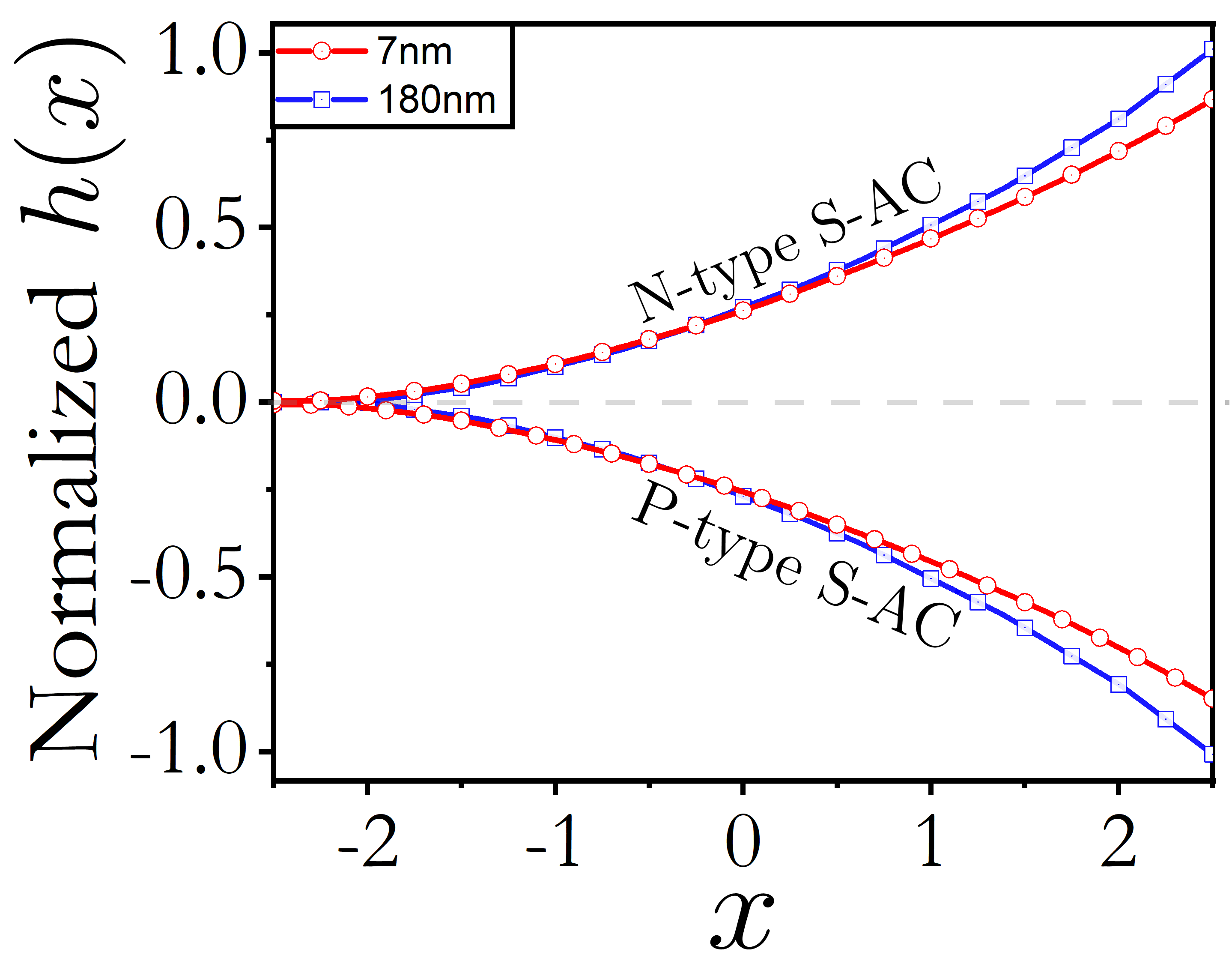}
	\label{protoS4}}
    \subfloat[]{\includegraphics[width=0.25\linewidth]{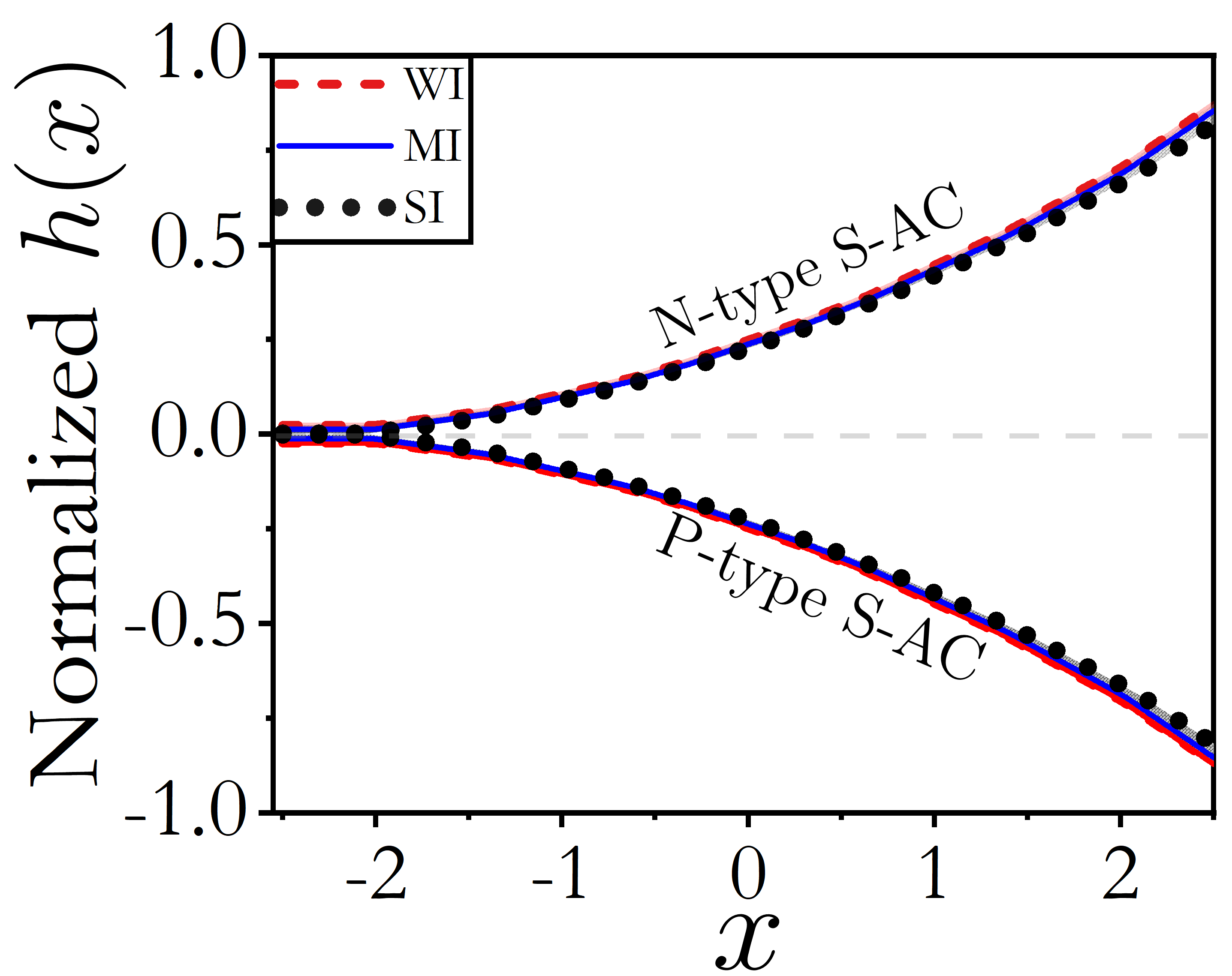}
    \label{protoS4_allregion_180nmvar}}  
        \subfloat[]{\includegraphics[width=0.25\linewidth]{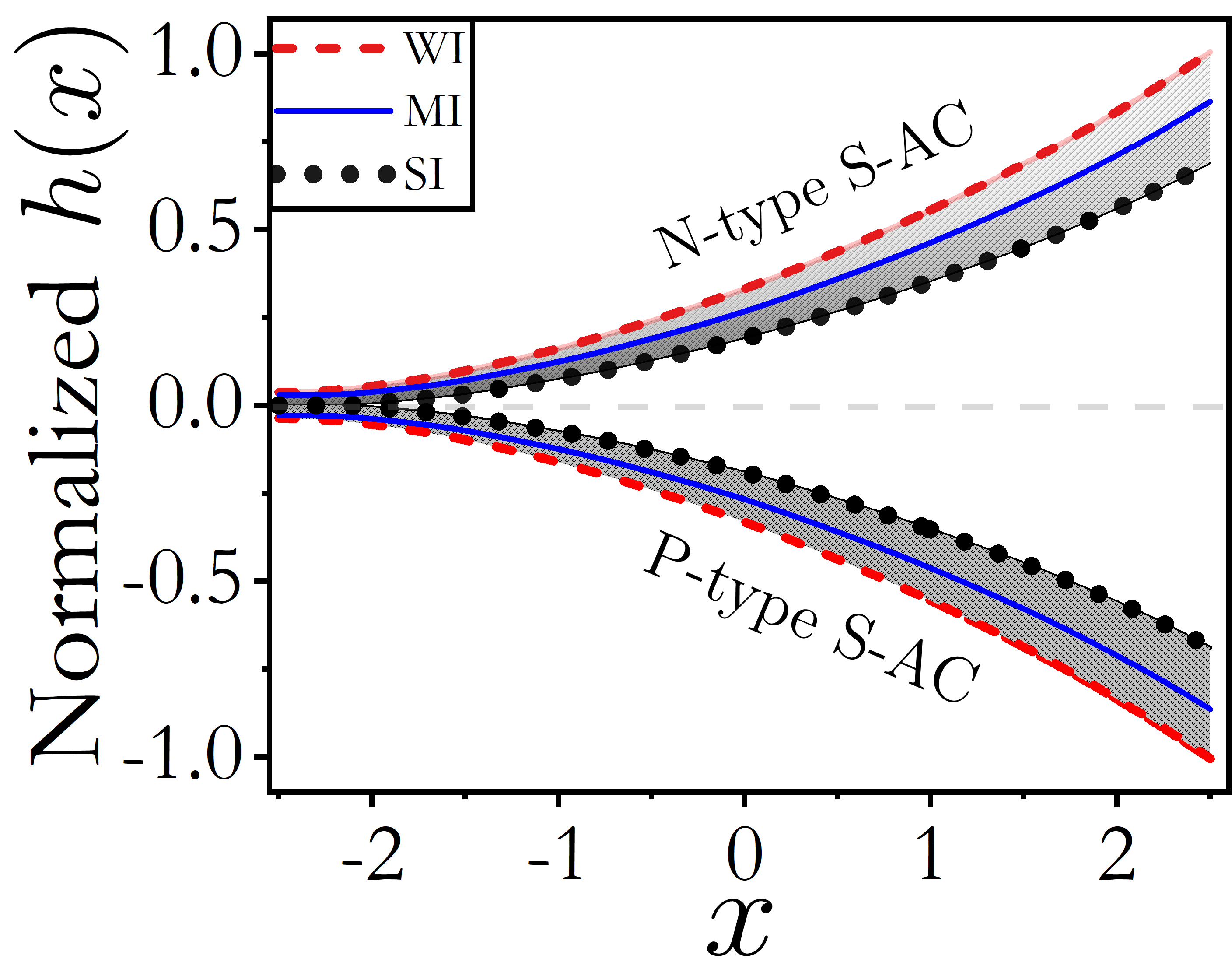}
	\label{protoS4_allregion_7nmvar}}

	\caption{Basic S-AC functions implemented by the N-type S-AC circuit and P-type S-AC circuit in different process technology nodes when a single input $x$ is varied: \protect\subref{protoS1} for spline-count $S = 1$;\protect\subref{protoS4} for spline-count $S = 3$; \protect\subref{protoS4_allregion_180nmvar}~{for different operating regimes in a 180nm process node; \protect\subref{protoS4_allregion_7nmvar}~for different operating regimes in a 7nm process node. The output current $h(x)$ shown in the plots have been normalized with respect to $I_{max}$, where $I_{max}$ is the maximum current for each biasing regime.}}
	\label{proto_shape}
\end{figure*}

\subsection{Generalized Margin Propagation and S-AC}\label{gmp} 

Both the log-sum-exp function $h_{\log}(.)$ and its multi-spline approximation $h(.)$ satisfy 
properties~\eqref{eq:lipschitz} and \eqref{eq:asymptote2} given by
\begin{equation}
	1 \ge \frac{\partial h_{\log}}{\partial x_i},\frac{\partial h}{\partial x_i} \ge 0, \forall i
	\label{eq:lipschitz}
\end{equation}
\begin{align}
	\begin{split}
		\mathop {\lim }\limits_{{x_i} \to \infty } \frac{{\partial h_{\log}}}{{\partial {x_i}}}, \frac{\partial h}{\partial x_i} = 1 \\
		\mathop {\lim }\limits_{{x_i} \to -\infty } \frac{{\partial h_{\log}}}{{\partial {x_i}}}, \frac{\partial h}{\partial x_i} = 0
	\end{split}
	\label{eq:asymptote2}
\end{align}
These properties indicate some common features of the shape of both $h_{\log}(.)$ and $h(.)$. It also indicates a mechanism to further generalize MP shapes $g(\cdot)$ instead of splines~\eqref{generic_msmp}. If a function $g: \mathbb{R} \rightarrow \mathbb{R}$ satisfies the following property 
\begin{itemize}
    \item $g(0)=0$ and $g(\cdot)$ is always positive or $g(\cdot) \ge 0$
    \item {$g'(\cdot) \ge 0$ and $g(-\infty)$ = 0}
    \item $g(\cdot)$ is monotonic function
\end{itemize}
then we propose a function $h$, which is computed as the solution to the following constraint 
\begin{equation}
   \sum\limits_{i = 1}^N \sum\limits_{j = 1}^S g(x_{i,j} - h) = C, \forall i = 1,..,N , \forall j = 1,..,S
    \label{char1_msmp_g}
\end{equation}
Since the choice of $g(.)$ is arbitrary, we refer to the function $h(.)$ as a shape-based function and any computation that exploits the constraint~\eqref{char1_msmp_g} as shape-based {analog} computing (S-AC).
In the next section, we show how $g(.)$ can be implemented using transistor and diode characteristics in a bias and process scalable manner.

\begin{figure*}[t]
	\centering
    \subfloat[]{\includegraphics[width=0.31\linewidth]{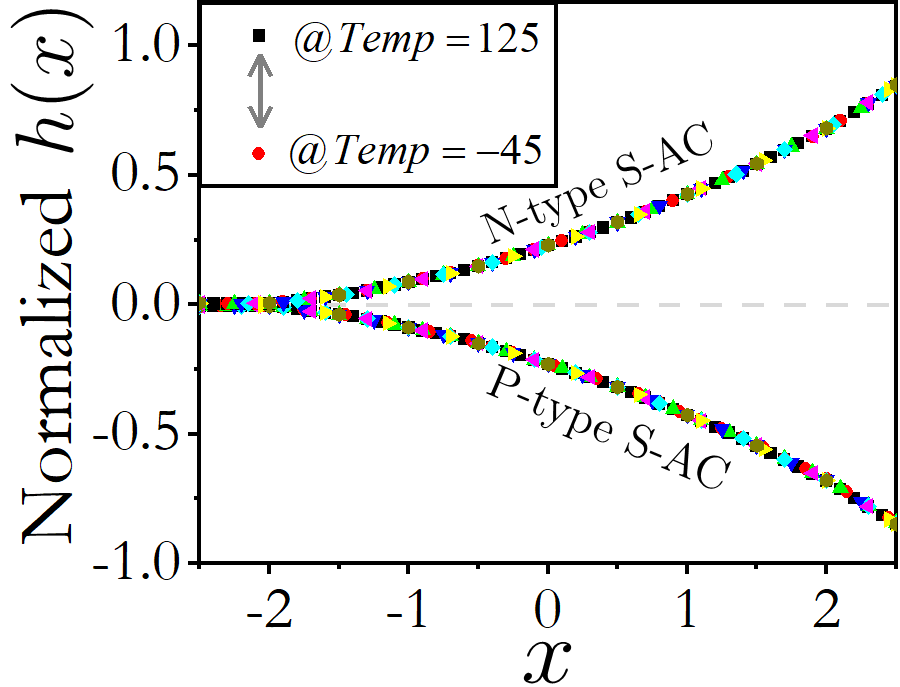}\label{temp_var}}
    \hspace{0.3 cm}
    \subfloat[]{\includegraphics[width=0.3\linewidth]{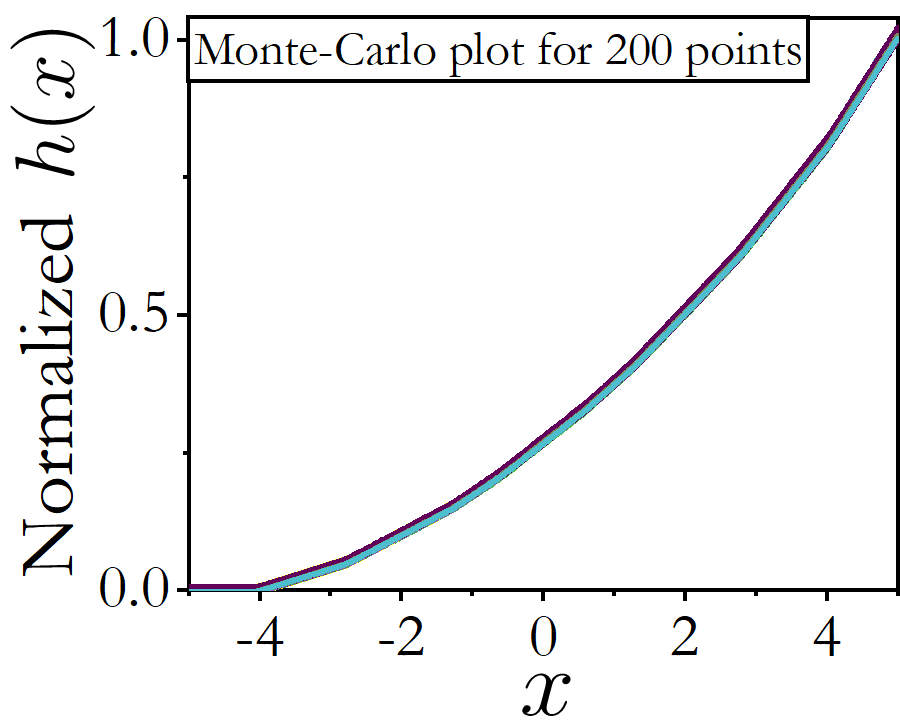}\label{monte_var}}
    \hspace{0.3 cm}
    \subfloat[]{\includegraphics[width=0.3\linewidth]{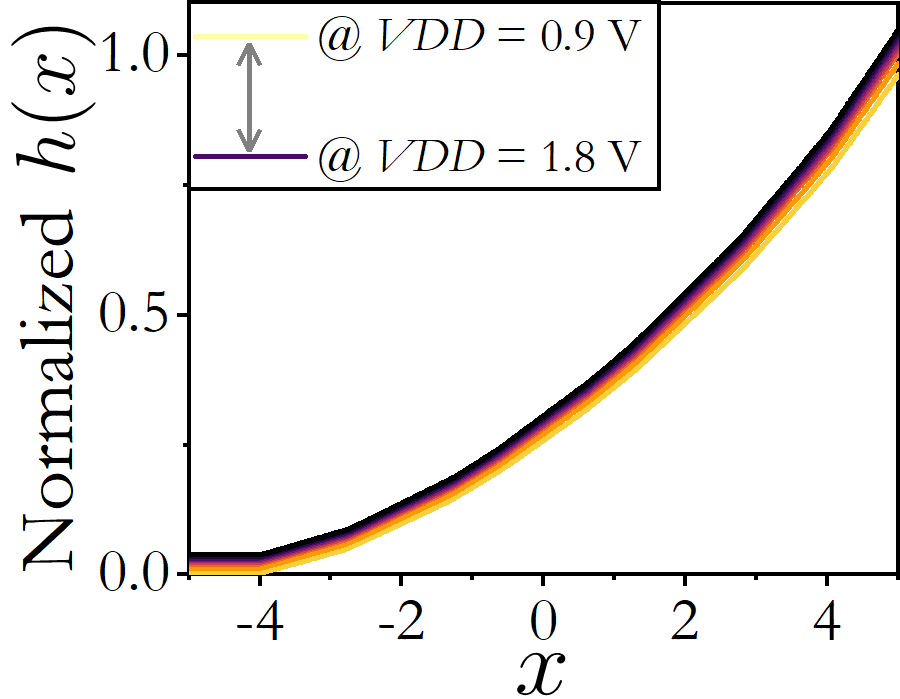}\label{power_var}}
    
	\caption{Basic S-AC functions implemented by the S-AC circuit in a 180nm process node when a single input $x$ is varied: \protect\subref{temp_var}~when the temperature is varied from $ - {45^ \circ }C$ to ${125^ \circ }C$ in 180nm; \protect\subref{monte_var}~in the presence of device mismatch (upto 5\% mismatch) - the plot shown here is for N-type S-AC circuit; \protect\subref{power_var}~in the presence of power supply voltage variation from $ {0.9 }V$ to ${1.8 }V$. The output current $h(x)$ shown in the plots have been normalized with respect to $I_{max}$, where $I_{max}$ is the maximum current for each of the biasing regime.}
	\label{proto_shape_var}
\end{figure*}
\begin{figure*}[t]
	\centering
	\subfloat[]{\includegraphics[width=0.3\linewidth]{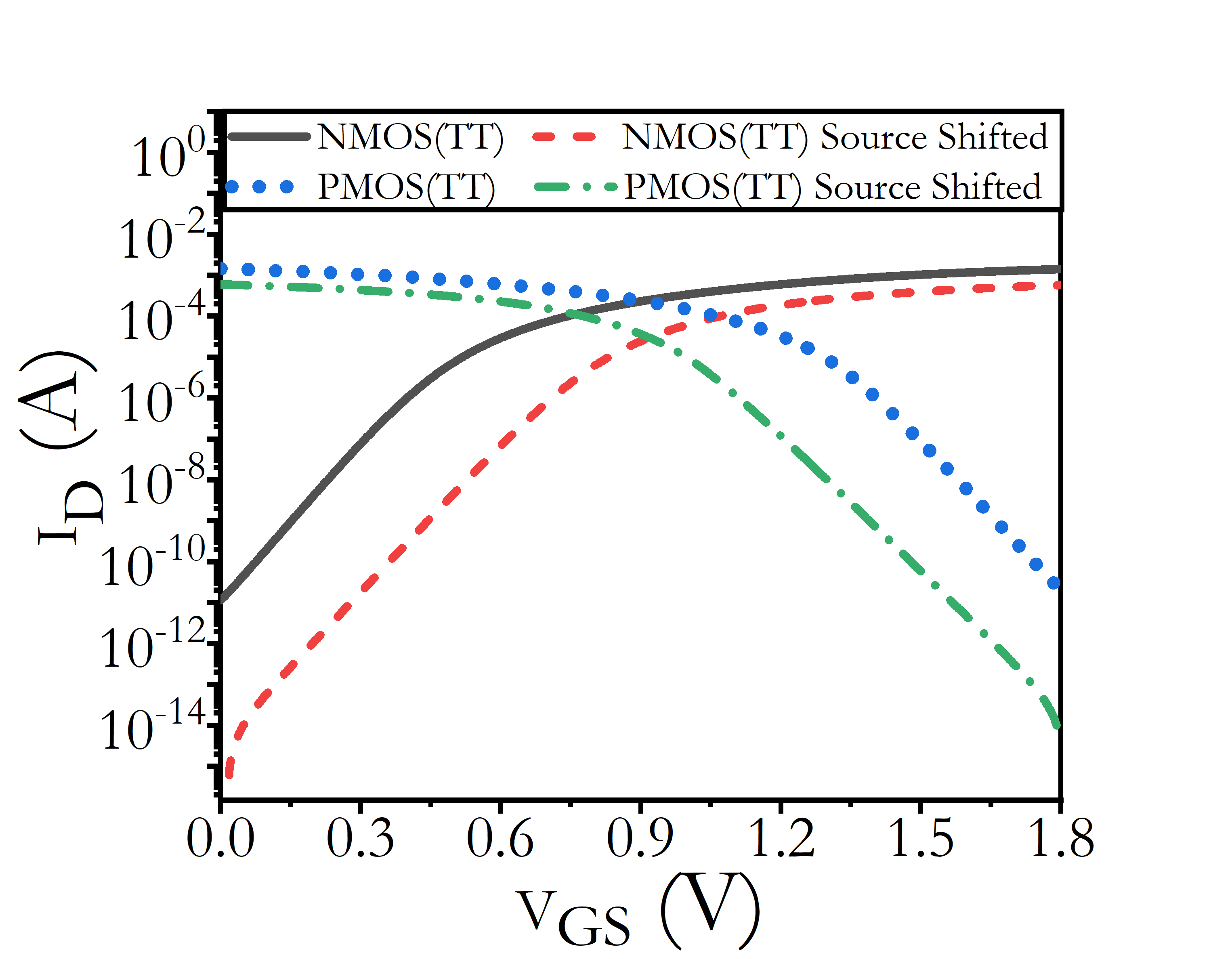}
		\label{sourceShifted_char}}
	\subfloat[]{\includegraphics[width=0.4\linewidth]{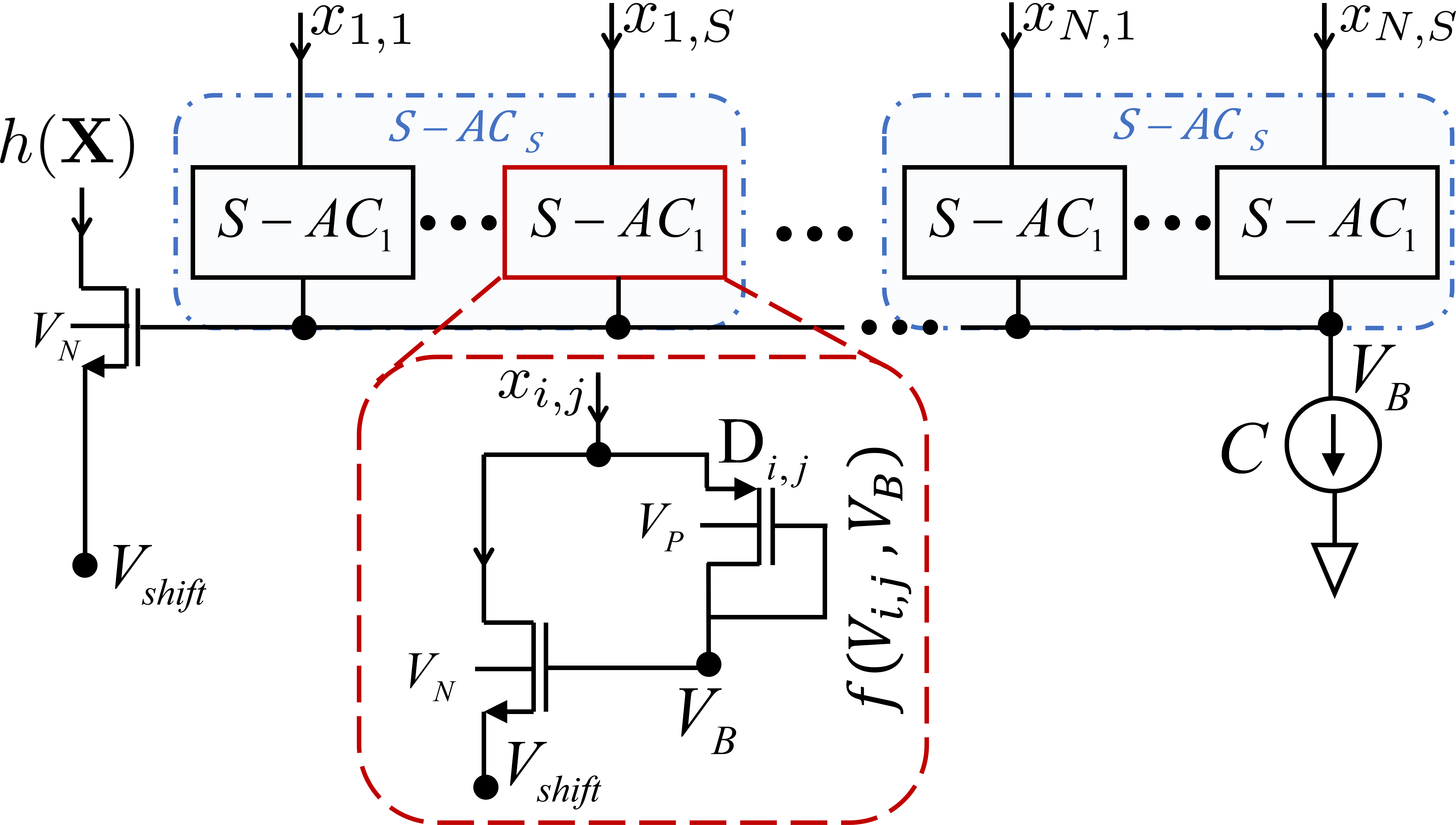}
		\label{sourceShifted_sac_ckt}}
	\subfloat[]{\includegraphics[width=0.3\linewidth]{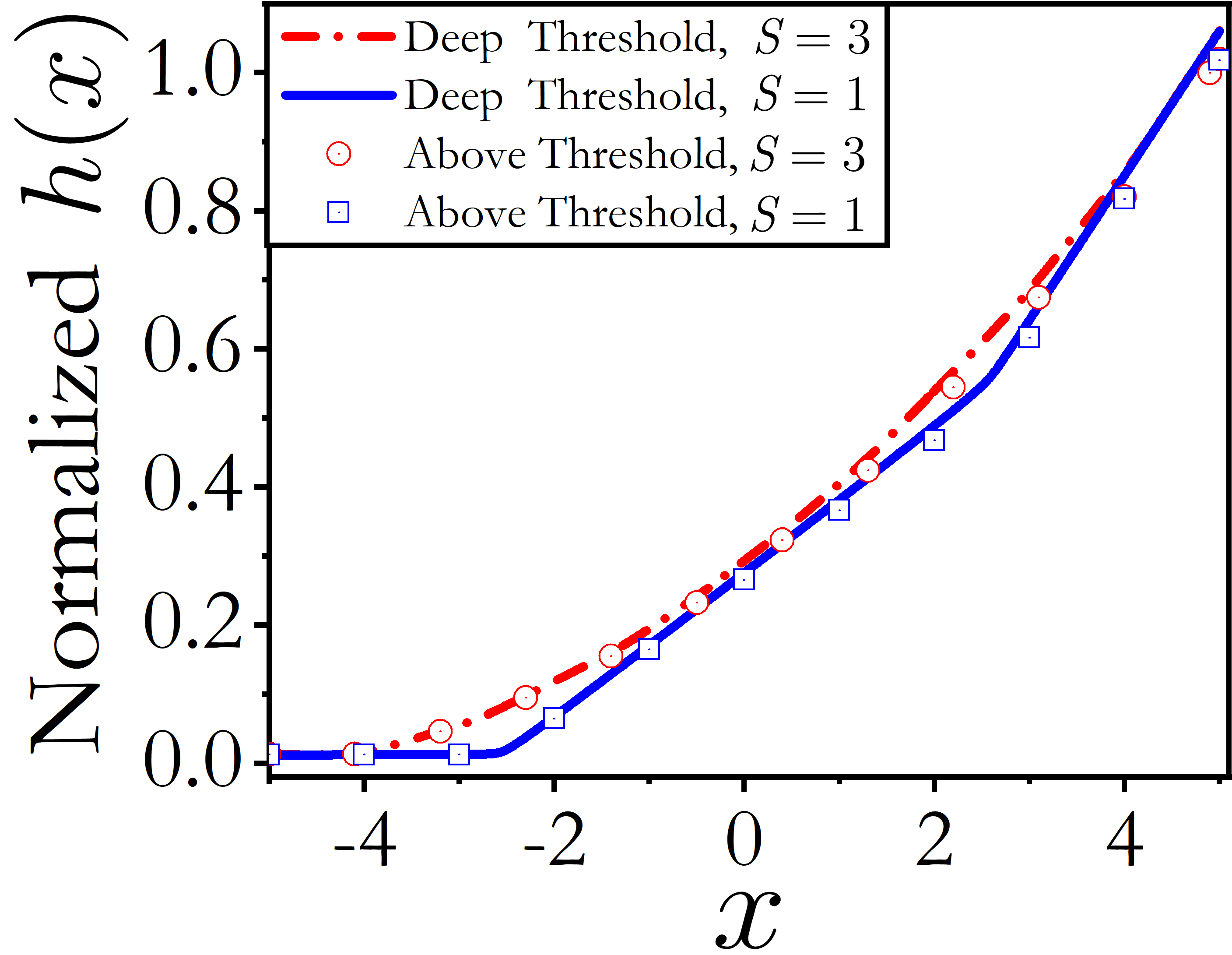}
		\label{sourceShiftedproto}}
	\caption{Results for S-AC circuit when operating in deep-threshold regime in a 180nm process: \protect\subref{sourceShifted_char}~{MOSFET I-V characteristics showing the effect of source shifting to lower the operating current into the diode leakage regime}; \protect\subref{sourceShifted_sac_ckt}~MOSFET implementation of the N-type S-AC circuit capable of operating in deep-threshold regime;  \protect\subref{sourceShiftedproto}~{Normalized output current response of the source-shifted S-AC circuit for $S=1,3$.}}
	\label{sourceShift}
\end{figure*}

\section{S-AC Circuit Implementation and Analysis}\label{sac_ckt_analy} 

\subsection{FET device characteristics and the basic S-AC Circuit}

In its most general form, the drain-to-source current ($I_{ds}$) flowing through an n-type MOSFET can be expressed as the difference between the forward and reverse currents~\cite{tsividis_mos,colinge2008finfets} as
\begin{equation}
    I_{ds} = I_s [f(V_g, V_s)-f(V_g, V_d)]
    \label{eq:mosfet}
\end{equation}
where $I_s$ is the specific current and $f:\mathbb{R}\times \mathbb{R} \rightarrow \mathbb{R}$ is a function that models the forward and reverse currents with respect to the gate ($V_g$), drain ($V_d$) and source ($V_s$) voltages respectively. A similar expression as~\eqref{eq:mosfet} also holds for a p-type MOSFET, except that the signs of the respective variables are reversed. Without any lack of generality, our analysis in this section will be based on the n-type MOSFET model; however, the formulation is applicable to p-type MOSFETs as well. It should also be noted that, as long as the source and the drain terminals are symmetric to each other, the expression in~\eqref{eq:mosfet} holds irrespective of the choice of transistor models such as EKV (Enz, Krummenacher, and Vittoz)~\cite{ekv}, ACM (Advanced Compact MOSFET)~\cite{acm}, etc. or operating regimes, i.e. weak-inversion, moderate-inversion or strong-inversion, or process nodes viz. MOSFET, FinFET, etc. The function $f(\cdot,\cdot)$ always satisfies the properties similar to $g(\cdot)$, and can be listed as:
\begin{itemize}
    \item $f(0,0)=0$ and $f(\cdot,\cdot)$ is always positive or $f(\cdot,\cdot) \ge 0$, by construction.
    \item {$f'(\cdot,\cdot) \ge 0$} 
    \item $f(\cdot,\cdot)$ is monotonic. For $V_{g1}>V_{g2}$, $f(V_{g1}, V_s)>f(V_{g2}, V_s)$ and for $V_{s1}>V_{s2}$, $f(V_{g},V_{s1})<f(V_{g}, V_{s2})$.
\end{itemize}
Thus, {$f(\cdot,\cdot)$} and hence the FETs could be used to implement S-AC function as follows: 

Given an input matrix $X\in\mathbb{R}^{N\times S}$ where $\textbf{x}_{i} \in \mathbb{R}^{N}$, $\forall i = 1,..,N$ is the input vector and $\textbf{x}_{j}\in \mathbb{R}^{S}$, $\forall j = 1,..,S$ is the number of splines, the basic shape of S-AC $h:\mathbb{R}^{N \times S} \rightarrow \mathbb{R}$ can be computed as a solution to the equation $h$(X)$ = f(V_B,0)$ where the variable $V_B$ is the solution to: 
\begin{equation}
    \sum\limits_{i = 1}^N {\sum\limits_{j = 1}^S f ({V_{i,j}},{V_B})}  = C, \forall i = 1,..,N, \forall j = 1,..,S
    \label{char1_msmp}
\end{equation}
\begin{equation}
    f({V_B},0) - f({V_B},{V_{i,j}}) + f({V_{i,j}},{V_B}) = {x_{i,j}}
    \label{char2_msmp}
\end{equation}
Here, $C$ is a hyper-parameter and $V_{i,j}$ is an internal variable. {Equation \eqref{char2_msmp} can also be written as
\begin{equation}
    f({V_{i,j}},{V_B}) - f( - {V_{i,j}},{V_B}) = {x_{i,j}} - f({V_B},0)
    \label{char3_msmp}
\end{equation}
A closer look at the derivatives of $f'({V_{i,j}} - {V_B})$ and $f'( - {V_{i,j}} + {V_B})$ show that these are intersecting curves with respect to ${x_{i,j}}$. The point of intersection defines a unique solution. Further, the left-hand side of \eqref{char4_msmp} is monotonic by definition of $f'( \cdot , \cdot )$, thus when $\frac{{d{x_{i,j}}}}{{d{V_{i,j}}}} \ge 0$,  $f({V_B})$ is the solution.
\begin{equation}
    \frac{{d{x_{i,j}}}}{{d{V_{i,j}}}} - f'({V_B,0}) = f'({V_{i,j}} - {V_B}) + f'( - {V_{i,j}} + {V_B})
    \label{char4_msmp}
\end{equation}
}

Equations~\eqref{char1_msmp} and \eqref{char2_msmp} can be implemented using CMOS circuits as shown in Fig.~\ref{n_sac_ckt}. Here, $x_{i,j}$ is the input current for the $i^{th}$ input and the $j^{th}$ spline and $h$(X) is the output current. $V_{i,j}$ and $V_B$ are the voltages across the $N_{i,j}^{th}$ transistor, $C$ is a constant current and $D_{i,j}$ denotes diode elements (Schottky, MOS diode or any other). {In addition, the S-AC circuit implementation with the GMP algorithm satisfies additional constraints such that the inputs must add up to a constant $C$. Thus, based on Fig.~\ref{n_sac_ckt}, the current through the diode $D_{i,j}$ as a function of the voltages can be given by $f(V_{i,j}-V_{B})$ where all the currents sum up to $C$. Then using KCL} at node $V_B$,~\eqref{char1_msmp} can be obtained while the current across diode $D_{i,j}$ gives~\eqref{char2_msmp}. A similar operation can be obtained in other quadrant using the PMOS variant shown in Fig.~\ref{p_sac_ckt}.

\subsection{Process and Temperature Scalability of Basic S-AC Circuit}

We first show that the basic S-AC shape-function implemented by circuits in Fig.~\ref{n_sac_ckt} and Fig.~\ref{p_sac_ckt} are robust to changes in biasing conditions and operating temperature. Fig.~\ref{protoS1} shows the proto-shape $h(x)$ obtained using the circuits shown in Fig.~\ref{n_sac_ckt} (N-type S-AC) and Fig.~\ref{p_sac_ckt} (P-type S-AC) for spline count $S=1$. Similar results are shown for spline count $S=4$ in Fig.~\ref{protoS4}. It can be noted that with the increase in the number of splines, the approximation accuracy increases while the basic S-AC function remains scalable across process-technology nodes. Fig.~\ref{protoS4_allregion_180nmvar} shows the similar shape-function (for $N=1$ and $S=3$ ) obtained for different MOSFET biasing regimes, i.e., WI, MI, and SI biasing regimes which correspond to different functions {$f(\cdot, \cdot)$} in~\eqref{char1_msmp}-\eqref{char2_msmp} in 180nm node. The plots show that the basic shape of S-AC remains robust to the biasing condition and is constrained within a well-defined "margin". This margin is determined by the design parameter $S$ and the inherent feedback from the hyper-parameter $C$. {The number of splines $S$ provides control over the precision and accuracy while $C$ is the constraint that is satisfied by \eqref{equation18wc_DUALsUM}  or its circuit equivalent given by \eqref{char1_msmp}.} Fig.~\ref{protoS4_allregion_7nmvar} shows similar response plots for 7nm process node. Fig.~\ref{temp_var} shows the effect of temperature on the shape function. It can be observed that the shape-function is almost immune to changes in temperature. Fig.~\ref{monte_var} shows the effect of Monte Carlo analysis on the basic shape of N-type S-AC, while Fig.~\ref{power_var} shows the effect of power supply variation in WI regime. It can be observed that the shape remains preserved despite variations in analysis.

\begin{figure*}[t]
	\centering
	\subfloat[]{\includegraphics[width=0.40\linewidth]{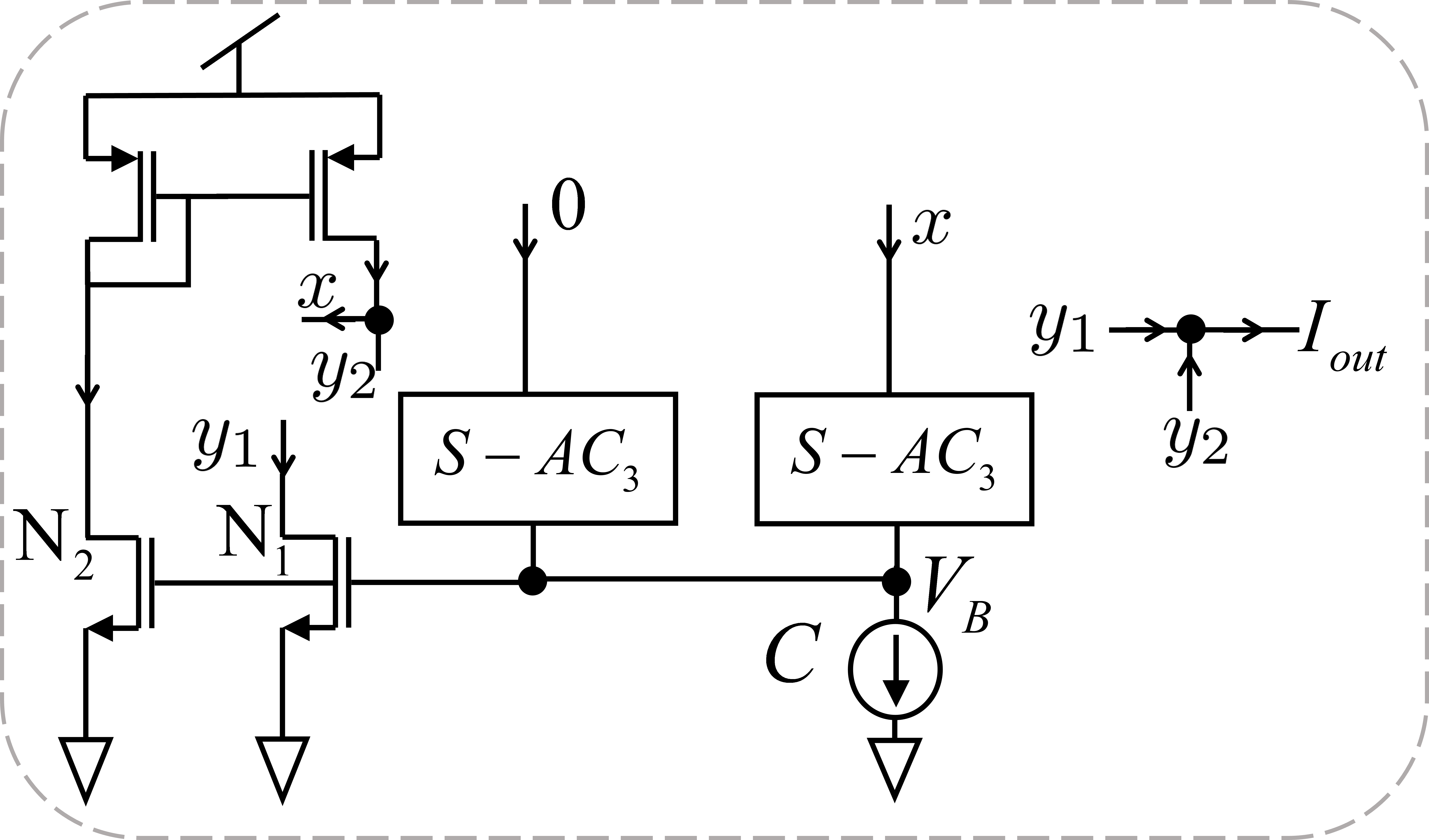}
		\label{cosh_ckt}}
	\subfloat[]{\includegraphics[width=0.28\linewidth]{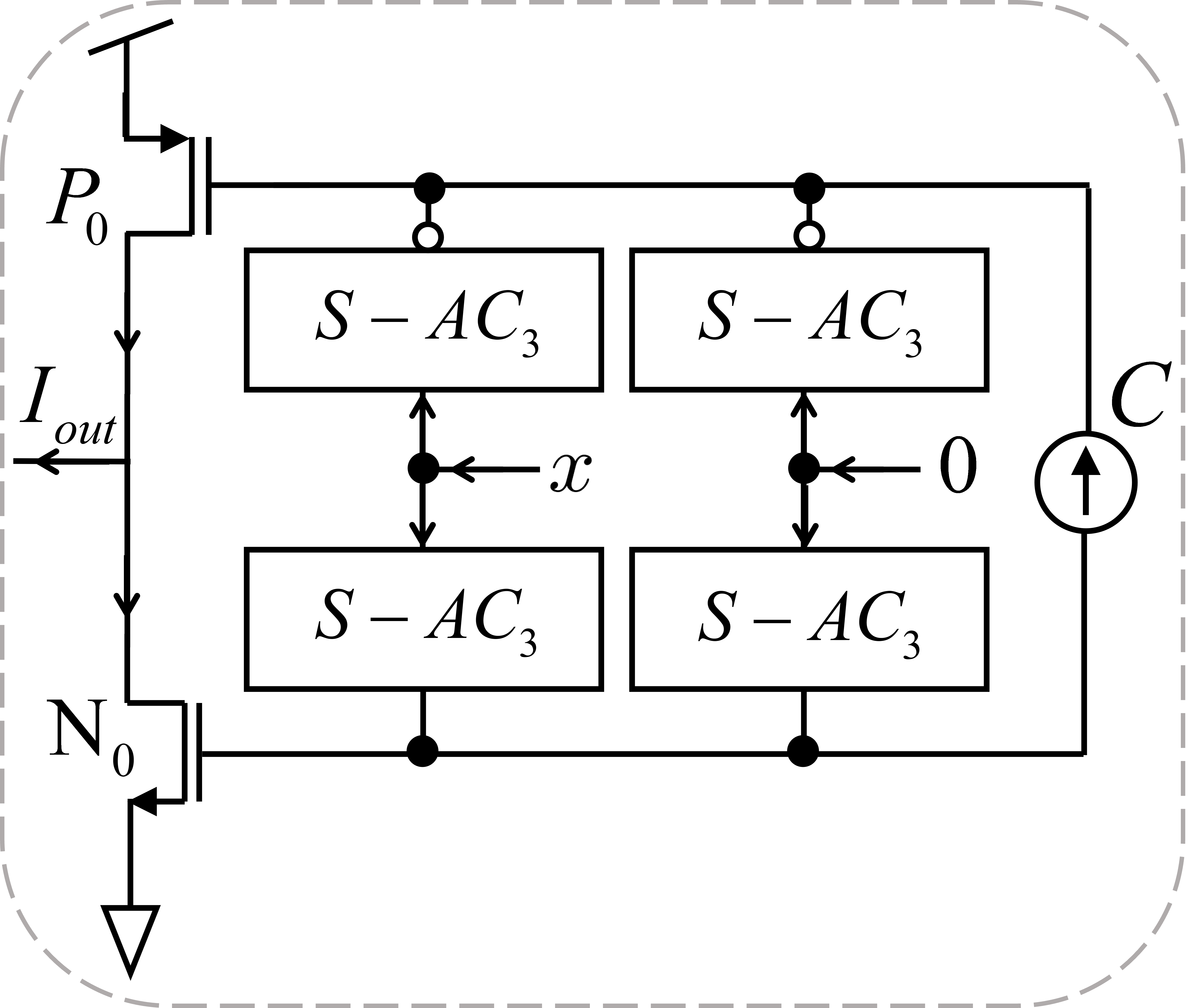}
		\label{sinh_ckt}}
	\subfloat[]{\includegraphics[width=0.30\linewidth]{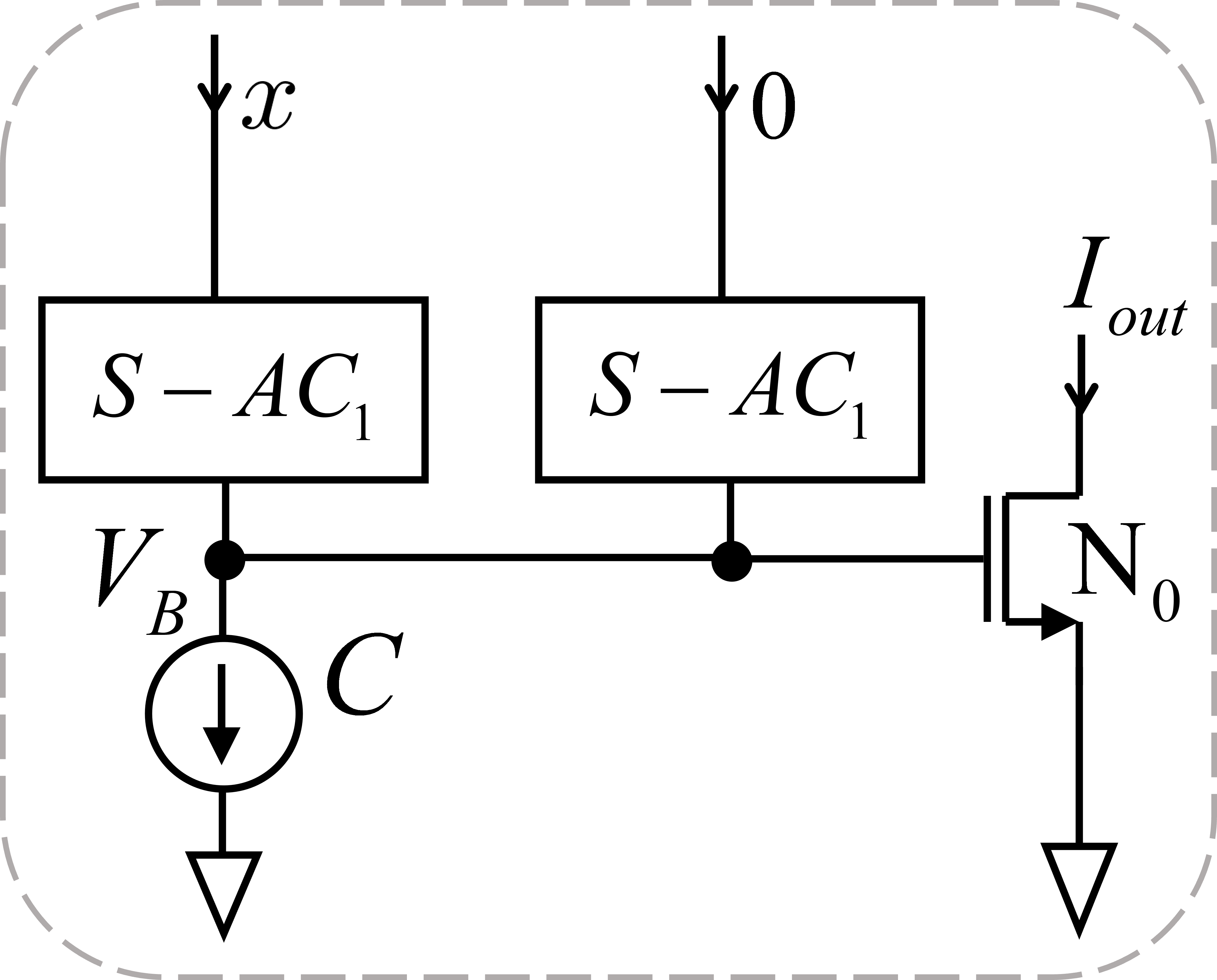}	\label{relu_ckt}}

	\subfloat[]{\includegraphics[width=0.66\linewidth]{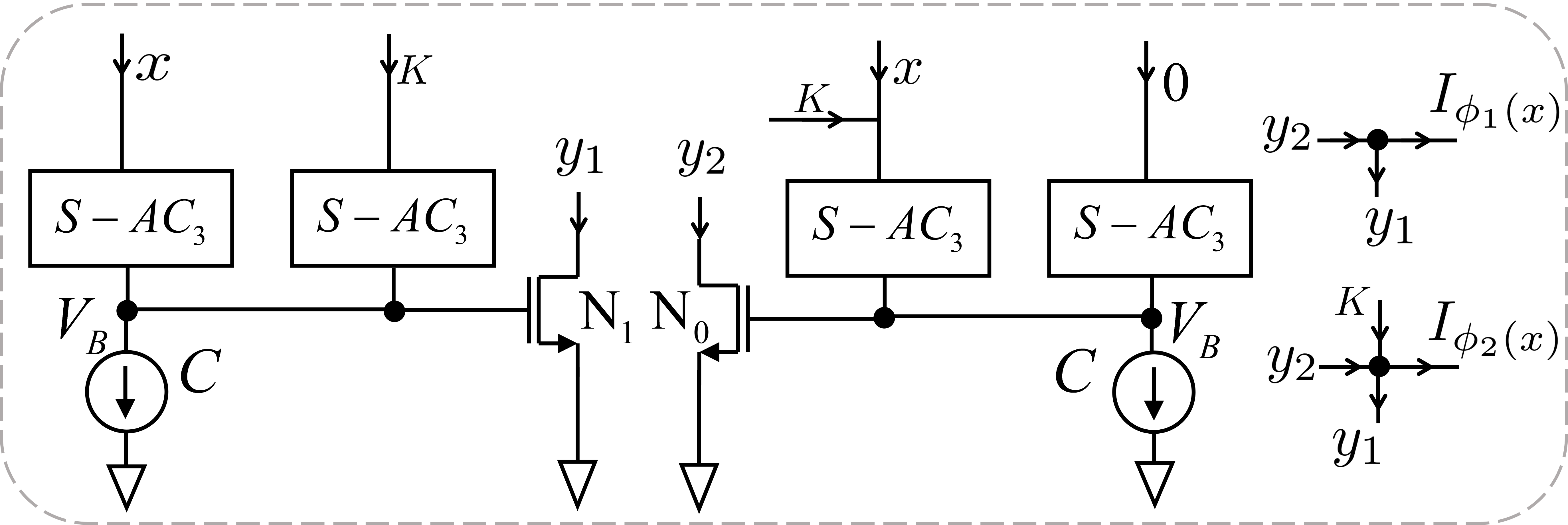}
		\label{tanh_ckt}}
		\hspace{0.15 cm}
	\subfloat[]{\includegraphics[width=0.275\linewidth]{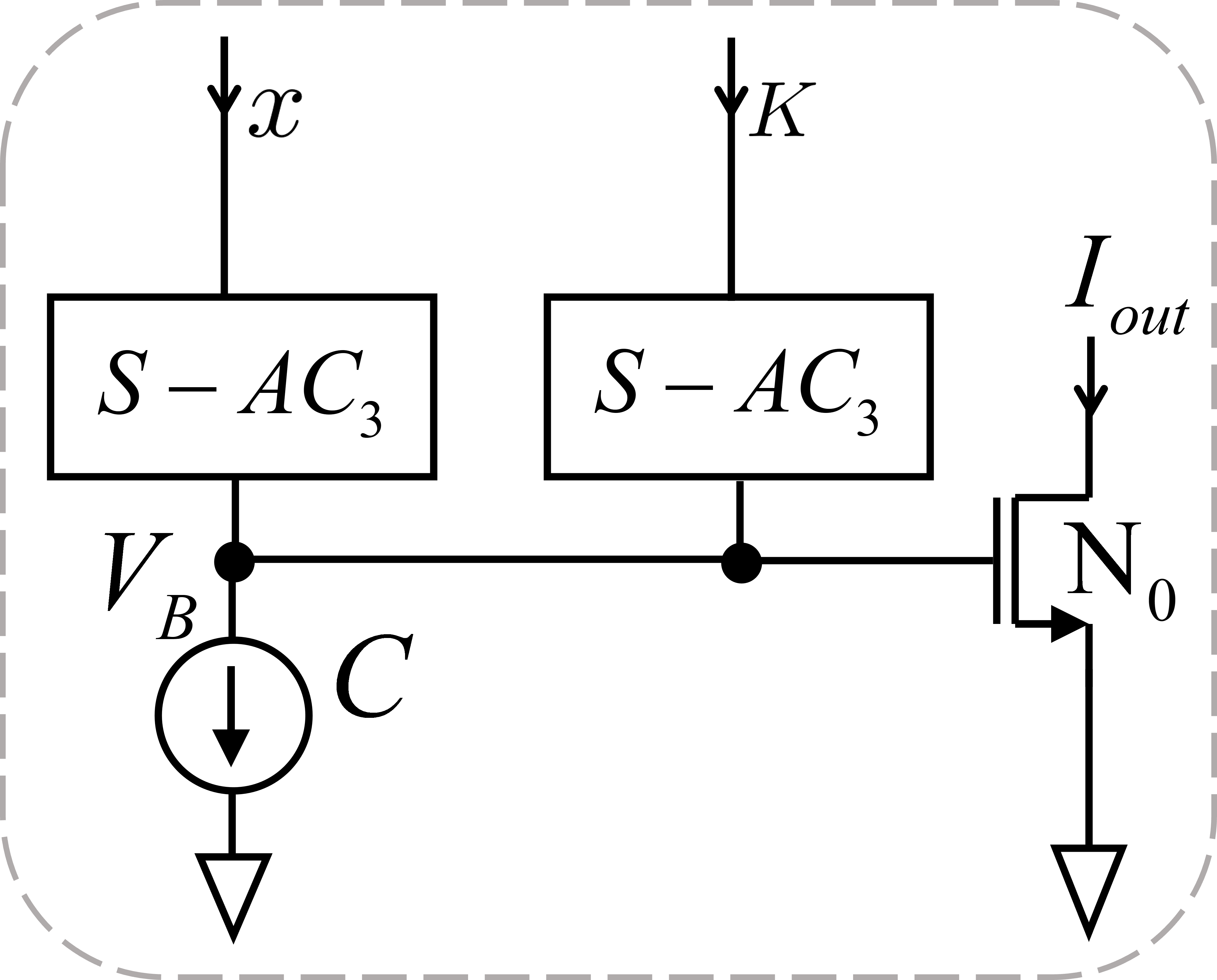}
		\label{softPlus_ckt}}
	\hfil
	\caption{Implementation of S-AC based analog activation standard cells for machine learning applications: \protect\subref{cosh_ckt}~$\cosh(\cdot)$;  \protect\subref{sinh_ckt}~$\sinh(\cdot)$; \protect\subref{relu_ckt}~ReLU;  \protect\subref{tanh_ckt}~Compressive non-linearity; $\phi_1(\cdot)$ similar to $\tanh(\cdot)$ and $\phi_2(\cdot)$ similar to Sigmoid functions respectively;  \protect\subref{softPlus_ckt}~Soft-Plus. Here $K$ represents a constant current.}
	\label{nonLin_ckt}
\end{figure*}

\subsection{Deep-threshold Operation of Basic S-AC Circuit}
To exploit the complete available current range of transistors in planar CMOS, two approaches can be followed. First, the $V_{GS}$ could be made negative; second, the threshold voltage, $V_{T0}$, could be increased in a way that the channel inversion occurs at higher $V_{GS}$ voltages~\cite{bernabe_fA,Baghtash_aA,azhari_fA}. Therefore, to bias the transistors at smaller currents, $V_{GS}$ should be biased to the reverse direction ($V_{GS} < 0V$ for NMOS and $V_{GS} > 0V$ for PMOS), which is defined as the deep sub-threshold region. For this approach, the source voltage shifting technique can be used. By shifting the source voltage slightly higher than the lowest potential ($V_{SS}$), the gate voltage can reach down to the lowest potential ($V_{SS}$). Fig.~\ref{sourceShifted_char} shows the $I_D$ vs. $V_{GS}$ characteristic plot in log scale for source shifted MOS transistors. The lowest value of current found with the source shifting technique was $1.97$fA for NMOS and $3.19$fA for PMOS in CMOS 180nm process node. On analyzing Fig.~\ref{sourceShifted_char} one can see that the source voltage shifting can exploit the complete available current range (down to the diffusion diodes reverse leakage currents).

For the second approach (so-called channel conduction manipulating technique), the body terminals of transistors are connected to the highest potential ($V_{DD}$). This will prevent the channel inversion from taking place at low $V_{GS}$ voltages which, along with source shifting, further lowers the lowest level of the operating current of the circuit. We used a technique that combines a fixed source shifted voltage with the channel conduction manipulation technique. Fig.~\ref{sourceShifted_sac_ckt} shows the circuit implementation of equations~\eqref{char1_msmp}-\eqref{char2_msmp} using the above two approaches combined, where source voltage modulation along with channel conduction manipulation techniques were used to shift the operation in femto-ampere (fA). Fig.~\ref{sourceShiftedproto} shows the response of the basic S-AC circuit operating in the deep-threshold at current levels down to femto-amperes (fA). The results show that if biased properly, S-AC can operate at ultra-low current levels and yet its characteristics are maintained.

\section{Synthesis of Analog Computing Modules using Basic Proto-shapes}\label{designs}
S-AC circuits shown in Fig.~\ref{n_sac_ckt} and Fig.~\ref{p_sac_ckt} can be used to synthesize mathematical functions required to perform complex machine learning tasks which can then be used as analog standard cells. As a proof of concept, we show the S-AC-based implementation of various machine learning computations such as activation functions, energy-efficient multiplication, Winner-Takes-all (WTA), N-of-M encoder, SoftArgMax, and Max circuits.  It can be noted that the proto-shape $h(x)$ can be mapped to \textit{exponential} for $e^x$ ranging from $\left[ { - \infty ,1} \right)$. Within this defined range i.e. $\left| {\frac{{\partial h}}{{\partial x}}} \right| < 1 $, if ~\eqref{eq:lipschitz} and ~\eqref{eq:asymptote2} are satisfied, then complex functions such as hyperbolic functions (\textit{Cosine, Sine}) and other such functions can also be constructed. The following text has a detailed formulation and S-AC based implementation of each of these circuits in both 180nm and 7nm process technology nodes.

\begin{figure*}[t]
	\centering
	\subfloat[]{\includegraphics[width=0.3\linewidth]{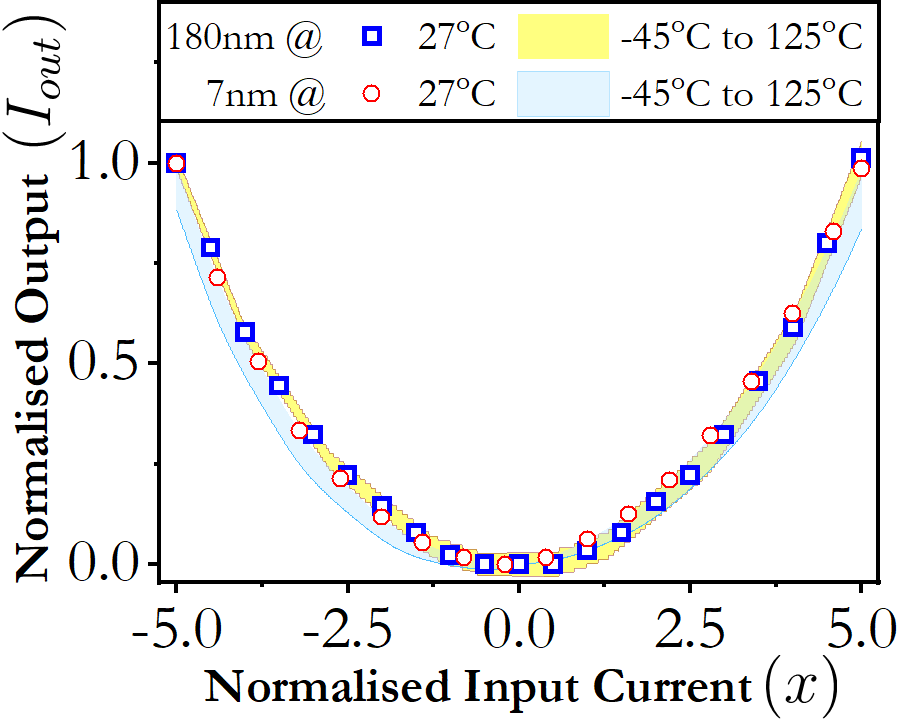}
		\label{cosh}}
	\subfloat[]{\includegraphics[width=0.3\linewidth]{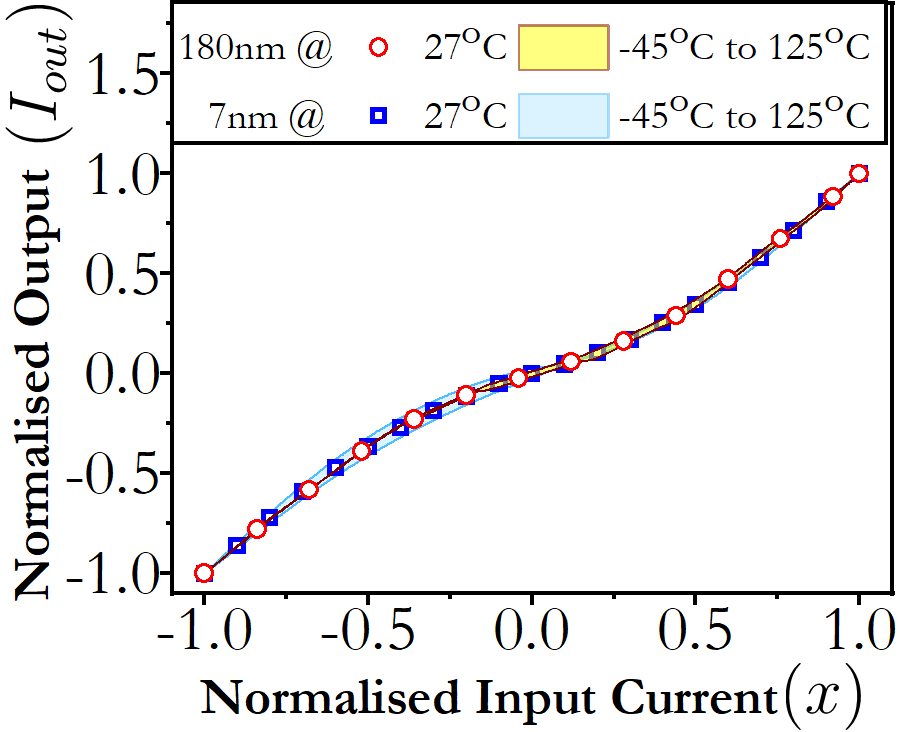}
		\label{sinh}}
	\subfloat[]{\includegraphics[width=0.3\linewidth]{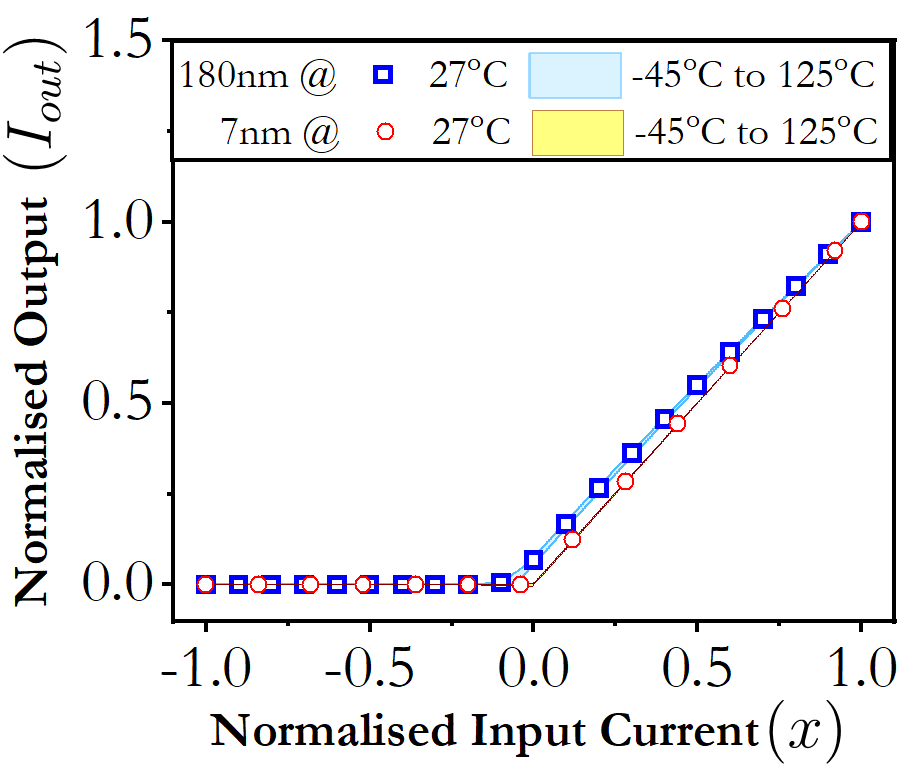}
		\label{relu}}

	\subfloat[]{\includegraphics[width=0.3\linewidth]{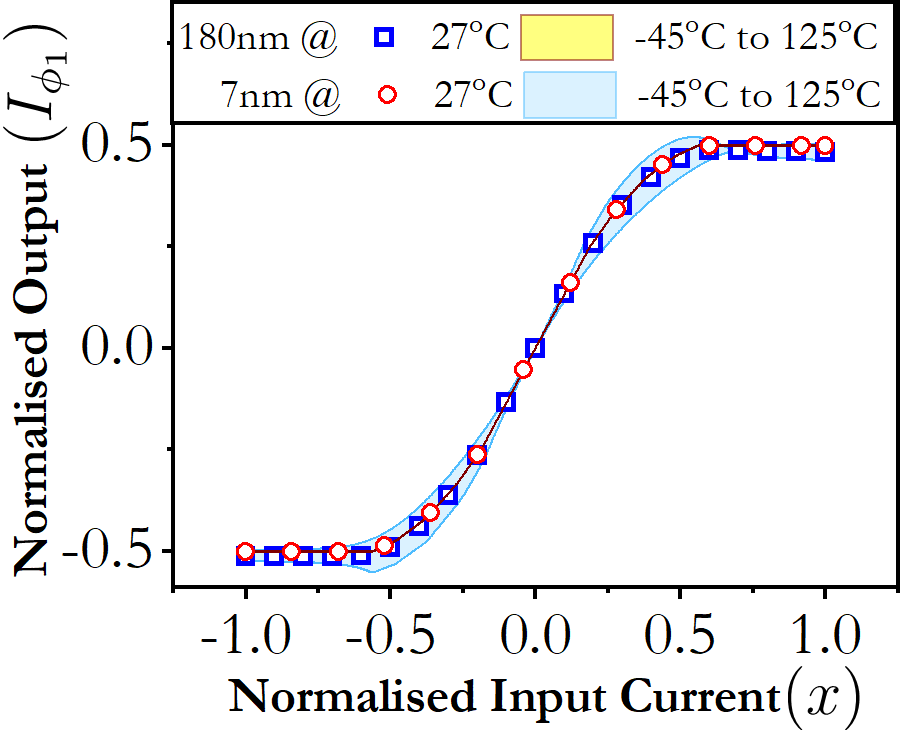}
		\label{tanh}}
	\subfloat[]{\includegraphics[width=0.3\linewidth]{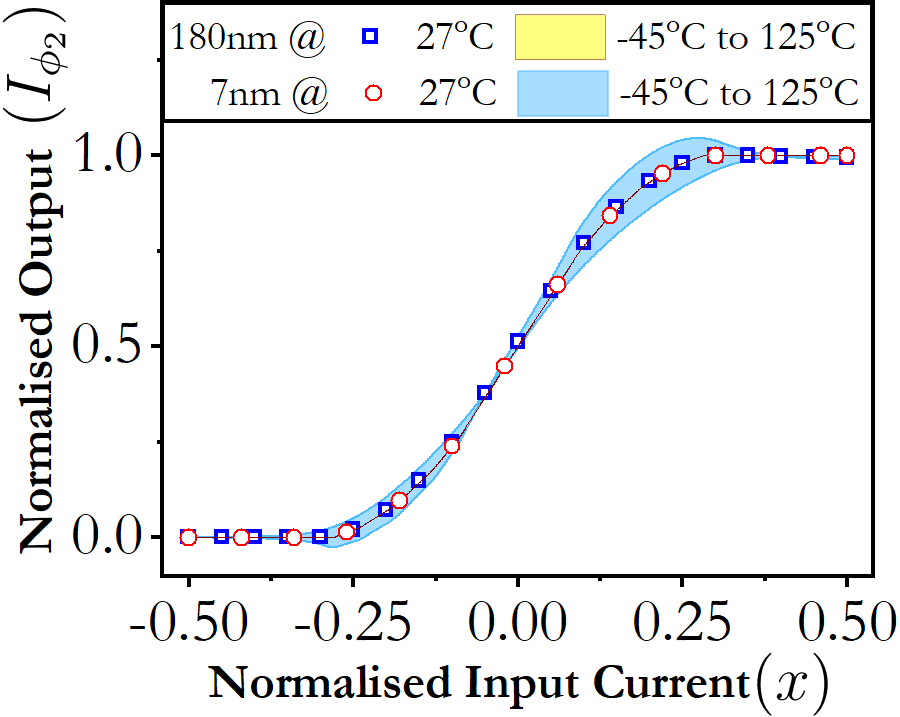}
		\label{sigmoid}}
	\subfloat[]{\includegraphics[width=0.3\linewidth]{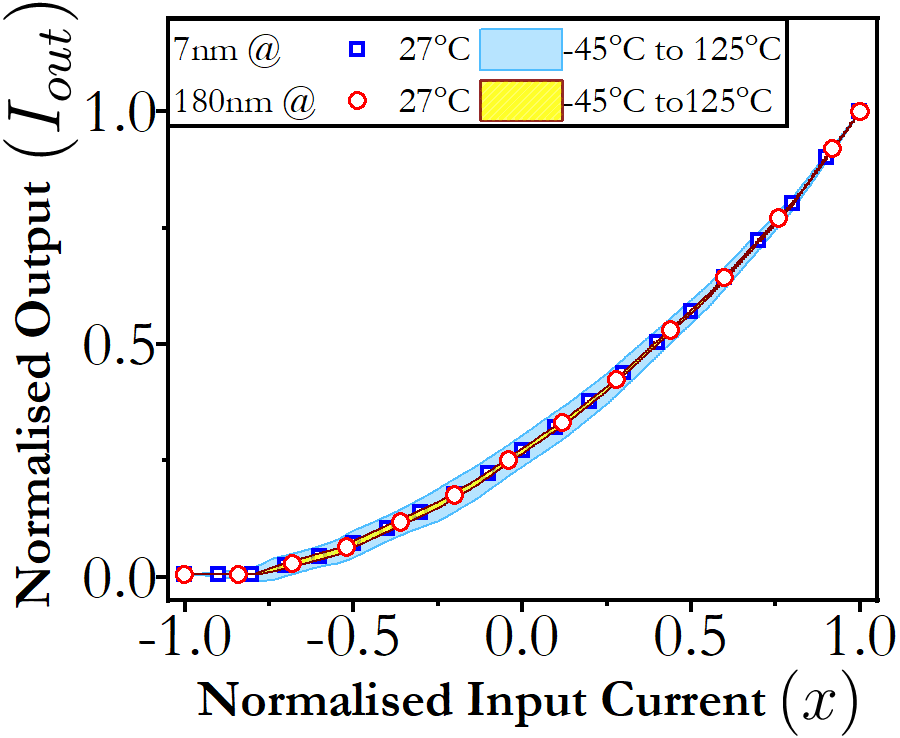}
		\label{softPlus}}
	\hfil
	\caption{Simulated output corresponding to the S-AC standard cells shown in Fig.~\ref{nonLin_ckt}:  \protect\subref{cosh}~$\cosh(\cdot)$; \protect\subref{sinh}~$sinh(\cdot)$; \protect\subref{relu_ckt}~ReLU; \protect\subref{tanh}~Compressive non-linearity $\phi_1(\cdot)$ equivalent to $\tanh(\cdot)$; \protect\subref{sigmoid}~Compressive non-linearity $\phi_2(\cdot)$ equivalent to Sigmoid; \protect\subref{softPlus}~Soft-Plus at FinFET (7nm) and CMOS (180nm) process nodes.}
	\label{nonLin_curve}
\end{figure*}

\begin{figure*}[t]
	\centering
	\subfloat[]{\includegraphics[width=0.30\linewidth]{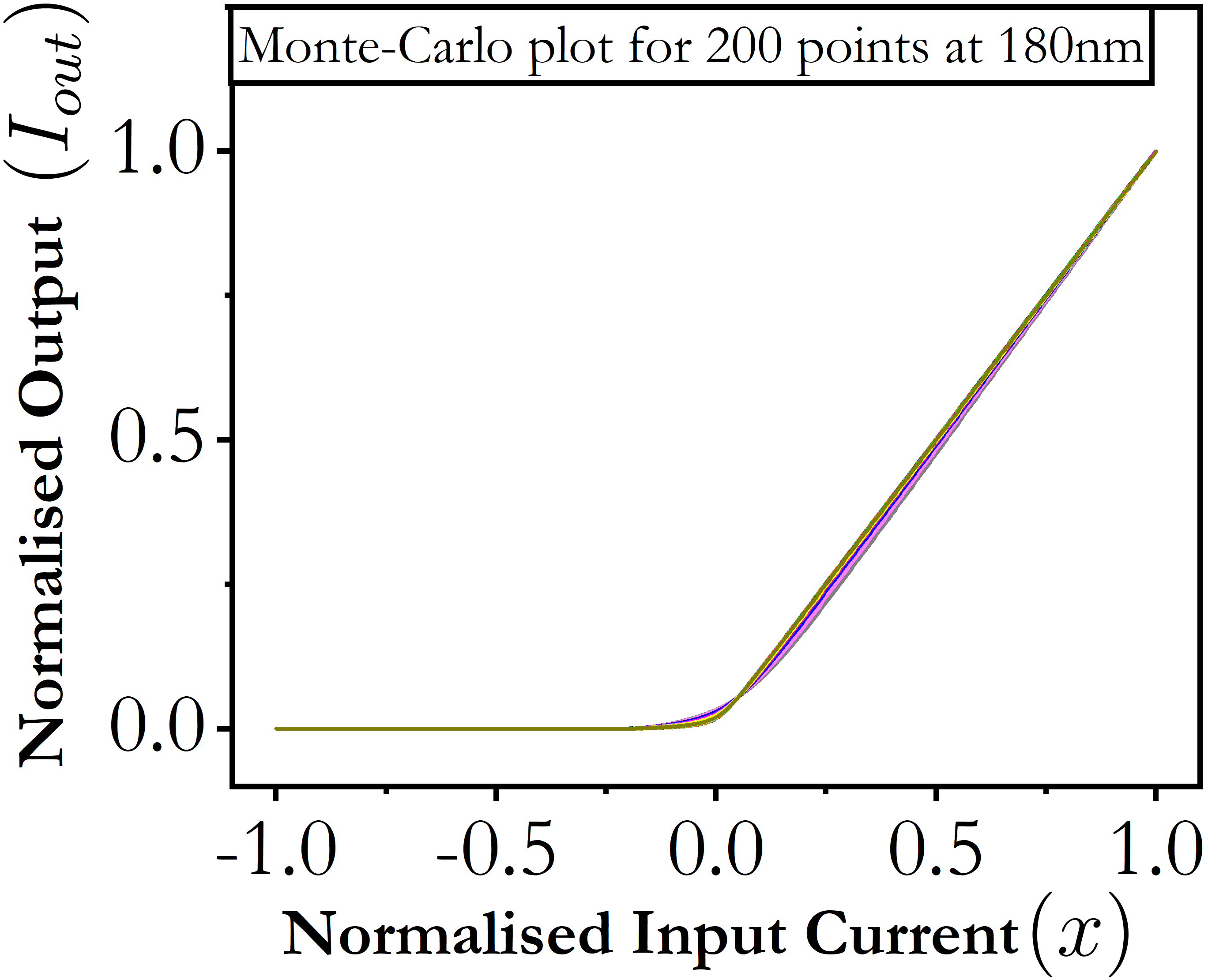}
		\label{mc_relu_180}}
	\subfloat[]{\includegraphics[width=0.30\linewidth]{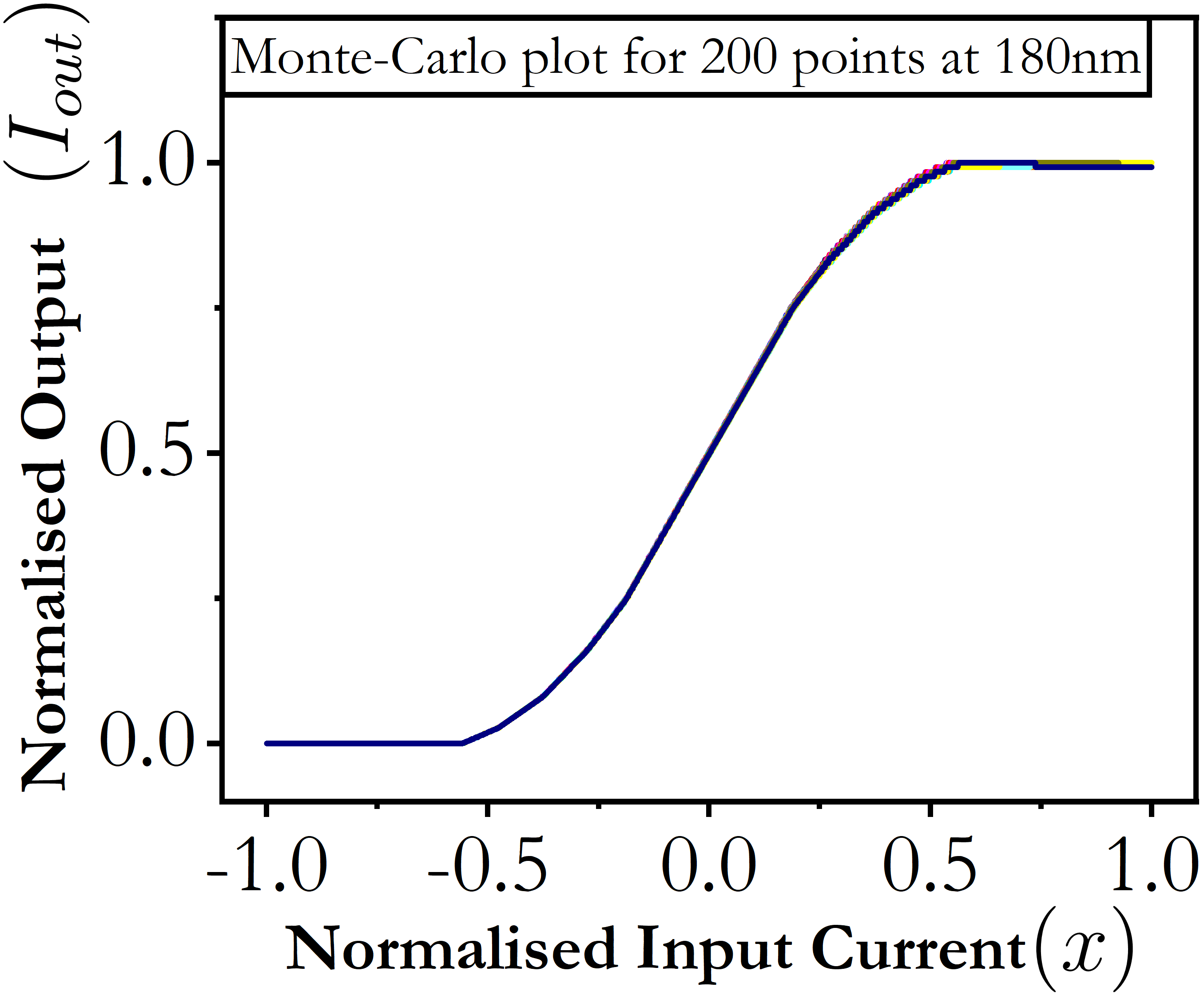}
		\label{mc_sigmoid_180}}
	\subfloat[]{\includegraphics[width=0.30\linewidth]{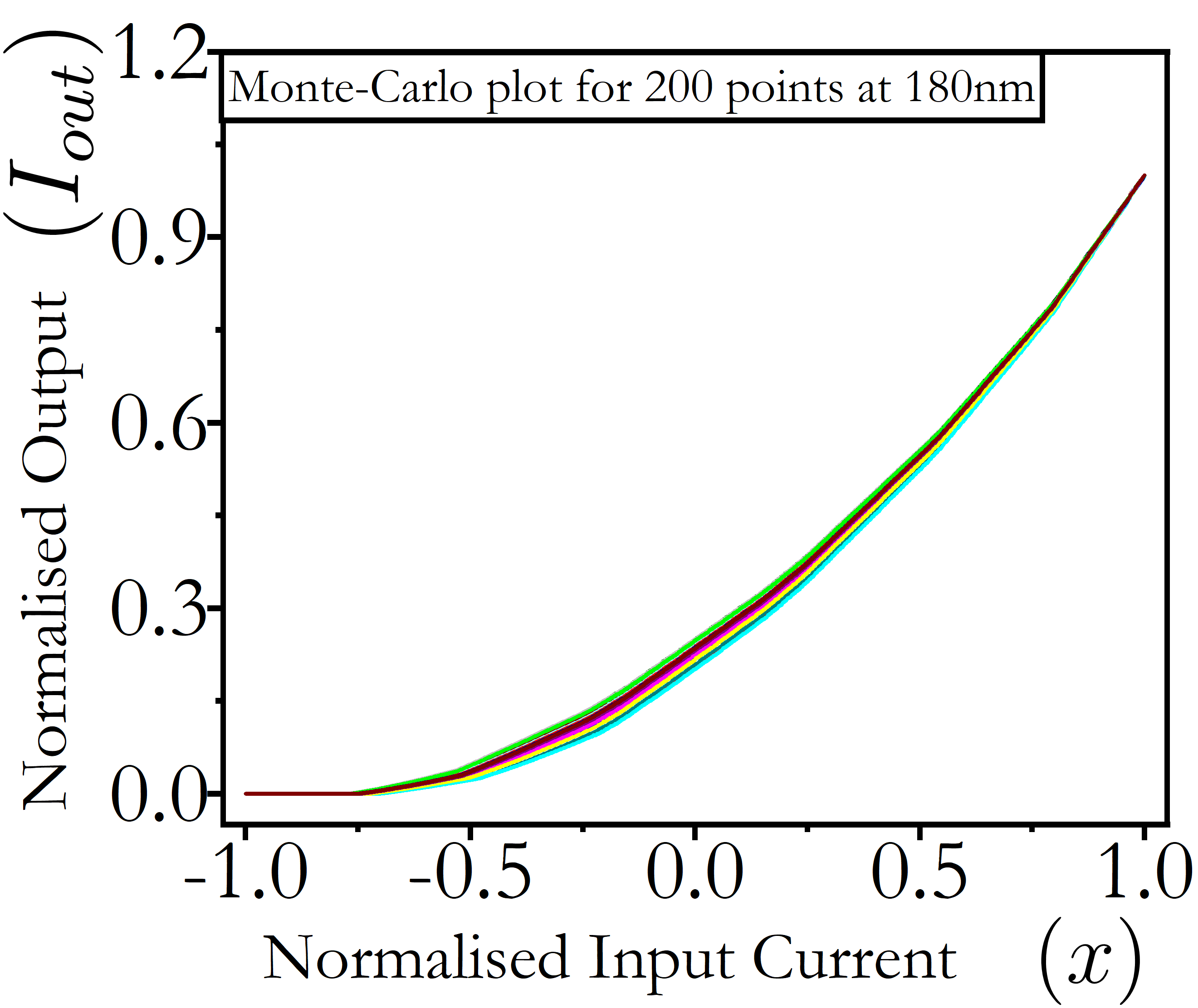}
		\label{mc_softrelu_180}}		
	
	\subfloat[]{\includegraphics[width=0.30\linewidth]{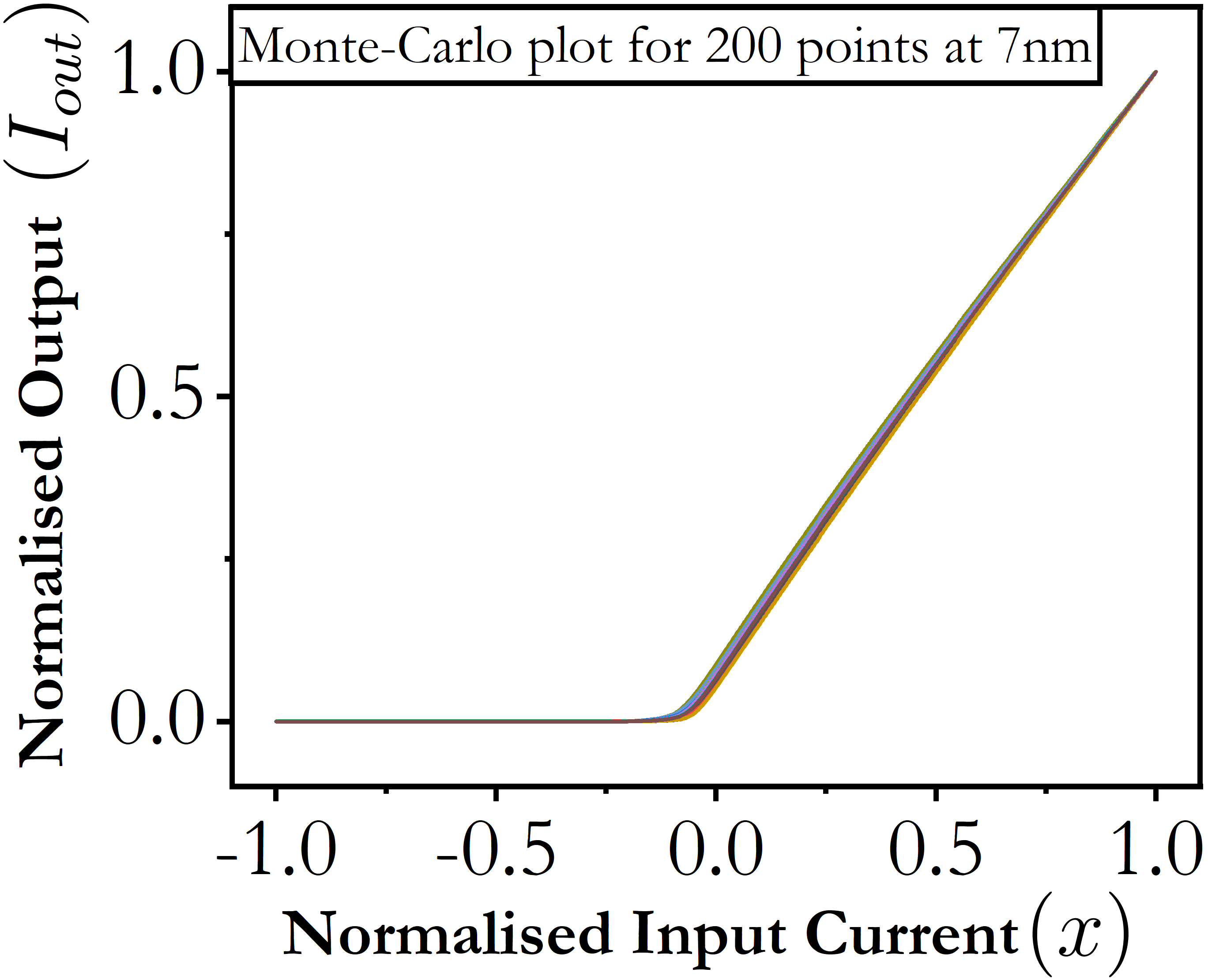}
		\label{mc_relu_7}}
	\subfloat[]{\includegraphics[width=0.30\linewidth]{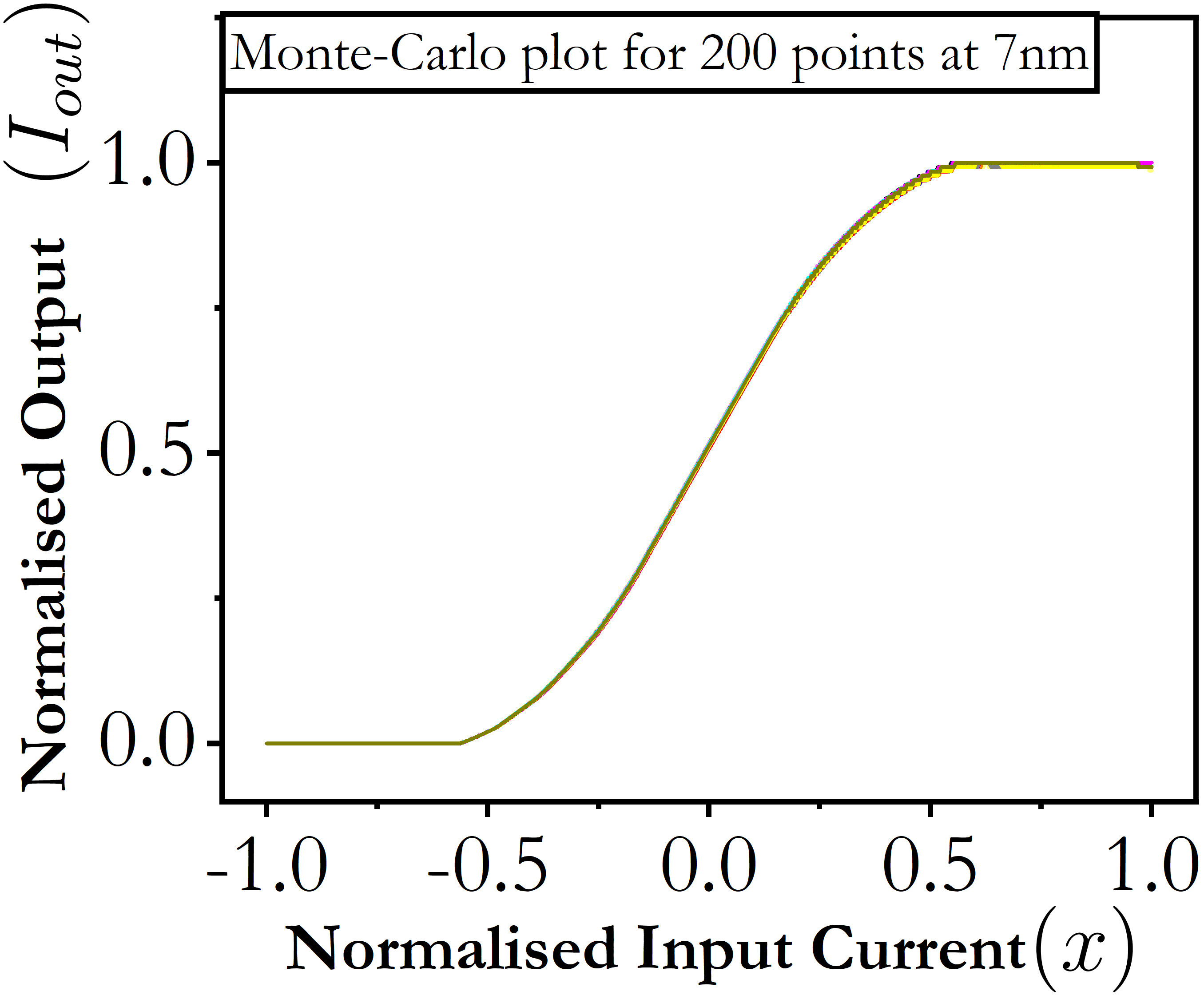}
		\label{mc_sigmoid_7}}
	\subfloat[]{\includegraphics[width=0.30\linewidth]{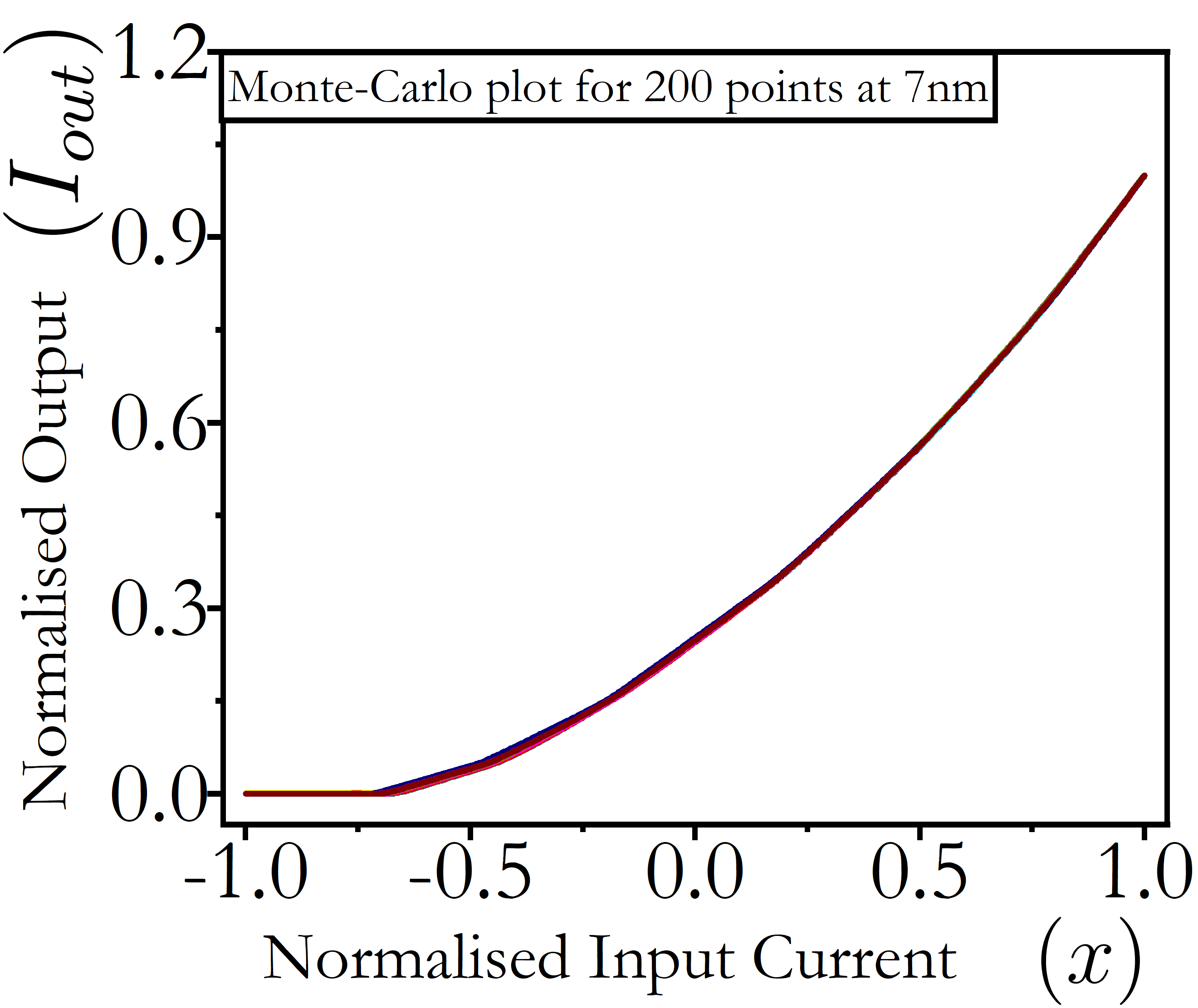}
		\label{mc_softrelu_7}}
	\hfil
	\caption{\textbf{{Monte Carlo plots along with Maximum $\%$ Deviation for S-AC based analog activation functions shown in Fig.~\ref{nonLin_ckt} when operated in Weak Inversion regimes:  \protect\subref{mc_relu_180}~ReLU in 180nm, $(3.11 \%)$;  \protect\subref{mc_softrelu_180}~Soft-plus in 180nm, $(2.44 \%)$; \protect\subref{mc_sigmoid_180}~Sigmoid in 180nm, $(7.31 \%)$; \protect\subref{mc_relu_7}~ReLU in 7nm, $(4.14 \%)$; \protect\subref{mc_sigmoid_7}~Sigmoid in 7nm, $(0.91 \%)$; \protect\subref{mc_softrelu_7}~Soft-plus in 7nm, $(3.86 \%)$.}}}
	\label{mc_act_curve}
\end{figure*}

\subsection{Cosine-Hyperbolic}The $\cosh(\cdot)$ function can be constructed using N-type S-AC standard cells for $S=3$ as shown in Fig.~\ref{cosh_ckt}. The $\cosh(\cdot)$ function is given by:
\begin{equation}
    \cosh(x)=\frac{e^{x}+e^{-{x}}}{2} \label{coshx_in_exp}
\end{equation}
Now, if in Fig.~\ref{exp}, the response of one S-AC unit is $h(x) \cong \frac{e^x}{2}$, then by tuning the offsets $O_1, \cdots{}, O_3$, we can get $\cosh(\cdot)$. In terms of S-AC computation, \eqref{coshx_in_exp} can be written as
\begin{equation}
    \cosh(x)=h(x) + h(-x) \label{coshx_in_sac}
\end{equation}
Thus the $\cosh(\cdot)$ function is the addition (KCL) of the response of S-AC ($y_1$ in Fig.~\ref{cosh_ckt}) and its vertically flipped response (shown by $y_2$ in Fig.~\ref{cosh_ckt}). Fig.~\ref{cosh} shows characteristics response $I_{out}$ obtained using Fig.~\ref{cosh_ckt} across different process nodes and different temperatures for the same values of offset.

\begin{figure}[t]
\centering
\includegraphics[width=1\linewidth]{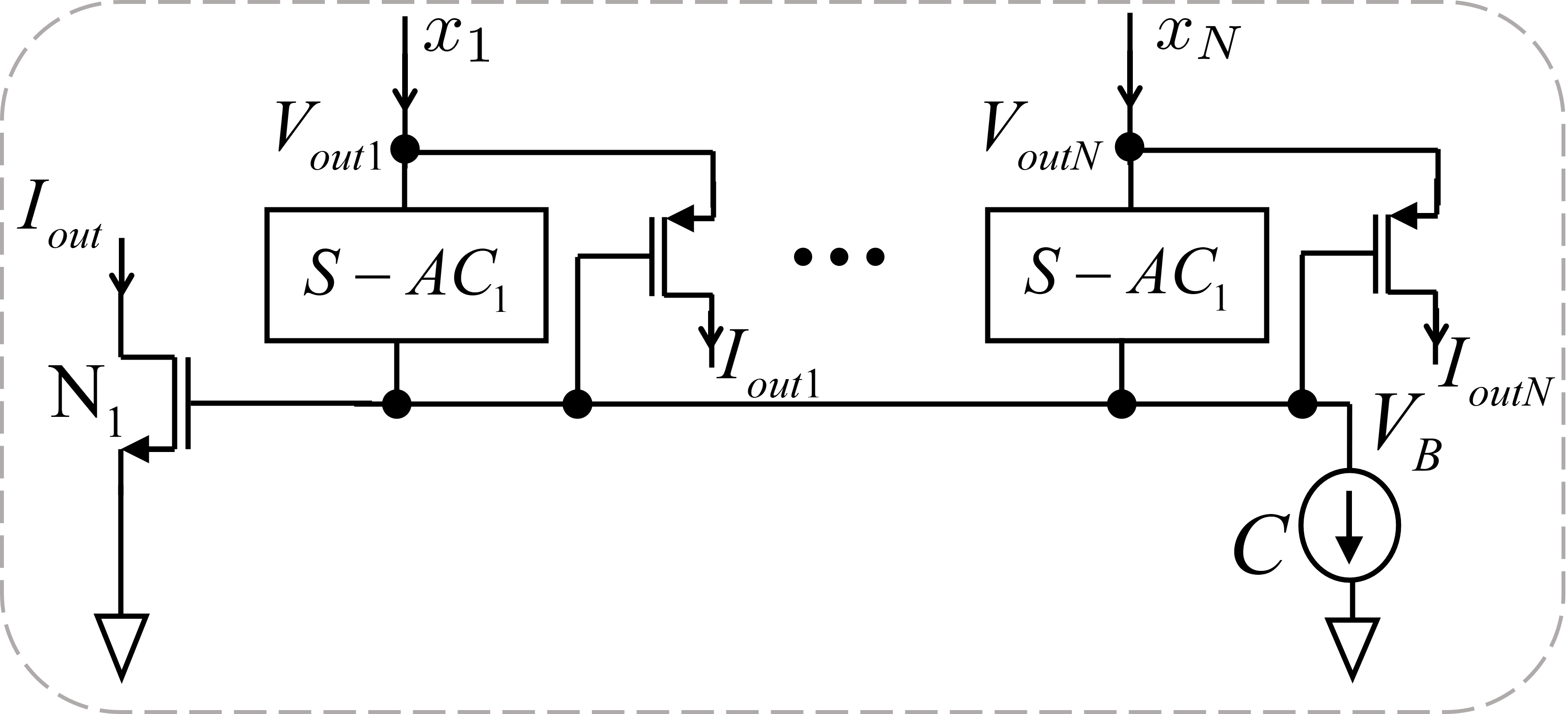}
\caption{Implementation of $N$-input winner takes all standard cells using S-AC units. This circuit can be tuned to function as a soft-WTA and Max circuit.}
\label{wta_ckt}
\end{figure}

\subsection{Sine-Hyperbolic} The $\sinh(\cdot)$ function can be constructed using both P- and N-type S-AC standard cells for $S=3$ as shown in Fig.~\ref{sinh_ckt}. The $\sinh(\cdot)$ function is given by:
\begin{equation}
    \sinh(x)=\frac{e^{x}-e^{-{x}}}{2} \label{sinhx_in_exp}
\end{equation}
Similar to $Cosine$ if the response of S-AC unit is $h(x) \cong \frac{e^x}{2}$, then \eqref{sinhx_in_exp} can be written in terms of S-AC computation as
\begin{equation}
    \sinh(x)=h(x) - h(-x) \label{sinhx_in_sac}
\end{equation}
Fig.~\ref{sinh} shows characteristics response $I_{out}$ obtained using Fig.~\ref{sinh_ckt} across different process nodes and temperatures.

\subsection{ReLU} ReLU can be implemented using two S-AC units as shown in Fig.~\ref{relu_ckt}. Fig.~\ref{relu_ckt} implements the function given by
\begin{equation}
{I_{out}} = \left\{ \begin{array}{l}
\max \left( {0,x} \right)\;\,\,\,\,\,\,\,\quad,C \to 0\\
\max \left( {0,C - x} \right)\;\;,else\;
\end{array} \right.
\end{equation}
Fig.~\ref{relu} shows the characteristic response of ReLU obtained from the circuit in Fig.~\ref{relu_ckt} at different process nodes and temperatures. {Fig.~\ref{mc_relu_180} and Fig.~\ref{mc_relu_7} show the Monte Carlo plot of S-AC \textit{ReLU} in 180nm and 7nm process technology nodes.}

\begin{figure*}[t]
	\centering
	\subfloat[]{\includegraphics[width=0.25\linewidth]{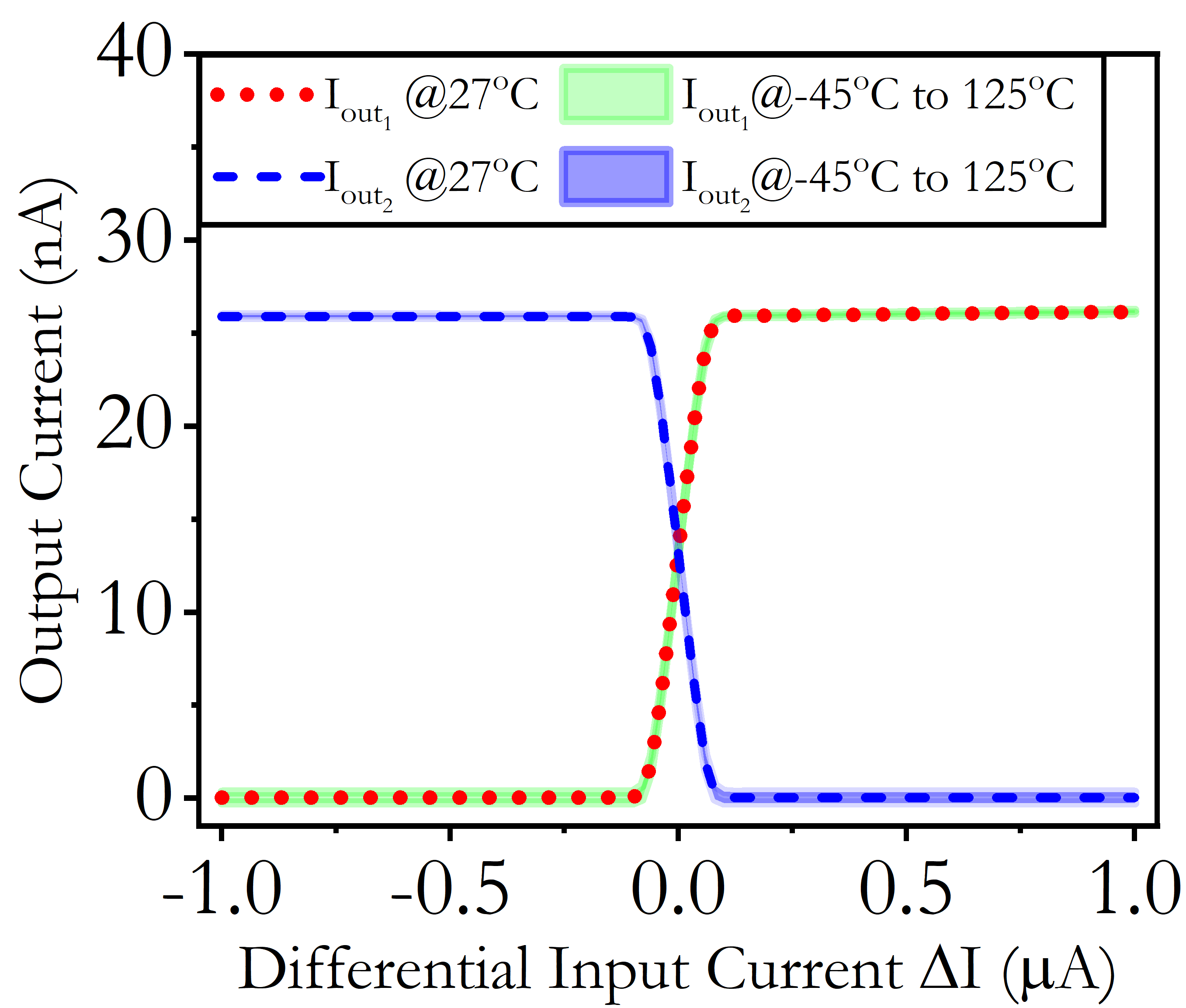}
		\label{wta_180_i}}
	\subfloat[]{\includegraphics[width=0.245\linewidth]{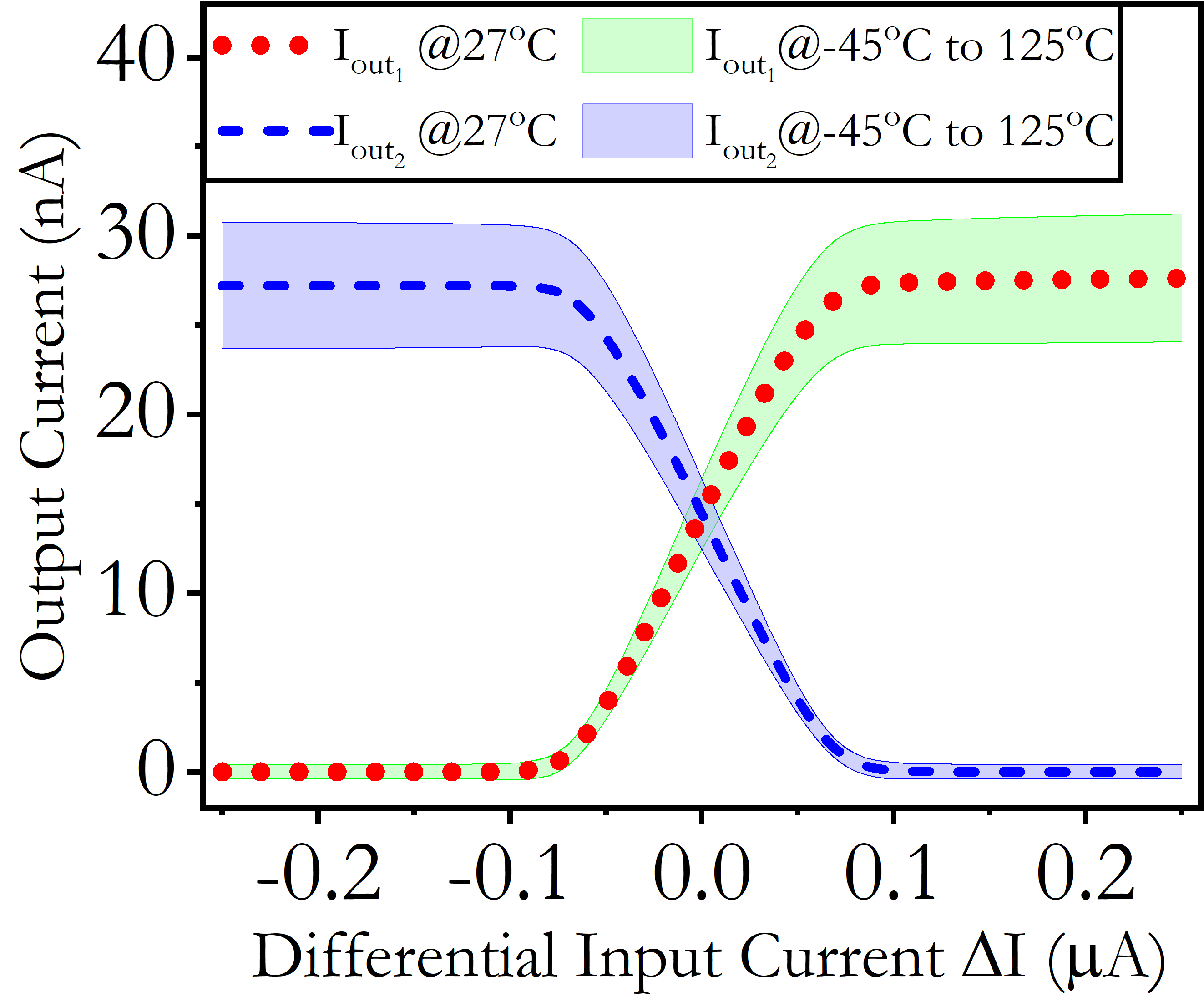}
		\label{wta_7_i}}
	\subfloat[]{\includegraphics[width=0.25\linewidth]{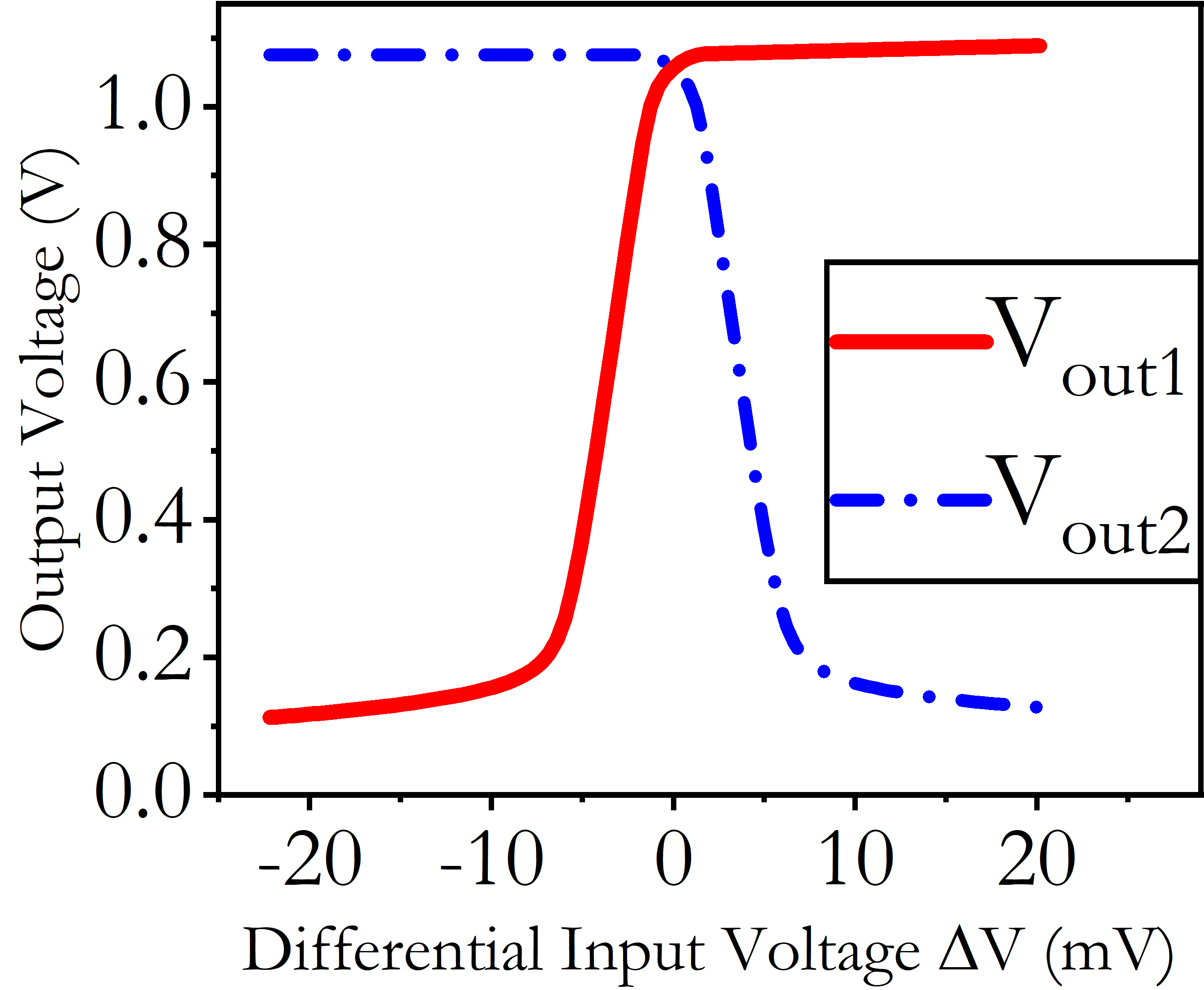}
		\label{wta_180_v}}
	\subfloat[]{\includegraphics[width=0.25\linewidth]{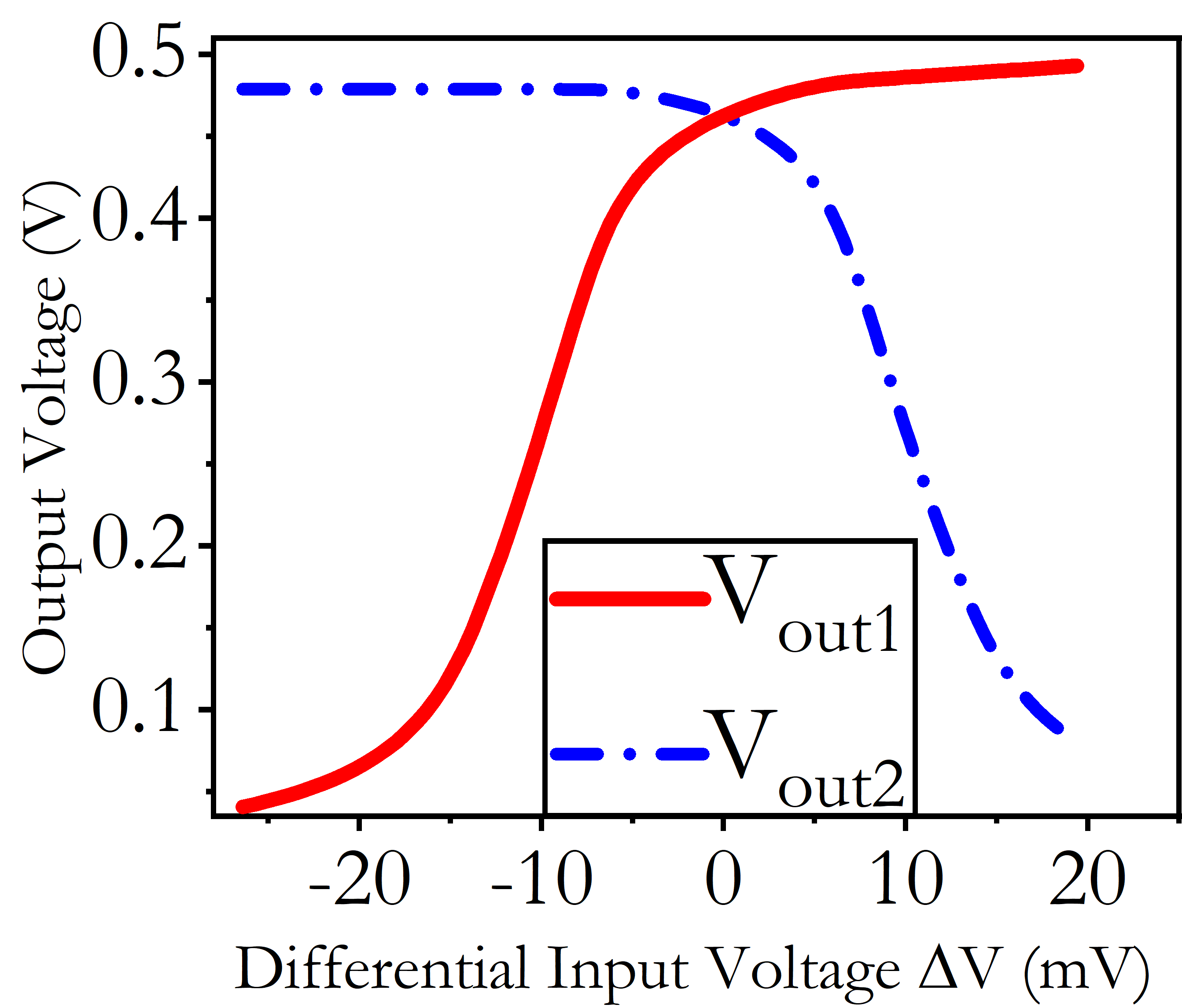}
		\label{wta_7_v}}
			\hfil

	\subfloat[]{\includegraphics[width=0.265\linewidth]{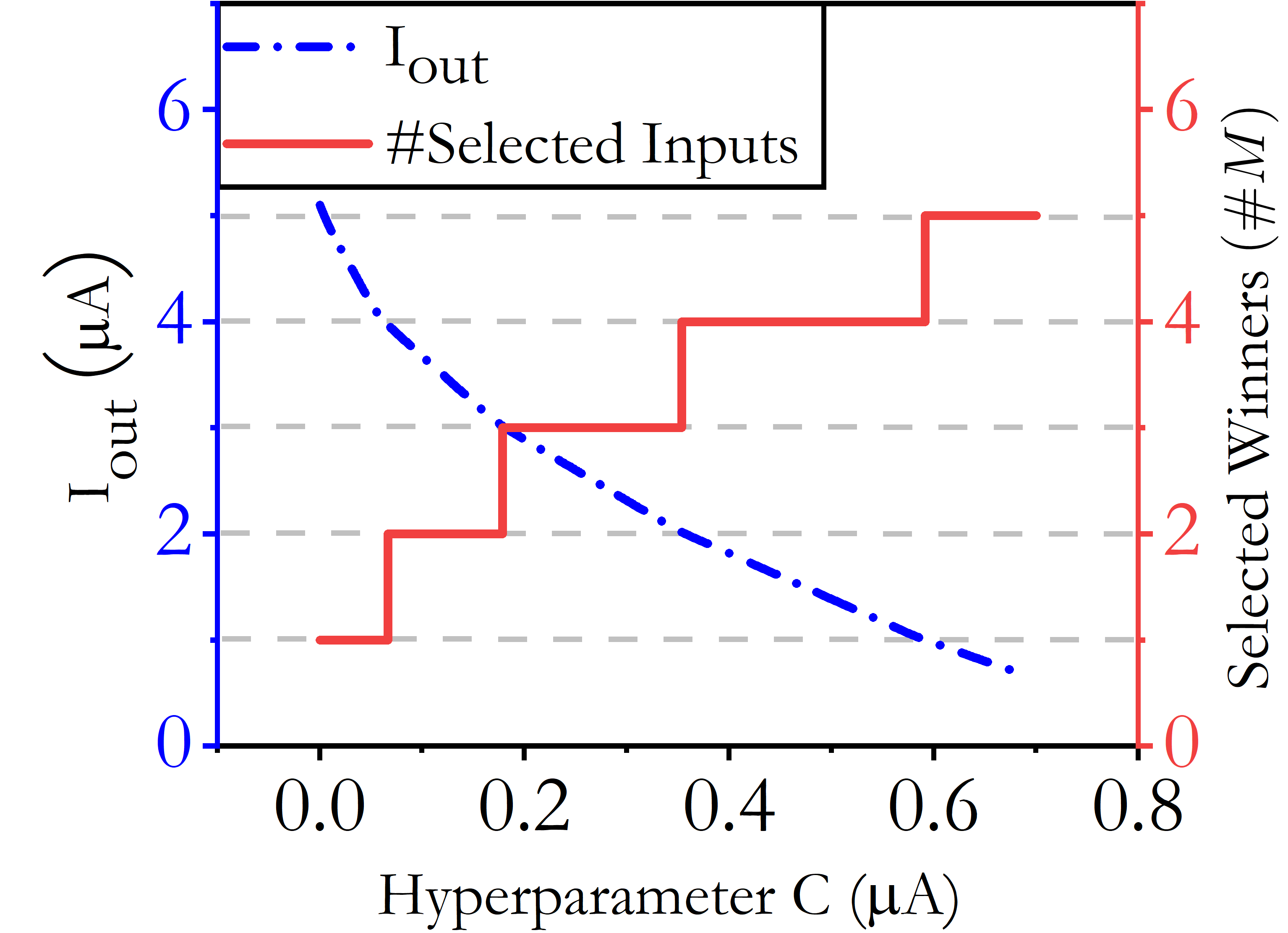}
		\label{wtaencoder180}}
	\subfloat[]{\includegraphics[width=0.268\linewidth]{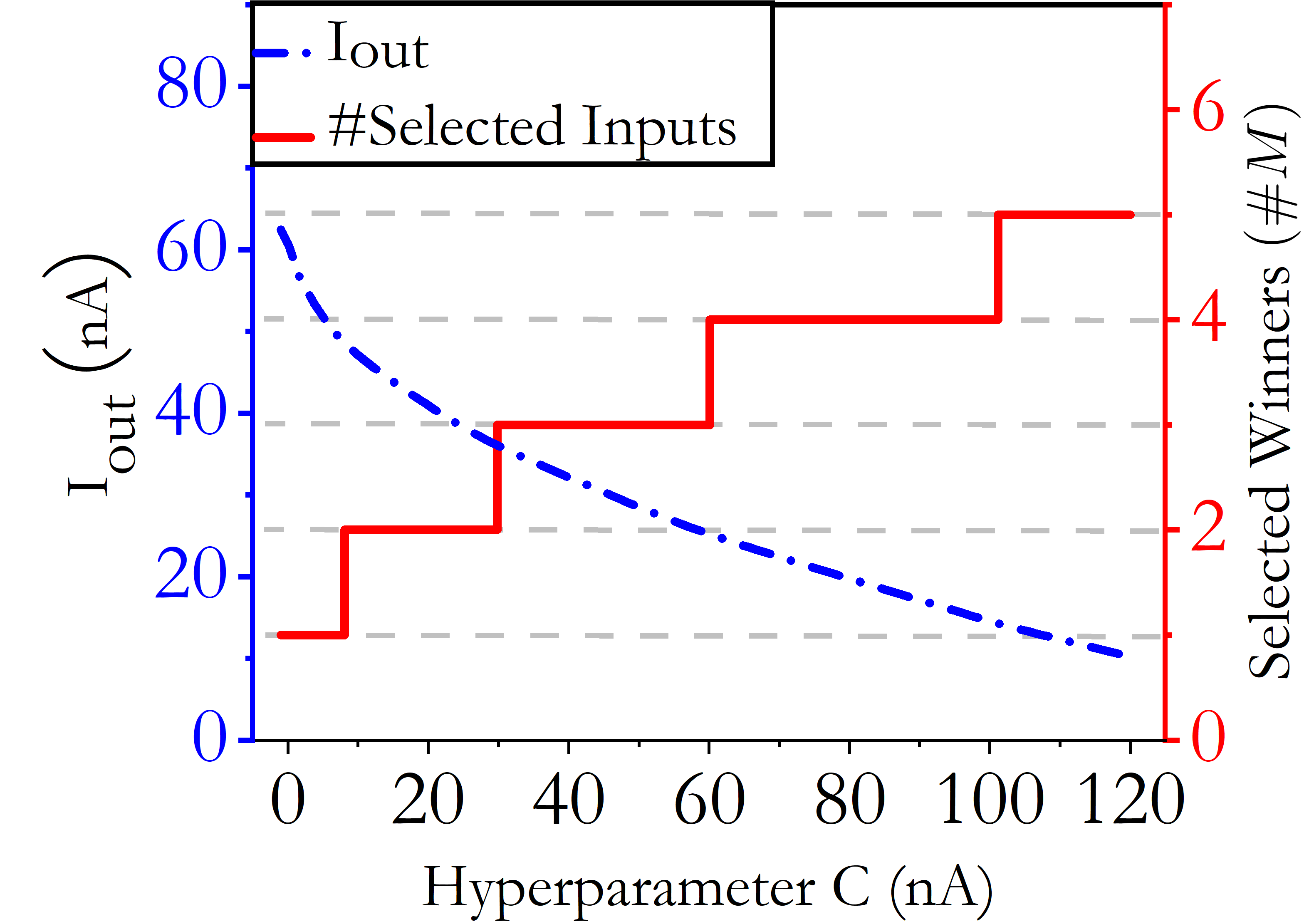}
		\label{wtaencoder7}}
			\subfloat[]{\includegraphics[width=0.23\linewidth]{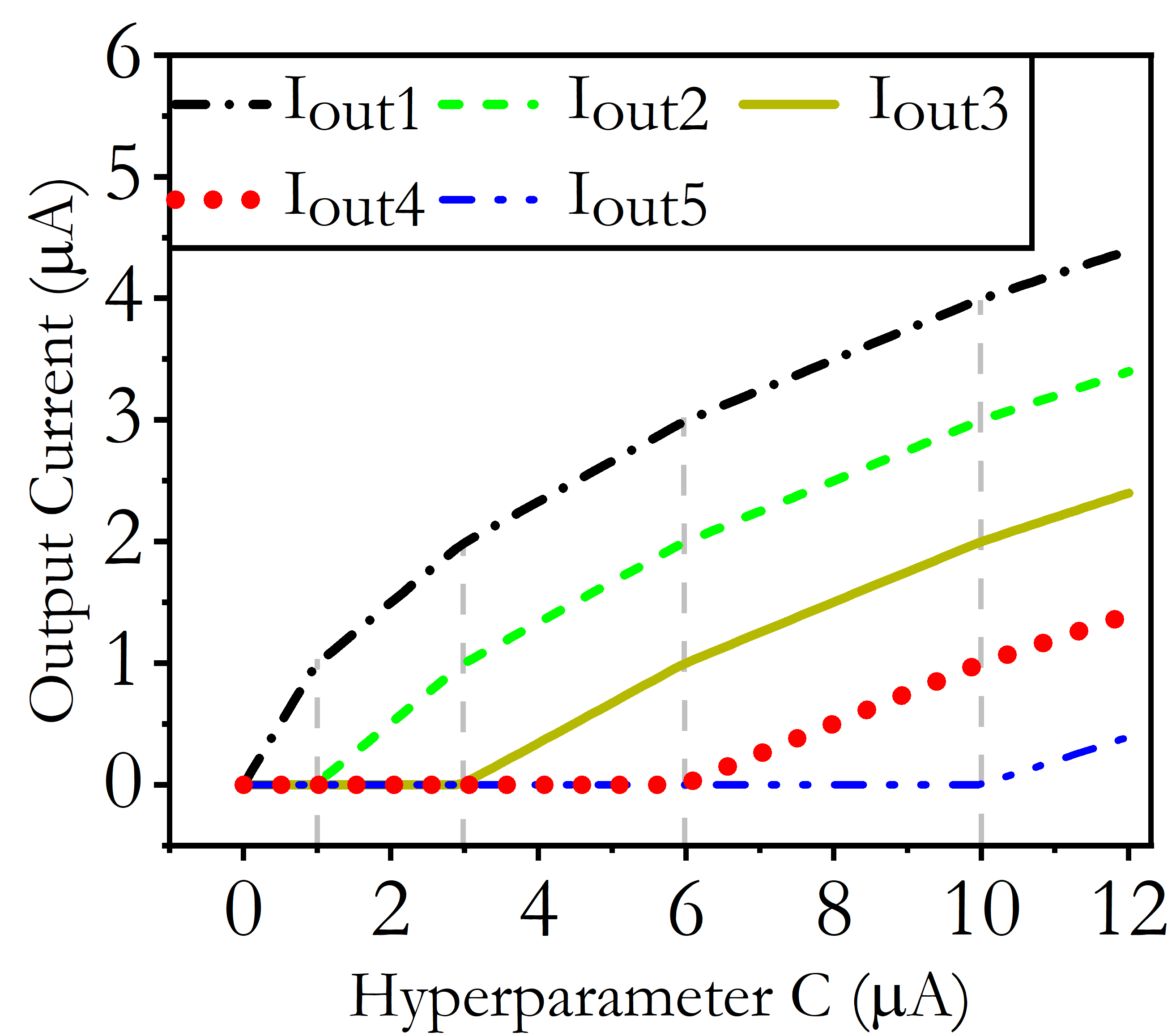}
	\label{wta_180_i_c}}
	\subfloat[]{\includegraphics[width=0.235\linewidth]{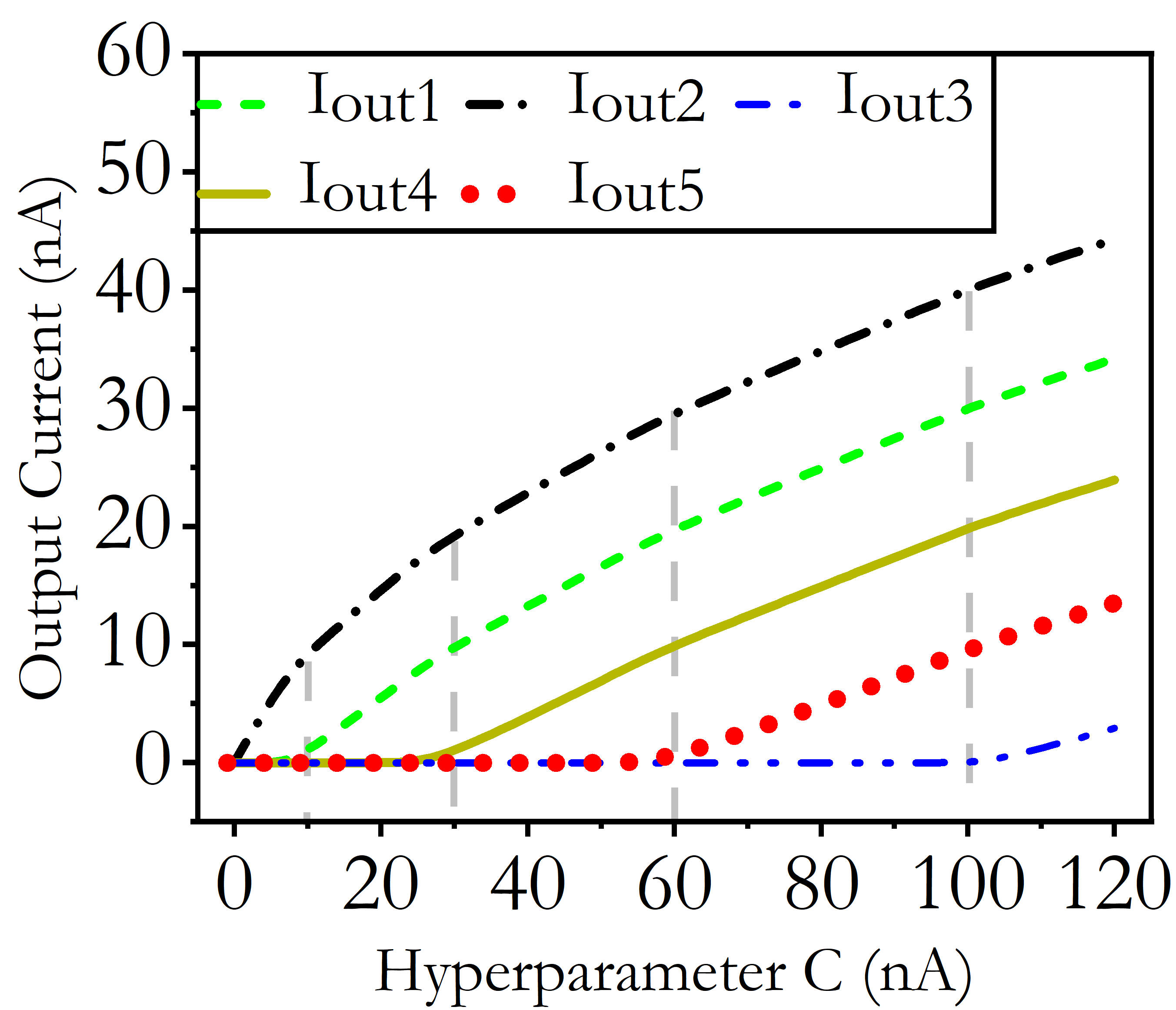}
		\label{wta_7_i_c}}	
	\hfil
	\caption{{Simulated output of a two input S-AC based WTA standard cell shown in Fig.~\ref{wta_ckt} showing: \protect\subref{wta_180_i}~current output ($I_{out1}$ and $I_{out2}$) versus  the differential input current for 180nm process node and \protect\subref{wta_7_i}~for 7nm process node; \protect\subref{wta_180_v}~Voltage output ($V_{out1}$ and $V_{out2}$) versus  the differential input voltage for 180nm process node and \protect\subref{wta_7_v}~for 7nm process node. The simulated output of a five-input S-AC WTA standard cell showing \protect\subref{wtaencoder180}~$M$ selected winners as a function of hyperparameter $C$ for 180nm process node and \protect\subref{wtaencoder7}~for 7nm process node where $I_{out}$ is the contribution of $M$ selected winners in the output. Simulated output of five-input S-AC WTA standard cell showing outputs $\left[ {{I_{out1}}, \cdots ,{I_{out5}}} \right]$ as a function of hyperparameter $C$ at \protect\subref{wta_180_i_c}~180nm and \protect\subref{wta_7_i_c}~7nm process node where responses of Fig.~\ref{soft_wta} are obtained for the inputs $\left[ {{x_{1,}}{x_{2,}}{x_{3,}}{x_{4,}}{x_5}} \right] = \left[ {\alpha ,2\alpha ,3\alpha ,4\alpha ,5\alpha } \right]$ where $\alpha  = 1\mu A $  for 180nm and $10nA$ for 7nm.}}
	\label{soft_wta}
\end{figure*}
    
\subsection{Compressive non-linearity} We formulate the compressive non-linearity function $\phi_1(\cdot)$ given by 
\begin{equation}
    \phi_1(x)=\log\frac{1+e^{x+K}}{e^x+e^{K}} \label{log_comp}
\end{equation}
where $K$ is a constant. This compressive function in \eqref{log_comp} can then be used to emulate $\tanh(\cdot)$ and sigmoid functions. Equation \eqref{log_comp} when mapped to S-AC computation can be written as:
\begin{equation}
    \phi_1(\cdot)=h(0,x+K)-h(x,K) \label{tanhx_in_sac}
\end{equation}
The circuit for emulating $\phi_1(\cdot)$ is shown in Fig.~\ref{tanh_ckt}. We show that the characteristics response ${I_{\phi 1}}$ shown in Fig.~\ref{tanh} can be used to emulate response equivalent to $\tanh(\cdot)$ function.

\subsection{Sigmoid} The sigmoid equivalent characteristic plot can be obtained from a shifted version of the $\phi_1(\cdot)$ function. Thus adding a constant current $K$ to ${I_{\phi 1}}$ gives a {s}igmoid equivalent response and is shown in Fig.~\ref{tanh_ckt}. We show that the characteristics response ${I_{\phi 2}}$ shown in Fig.~\ref{sigmoid} can be used to emulate a response equivalent to the {$s$}$igmoid$ function. {The Monte Carlo plot of S-AC \textit{sigmoid} in both 180nnm and 7nm process node is shown in Fig.~\ref{mc_sigmoid_180} and Fig.~\ref{mc_sigmoid_7} respectively.}

\subsection{Soft-Plus} The soft plus activation function can be obtained using two $S=3$, S-AC units. The circuit for emulating the $Soft$-$Plus$ response is shown in Fig.~\ref{softPlus_ckt}. We show that the characteristics response ${I_{out}}$ shown in Fig.~\ref{softPlus} are robust across different process nodes and different temperatures. {Fig.~\ref{mc_softrelu_180} and Fig.~\ref{mc_softrelu_7} show the Monte Carlo plot of S-AC \textit{soft-plus} in 180nm and 7nm process technology node.}

\subsection{S-AC based Winner-Take-All}
A winner-take-all (WTA) circuit is designed to emulate the $max(\cdot)$ function. The proposed S-AC-based WTA circuit is shown in Fig.~\ref{wta_ckt}. It is modular, i.e., it can be extended for N inputs like the original circuit proposed by Lazzaro et al. (1989)~\cite{lazzaro1988winner}. Fig.~\ref{wta_180_i} and Fig.~\ref{wta_7_i} show the current characteristics plot of two-input S-AC based WTA at 180nm and 7nm for an input differential current. It can be observed that when the differential current is $0$, the output currents are equal. The corresponding voltage outputs for both the process nodes are shown in Fig.~\ref{wta_180_v} and Fig.~\ref{wta_7_v} respectively. 

\subsection{S-AC based N-of-M Encoder}
The N-of-M encoder extends the computational capabilities of the standard WTA circuit and has found profound importance in sparsely distributed memory~\cite{FURBER20041437}. N-of-M encoder allows the user to obtain the max current that takes into account the influence due to the contribution of top M winners out of N inputs ($M/N$). For the specific case of $M = 1$, the N-of-M encoder behaves like a simple WTA circuit.  Fig.~\ref{wtaencoder180} shows the response of the five-input S-AC-based WTA circuit shown in Fig.~\ref{wta_ckt} as a function of hyper-parameter $C$. It can be noted that with the increase in hyper-parameter value, the output current $I_{out}$ decreases and is the result of more than one winner. Using \eqref{char1_msmp} for $S=1$, $I_{out}$ is given by
\begin{align}
{I_{out}} = \frac{{\sum\nolimits_{i = 1}^M {{x_i} - C} }}{M}
\label{wta_eqn}
\end{align}
where $M$ is the number of winners. Similar results for 7nm are shown in Fig.~\ref{wtaencoder7}. It can be analyzed  that depending on the hyper-parameter $C$, the circuit can select the top $M$ winners.

\subsection{S-AC based SoftArgmax}
In machine learning, SoftArgmax offers an improvement over Argmax to support backpropagation and gradient operation. The S-AC-based WTA circuit can be configured to implement SoftArgmax. Fig.~\ref{wta_180_i_c} and Fig.~\ref{wta_7_i_c} show the response curve of outputs $I_{out1}, \cdots, I_{out5}$ for the variation in hyper-parameter $C$ for 180nm and 7nm respectively. It can be observed that with the increase in hyper-parameter, outputs corresponding to the maximum input along with other inputs are activated and can be given by
\begin{align}
{I_{ou{t_i}}} = {x_i} - C ,\qquad \forall x_i > C
\label{soft-argmax}
\end{align}

\subsection{S-AC based Max Circuit}
The S-AC based \textit{winner-take-all} circuit can also be configured to select the maximum input among the given set of $N$ inputs. For hyper-parameter, $C \to 0$, the circuit starts behaving as a max input selector.

\subsection{S-AC based Four Quadrant Multiplier}

\begin{figure}[t]
\centering
\includegraphics[width=1\linewidth]{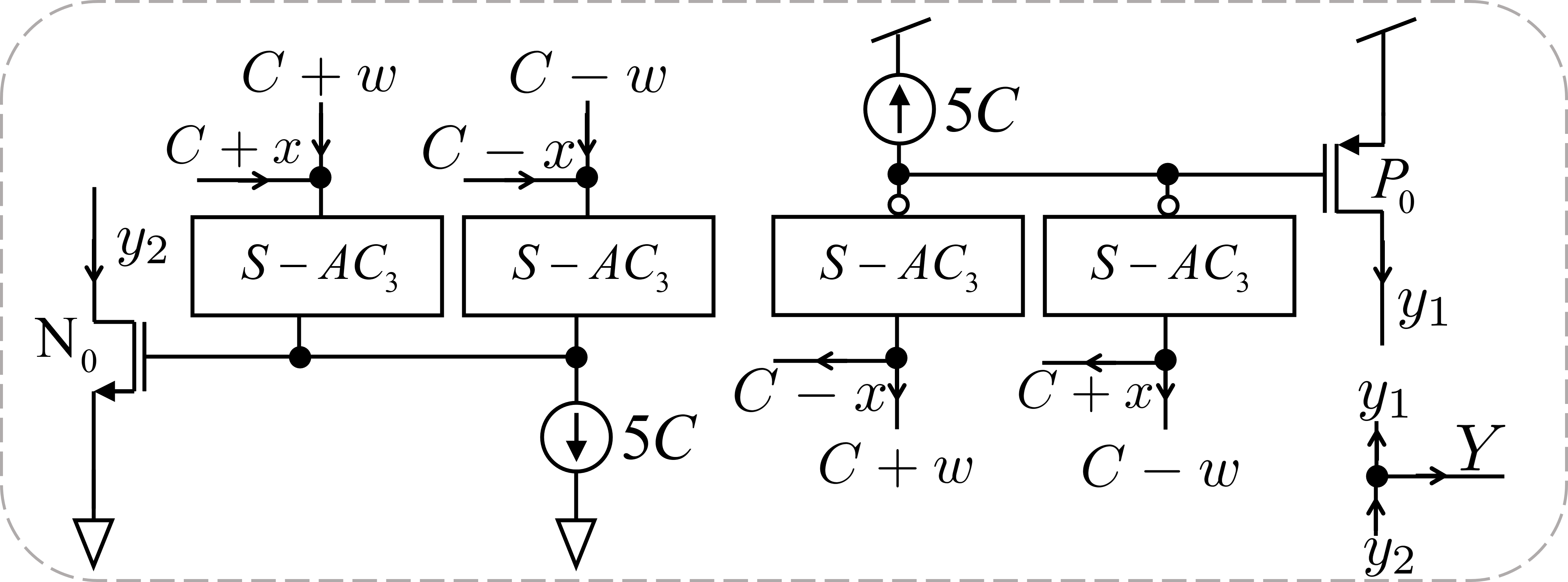}
\caption{Implementation of a four-quadrant multiplier using P-type and N-type S-AC circuits.}
\label{mul_ckt}
\end{figure}

The four-quadrant multiplier is proposed in~\cite{sac1}. Owing to the Lipchitz behavior of the S-AC function, the design is formulated such that the multiplication satisfies the Lipchitz condition. Fig.~\ref{mul_ckt} shows the S-AC cell-based implementation of a multiplier circuit. Consider the following equation, where $y$ is given by 
\begin{multline}
y = h(C + w + C + x) - h(C + w + C - x) +  \ldots \\
h(C - w + C - x) - h(C - w + C + x)
\label{exp0}
\end{multline}

The goal is to implement scalar multiplication between two variables $x$ and $w$. Here ${x,w,y} \in$ $\mathbb{R}$, $h$ is a non-linear monotonic function and $C$ is a hyperparameter. If we write the Taylor expansion of $h(x)$ around $w$, we get

\begin{multline} 
        h(C + w + C + x) = h(0) + \frac{h'(0)}{1!}(C + w + C + x) \ldots \\+ \frac{h''(0)}{2!}(C + w + C + x)^2 + \ldots
    \label{exp1}
\end{multline}
\begin{multline}
      h(C+w+C-x) = h(0) + \frac{h'(0)}{1!}(C+w+C-x) \ldots \\+ \frac{h''(0)}{2!}(C+w+C-x)^2 + \ldots
    \label{exp2}
\end{multline}
\begin{multline}
      h(C-w+C-x) = h(0) + \frac{h'(0)}{1!}(C-w+C-x) \ldots \\+ \frac{h''(0)}{2!}(C-w+C-x)^2 + \ldots
    \label{exp3}
\end{multline}
\begin{multline}
        h(C-w+C+x) = h(0) + \frac{h'(0)}{1!}(C-w+C+x) \ldots \\+ \frac{h''(0)}{2!}(C-w+C+x)^2 + \ldots
    \label{exp4}
\end{multline}

Substituting \eqref{exp1} - \eqref{exp4} in \eqref{exp0} and ignoring the higher order terms, we get,
\begin{align}
y \cong 4 h''(0)  x \times w
\label{taylor1}
\end{align}
where $4 h''(0) $ is a scaling factor. For $h(\cdot)$ assumed to be a non-linear function such that $h''(0) \neq 0$, \eqref{taylor1} simplifies to 
\begin{align}
    y \cong x \times w
\label{multipliyeqn}
\end{align}
\eqref{exp0} shows that the multiplication is then reduced to addition and subtraction operations, thus altering the use of a bulky multiplier.

\begin{figure*}[t]
	\centering
	\subfloat[]{\includegraphics[width=0.28\linewidth]{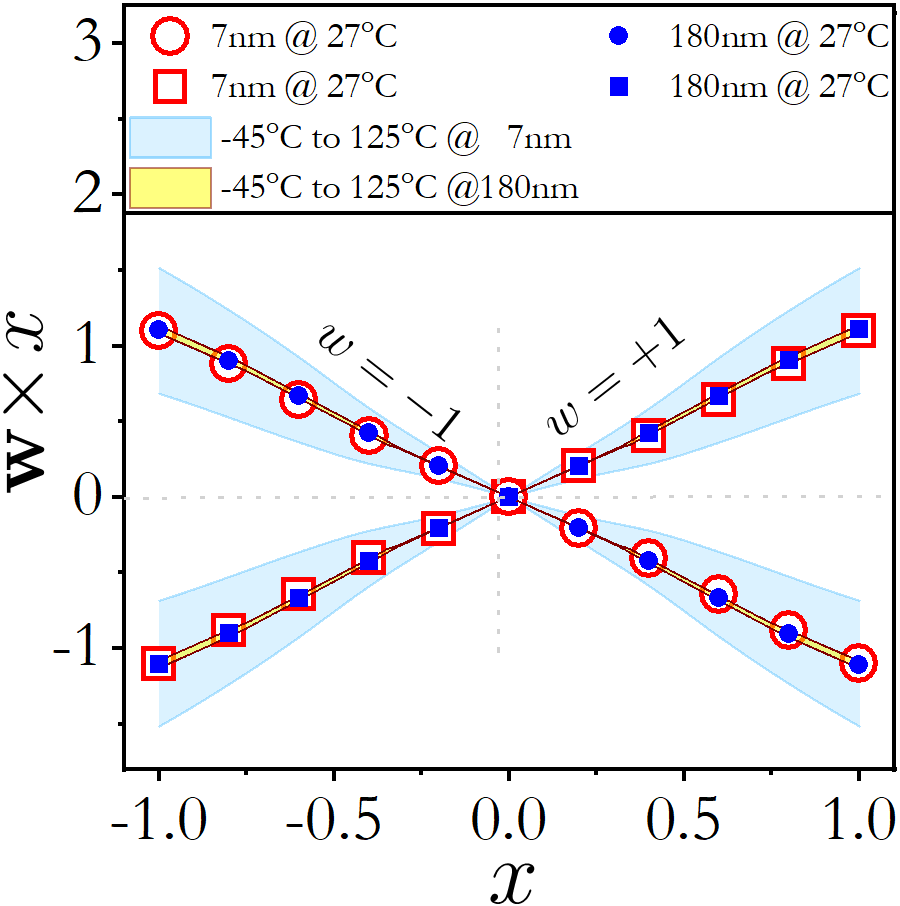}
		\label{mul}}
		\subfloat[]{\includegraphics[width=0.31\linewidth]{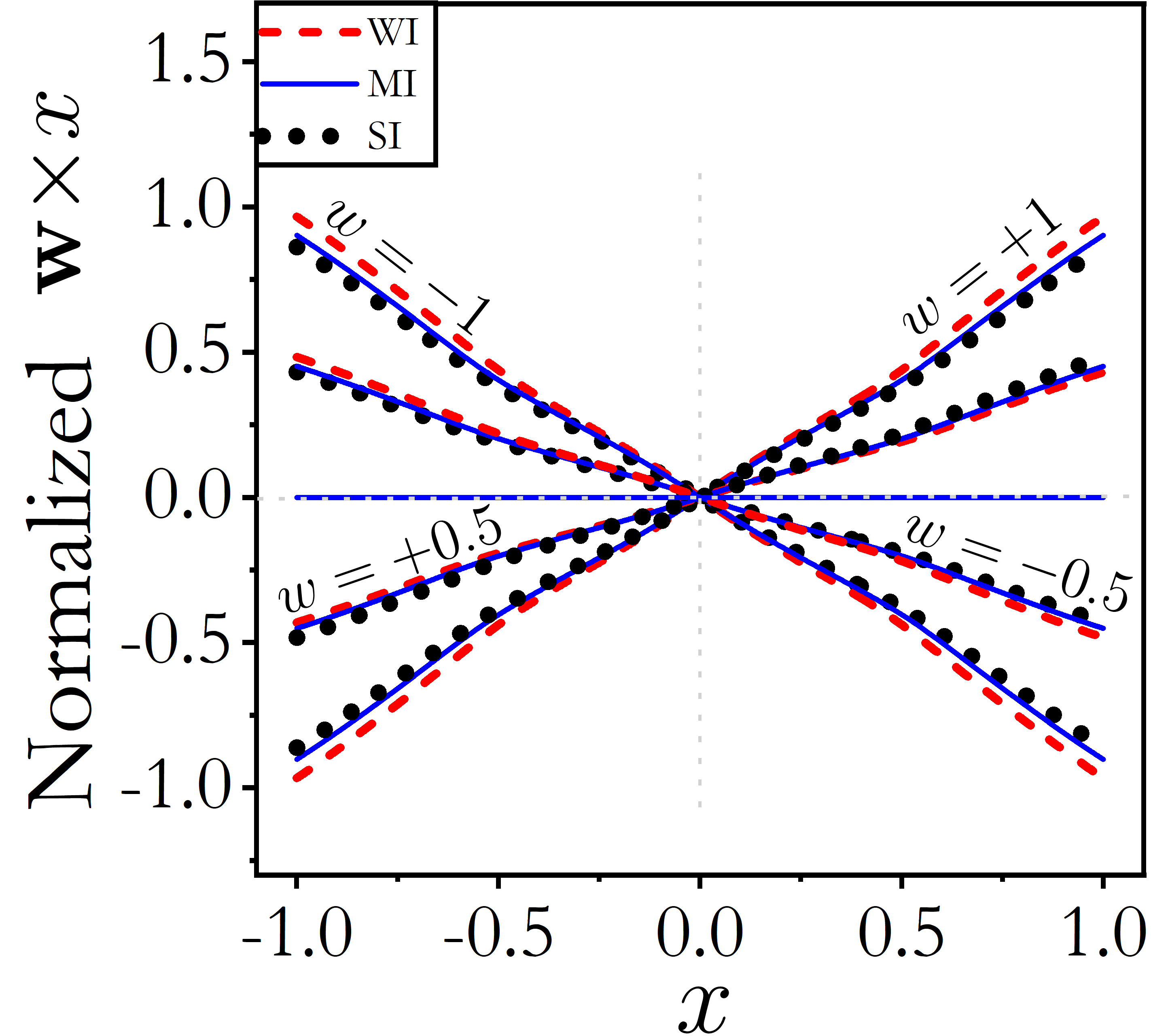}
		\label{multiply_7nm}}
	\subfloat[]{\includegraphics[width=0.31\linewidth]{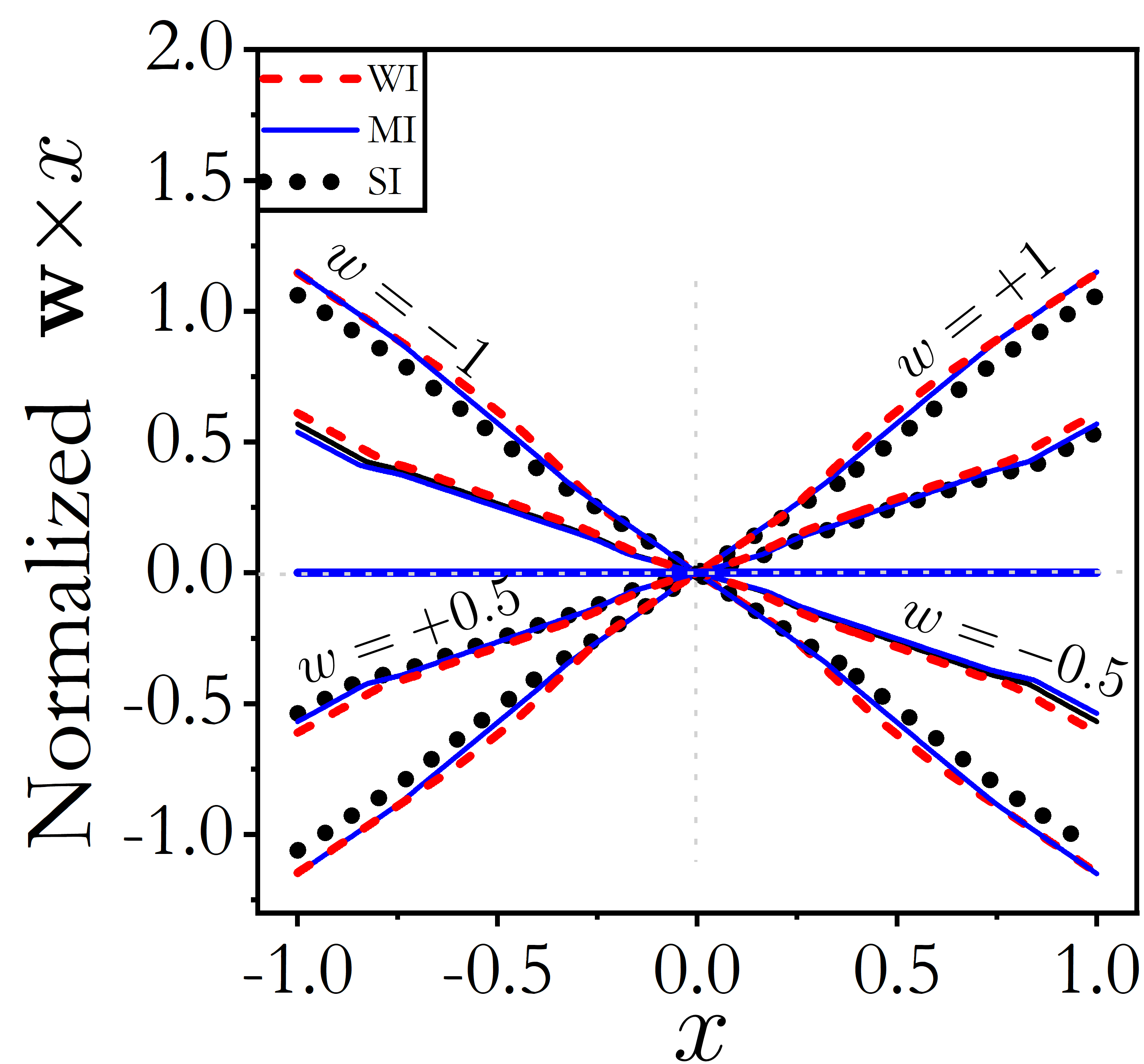}
		\label{multiply_180nm}}
	\caption{Simulated multiplier characteristics obtained for $S = 3$, 
	\protect\subref{mul}~at different process nodes and across temperature; \protect\subref{multiply_7nm}~{at 7nm process node for  different operating regimes; \protect\subref{multiply_180nm}~at 180nm process node for  different operating regimes.}}
	\label{wta_ckt_mult}
\end{figure*}

In Fig.~\ref{mul}, we show that the behavior of the four-quadrant S-AC multiplier is scalable across the process technology nodes. Fig.~\ref{multiply_7nm} shows the response of the four-quadrant S-AC multiplier at different operating regimes for the 7nm process node. A similar response can be observed in Fig.~\ref{multiply_180nm} at the 180nm process node. It can be analyzed from Fig.~\ref{multiply_7nm} and Fig.~\ref{multiply_180nm} that the shape of the multiplier characteristic curve remains preserved when the circuit operation moves from WI to SI.

\subsection{Performance and Trade-off Analysis}\label{perf_analysis}
This section shows the analysis of the S-AC computational blocks at different process technology nodes, viz. 180nm and 7nm and at different operating conditions. The inter-dependence of power, throughput, accuracy and area trade-off for designing a performance scalable system is explained.  

\begin{figure*}[t]
	\centering
\subfloat[]{\includegraphics[width=0.38\linewidth]{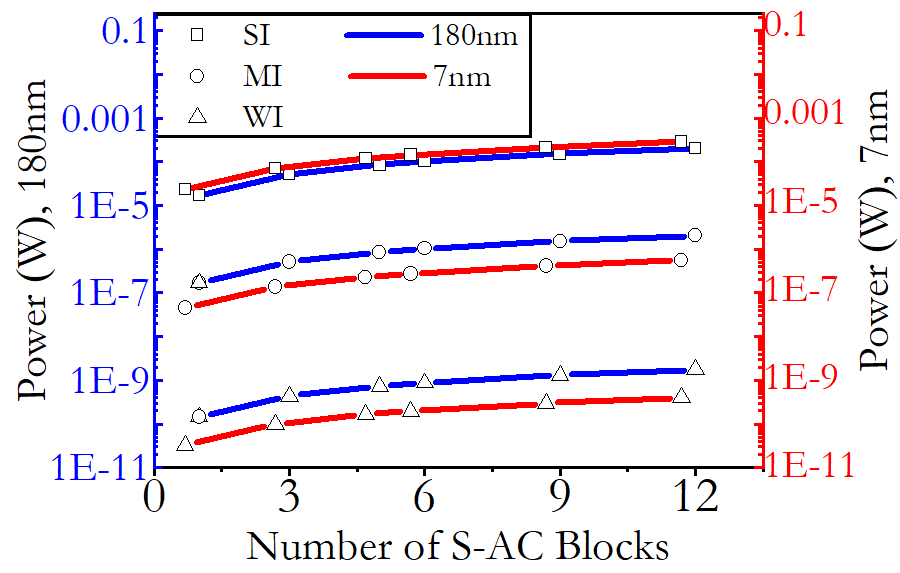}\label{power}}
\subfloat[]{\includegraphics[width=0.3\linewidth]{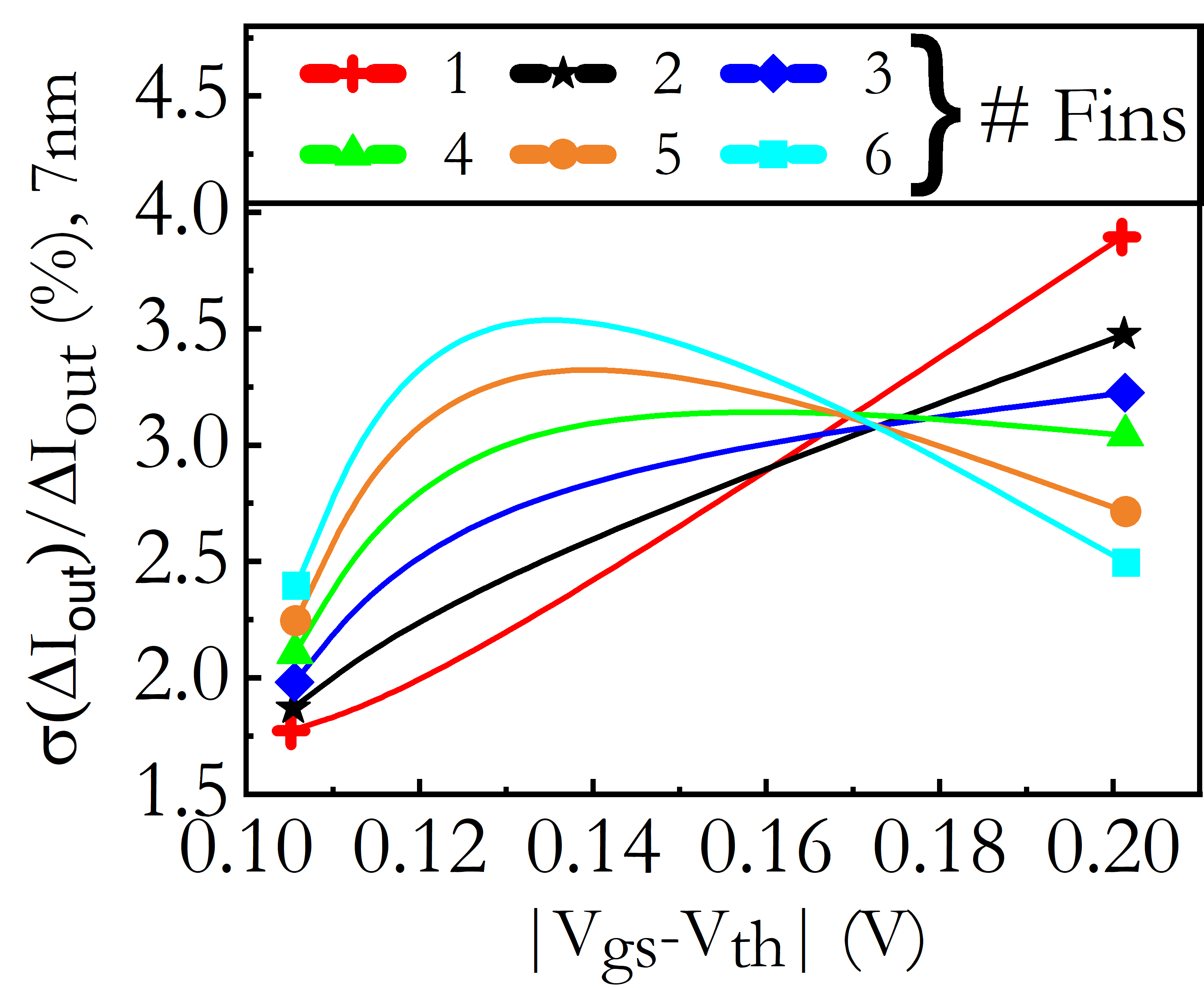}\label{mismatch7}}
\subfloat[]{\includegraphics[width=0.3\linewidth]{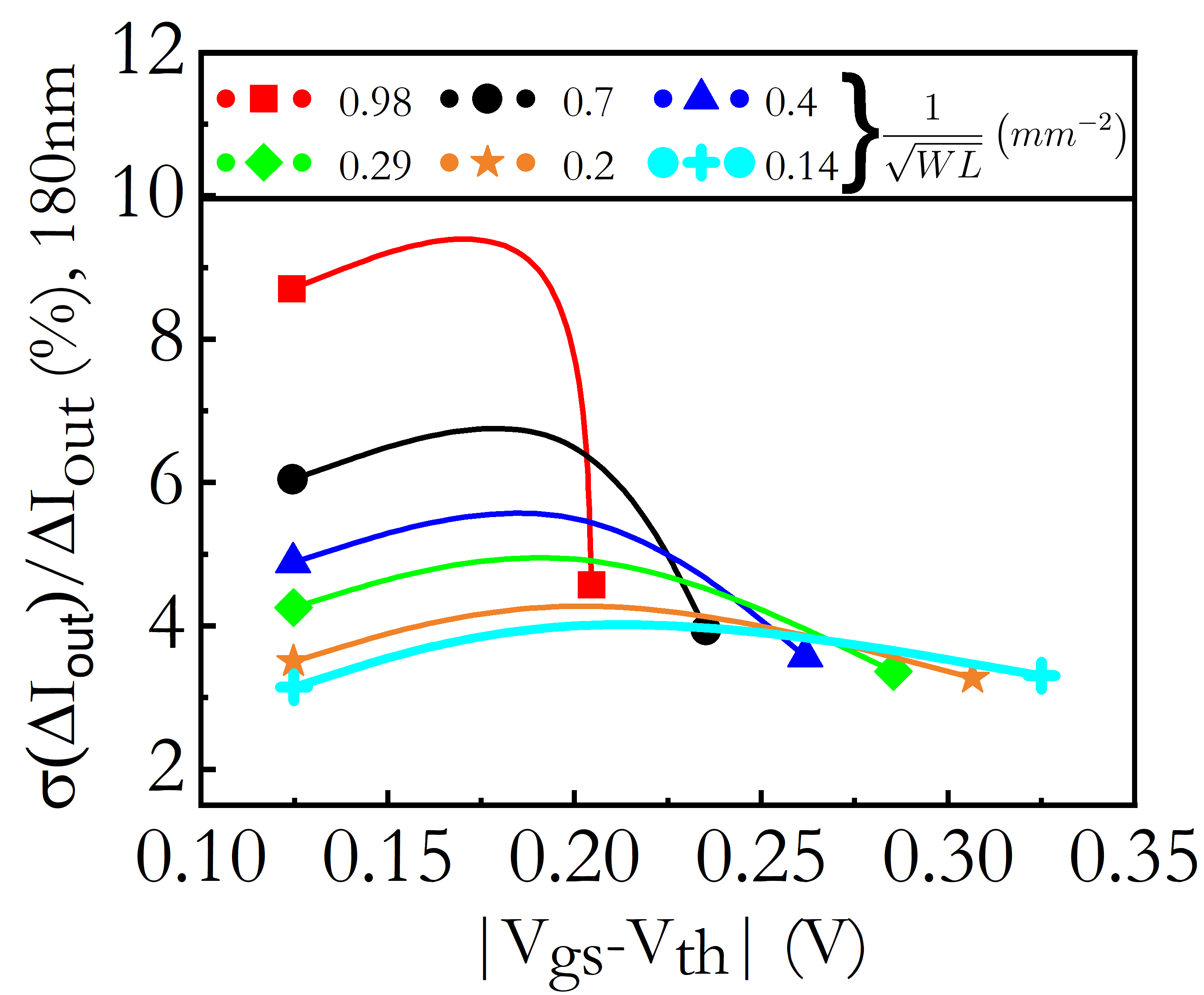}\label{mismatch180}}
 		
\caption{Simulations showing \protect\subref{power}~average power consumption at different process technology nodes as the function of $S$ (number of inputs $N=1$); \protect\subref{mismatch7}~{Standard deviation of output current $h(\textbf{x})$ as a function of change in Fins count and gate-source voltage for 7nm; \protect\subref{mismatch180}~change in aspect ratio and overdrive voltage for 180nm.}}
\label{performance_par}
\end{figure*}

\begin{table}[t]
\centering
	\caption{Operation Performance Parameter for S-AC System}
	\scalebox{0.83}{
\begin{tabular}{|c|cccccc|}
\hline
\multirow{3}{*}{\textbf{\begin{tabular}[c]{@{}c@{}}Parameter\\ (@ $S=1$)\end{tabular}}}                                                    & \multicolumn{6}{c|}{\textbf{Technology and Operating Regimes}}                                                                                                                             \\ \cline{2-7} 
                                                                                       & \multicolumn{3}{c|}{\textbf{CMOS 180nm}}                                                               & \multicolumn{3}{c|}{\textbf{FinFET 7nm}}                                          \\ \cline{2-7} 
                                                                                       & \multicolumn{1}{c|}{\textbf{SI}} & \multicolumn{1}{c|}{\textbf{MI}} & \multicolumn{1}{c|}{\textbf{WI}} & \multicolumn{1}{c|}{\textbf{SI}} & \multicolumn{1}{c|}{\textbf{MI}} & \textbf{WI} \\ \hline
\textbf{\begin{tabular}[c]{@{}c@{}}Computational Efficiency\\ (TOPS/mm2)\end{tabular}} & \multicolumn{1}{c|}{5}        & \multicolumn{1}{c|}{0.082}        & \multicolumn{1}{c|}{1.83e-4}     & \multicolumn{1}{c|}{5100}       & \multicolumn{1}{c|}{3460}      & 50.28       \\ \hline
\textbf{\begin{tabular}[c]{@{}c@{}}Power Efficiency\\ (TOPS/W)\end{tabular}}           & \multicolumn{1}{c|}{13.49}        & \multicolumn{1}{c|}{27.74}        & \multicolumn{1}{c|}{73.36}        & \multicolumn{1}{c|}{69.4}       & \multicolumn{1}{c|}{2.9e4}     & 3.6e5     \\ \hline
\textbf{\begin{tabular}[c]{@{}c@{}}System Efficiency\\ (pJ/MAC)\end{tabular}}          & \multicolumn{1}{c|}{0.19}     & \multicolumn{1}{c|}{0.24}      & \multicolumn{1}{c|}{0.67}     & \multicolumn{1}{c|}{0.03}     & \multicolumn{1}{c|}{0.14e-3}      & 6.12e-3     \\ \hline
\end{tabular}}
	\label{operationalperformance_table}
\end{table}

\subsubsection{Power \& Operation Performance Analysis} Fig.~\ref{power} shows the plot of the average power consumption plot with the increase in the number of S-AC units for different technology nodes and at different biasing regimes. It can be observed that with the increase in S-AC units in parallel, power consumption increases for a fixed value of hyperparameter $C$. 

Table~\ref{operationalperformance_table} shows the operational performance parameters~\cite{20inxs,21isaac} of the S-AC analog cells at different operating regimes and at different process nodes. We here computed the peak capabilities of the S-AC architectures for some of the widely used performance metrics such as\textit{ Computational EfficiencyPower Efficiency, System Efficiency}. It can be observed that the \textit{Computational Efficiency} is highest in SI for planar CMOS and Finfet node. It also shows that the best \textit{System Efficiency} and \textit{Power Efficiency} and lowest \textit{pJ/MAC} operation can be obtained in weak inversion.

\subsubsection{Mismatch and Process Variation Analysis} In the sub-micron technology nodes, the threshold voltage $(V_{T0})$ and current factor $(\beta)$ differences are the dominant sources underlying the drain-source current or gate-source voltage mismatch for a matched pair of MOS transistors \cite{mismatch1,mismatch2}. These mismatches affects the speed, accuracy, and other performance parameters of analog circuits. Fig.~\ref{mismatch7} shows the variation of output current due to variation in the \textit{Fin} count and overdrive voltage in the 7nm FinFET node. It can be observed that the variations are within $5\%$. Similar variations can be observed in Fig.~\ref{mismatch180} for planar CMOS for change in area and over-drive voltage.

\subsubsection{SNR Analysis} In a standard analog system (generally amplifiers) the total noise (input noise $n_{in}$, and the noise contributed by the circuit itself $n_{ckt}$) gets multiplied by the system gain and appears at the output. Thus resulting in no improvement in the signal-to-noise ratio (SNR). S-AC architectures exploit the inherent parallelism in the structure to overcome this limitation. In S-AC parallel current-mode configuration, because we are summing two uncorrelated noise sources, the overall noise increases as $\sqrt{2}$, while the correlated input signal amplitude increases by 2. Mathematically, for a single S-AC block having input signal $x_{1}$ and gain $G_1$, the total combined input signal is given by $x_{1} + n_{in_{1}}$ . The output includes the output signal $Z$ plus the total output noise $n_{out_{1}}$.
\begin{align}
	{n_{out_{1}}} &= {n_{in_{1}}} \times G_1 + {n_{ckt_{1}}}
	\label{equation63}\\
	{Z} &= {x_{in_{1}}} \times G_1
	\label{equation64}
\end{align}
The SNR is calculated by dividing the output RMS signal power by the output RMS noise power. This comes out to be
\begin{align}
	SNR_1=\frac{{{Z}}}{{{n_{out_{1}}}}} = \frac{{{{\left( {{x_{in_{1}}} \times G_1} \right)}^2}}}{{{{\left( {{n_{in_{1}}} \times G_1} \right)}^2} + n_{ckt1}^2}}
	\label{equation65}
\end{align}
Assuming the external noise input power is minimal (for simplicity), then \eqref{equation64} reduces  to
\begin{align}
	SNR_1= \frac{{{{\left( {{x_{in_{1}}} \times G_1} \right)}^2}}}{ n_{ckt1}^2}
	\label{equation66}
\end{align}
Now, adding a second S-AC block in parallel increases the RMS signal power by $2 \times$. However, it only increases the RMS circuit noise by $\sqrt{2}$ because the circuit adds uncorrelated noise. So instead of the noise doubling, we obtain a noise of $\sqrt{2} \times n_{ckt}$. The SNR equation for two interconnected S-AC blocks (assuming the composite gain remains almost the same)
becomes 
\begin{align}
	SNR_2= \frac{{{{\left( {2{x_{in}} \times G} \right)}^2}}}{{{{\left( {2{n_{in}} \times G} \right)}^2} + \left[ {\sqrt 2 \left( {{{\left( {{n_{ck{t_1}}}} \right)}^2} + {{\left( {{n_{ck{t_1}}}} \right)}^2}} \right)} \right]}}
	\label{equation67}
\end{align} 
For ${n_{ck{t_1}}} = {n_{ck{t_2}}} = {n_{ckt}}$ and  assuming the external input noise power is minimal \eqref{equation67} changes to
\begin{align}
	SNR_2  = \frac{{{{\left( {{x_{in}} \times G} \right)}^2}}}{{\left( {0.5{{\left( {{n_{ckt}}} \right)}^2}} \right)}}
	\label{equation68}
\end{align}
On comparing \eqref{equation66} and \eqref{equation68} one can analyze that for each increase in the number of connected S-AC blocks in parallel, the circuit SNR increases by twice.


\begin{table}[t]
\centering
\caption{ S-AC multiplier parameters for different $S$}
\scalebox{1}{
\begin{tabular}{|c|c|c|c|}
\hline
Error (\%)              & S=1      & S=2      & S=3      \\ \hline
Max Error          & 50      & 33.333   & 11.111   \\ \hline
Average abs   error  & 22.29 & 9.31 & 3.66  \\ \hline
Error bias           & 22.29 & -9.31 & -2.57 \\ \hline
Standard   Deviation & 11.78\%  & 5.89\%  & 1.68\%   \\ \hline
Approx. Area Savings$^{*}$ & 68.7\%  & 49.9\%  & 31.3\%   \\ \hline
Approx. Power Savings$^{*}$ & 68.4\%  & 52.7\%  & 37.2\%   \\ \hline
\end{tabular}}
 \begin{tablenotes}
 \item $^{*}$ Compared to full-precision multiplier~\cite{ml3}.
 \end{tablenotes}
\label{mul_varS_table}
\end{table}

\subsubsection{S-AC Area and Power Saving Analysis}
S-AC design offers a range of trade-offs between accuracy, area, and power benefits by changing the design parameter ($S$). The choice of $S$ is determined based on the application requirements. Theoretically, the number of splines selected can vary from $1$ to $S$, therefore offering a wide range of trade-offs to choose from. We here evaluate the performance of the S-AC multiplier for input dimension $N = 2$ as a function of $S$. The results summarized in Table~\ref{mul_varS_table} show the maximum error, average absolute error, error bias, and standard deviation. It can be analyzed that with the increase in the value of $S$, all error metrics decrease. Furthermore, the errors reduce to roughly half for each increase in the value of $S$. Table~\ref{mul_varS_table} shows the design savings offered by S-AC methodology when compared with a similar full-precision state-of-the-art multiplier implemented in~\cite{ml1}. As expected, the design area and average power consumption increase as a factor of $S$. As analyzed, significant savings can be achieved while introducing insignificant amounts of error. As an example, with an average absolute error of 3.66\%, $S = 3$ offers up to 31.3\% in area savings and up to 37.2\% in power savings.

\begin{table}[t]
\caption{Energy/Operation and Mean Absolute Deviation}
 \scalebox{0.9}{
\centering
\begin{tabular}{|c|c|c|ccc|}
\hline
\multirow{3}{*}{Operation} & \multirow{3}{*}{Err*} & \multirow{3}{*}{\begin{tabular}[c]{@{}c@{}}Process\\      Node\end{tabular}} & \multicolumn{3}{c|}{\multirow{2}{*}{Energy per Operation (fJ)}} \\
 &  &   & \multicolumn{3}{c|}{} \\ \cline{4-6} 
 &  &   & \multicolumn{1}{c|}{WI} & \multicolumn{1}{c|}{MI} & SI \\ \hline
\multirow{2}{*}{Cosh} & \multirow{2}{*}{0.0599}  & 180nm & \multicolumn{1}{c|}{40.86} & \multicolumn{1}{c|}{108.12} & 222 \\ \cline{3-6} 
 &  &  7nm & \multicolumn{1}{c|}{0.0196} & \multicolumn{1}{c|}{0.612} & 23.1 \\ \hline
\multirow{2}{*}{Sinh} & \multirow{2}{*}{0.0098} & 180nm & \multicolumn{1}{c|}{81.72} & \multicolumn{1}{c|}{216.24} & 444 \\ \cline{3-6} 
 &  &  7nm & \multicolumn{1}{c|}{0.0392} & \multicolumn{1}{c|}{1.224} & 46.2 \\ \hline
\multirow{2}{*}{ReLU} & \multirow{2}{*}{0.0337}  & 180nm & \multicolumn{1}{c|}{11} & \multicolumn{1}{c|}{24.42} & 75.94 \\ \cline{3-6} 
 &  &  7nm & \multicolumn{1}{c|}{0.0035} & \multicolumn{1}{c|}{0.101} & 2.97 \\ \hline
\multirow{2}{*}{\begin{tabular}[c]{@{}c@{}}Compressive\\ Non-Linearity\end{tabular}} & \multirow{2}{*}{0.0054} & 180nm & \multicolumn{1}{c|}{81.72} & \multicolumn{1}{c|}{216.24} & 444 \\ \cline{3-6} 
 &  &  7nm & \multicolumn{1}{c|}{0.0392} & \multicolumn{1}{c|}{1.224} & 46.2 \\ \hline
\multirow{2}{*}{Soft-plus} & \multirow{2}{*}{0.0321}  & 180nm & \multicolumn{1}{c|}{40.86} & \multicolumn{1}{c|}{108.12} & 222 \\ \cline{3-6} 
 &  &  7nm & \multicolumn{1}{c|}{0.0196} & \multicolumn{1}{c|}{0.612} & 23.1 \\ \hline
\multirow{2}{*}{\begin{tabular}[c]{@{}c@{}}WTA\\(N-Input)\end{tabular}} & \multirow{2}{*}{0.1760} & 180nm & \multicolumn{1}{c|}{N $\times$ 6.81} & \multicolumn{1}{c|}{N $\times$ 18.02} & N $\times$ 37 \\ \cline{3-6} 
 &  &  7nm & \multicolumn{1}{c|}{N $\times$ 0.0033} & \multicolumn{1}{c|}{N $\times$0.102} & N $\times$ 3.85 \\ \hline
\multirow{2}{*}{\begin{tabular}[c]{@{}c@{}}Multiply/\ \\(Divide)\end{tabular}} & \multirow{2}{*}{0.0139} & 180nm & \multicolumn{1}{c|}{190} & \multicolumn{1}{c|}{240} & 670 \\ \cline{3-6} 
 &  &  7nm & \multicolumn{1}{c|}{0.018} & \multicolumn{1}{c|}{0.42} & 90\\ \hline
\end{tabular}
}
\label{computationalcomplexity_table}
 \begin{tablenotes}
 \item $^{*}Err = MA{X_{\forall x}}\left| {Mean\,Absolute\,Deviation} \right|$ between 180nm and 7nm process node at room temperature.
 \end{tablenotes}
\end{table}

\subsubsection{Task-Energy Efficiency Analysis}
Table~\ref{computationalcomplexity_table} summarizes the energy requirement and Mean-Deviation for the S-AC block for implementing basic computational operations at different operating regimes and at different process nodes. We reported the maximum mean absolute deviation obtained from the resultant functional shapes at 180nm and 7nm when similar architecture was used in both process nodes. Table~\ref{computationalcomplexity_table} also summarizes the Energy/Operation obtained at different operating regimes and at different process nodes for different S-AC computations. The least energy consumption is seen in the WI regime and the worst in the SI regime.

\section{Case Study: S-AC based Neural Network}\label{analog_system}
In this section we show the design flow optimization of a S-AC based neural network synthesized using S-AC standard cells. Note that the approach presented here can also be generalized to implement other machine learning architectures.

\subsection{Algorithm to S-AC Hardware Mapping}\label{designs2}
Consider a vector ${\textbf{x}} \in \mathbb{R}^N$ where the output $y$ for a standard MLP~\cite{cristianini2000introduction} is given as, 
\begin{align}
    y = \varphi \left( {\eta \left( \textbf{x} \right)} \right)
    \label{perceptron_relu}
\end{align}

\begin{figure}[t]
\centering
\includegraphics[width=1\linewidth]{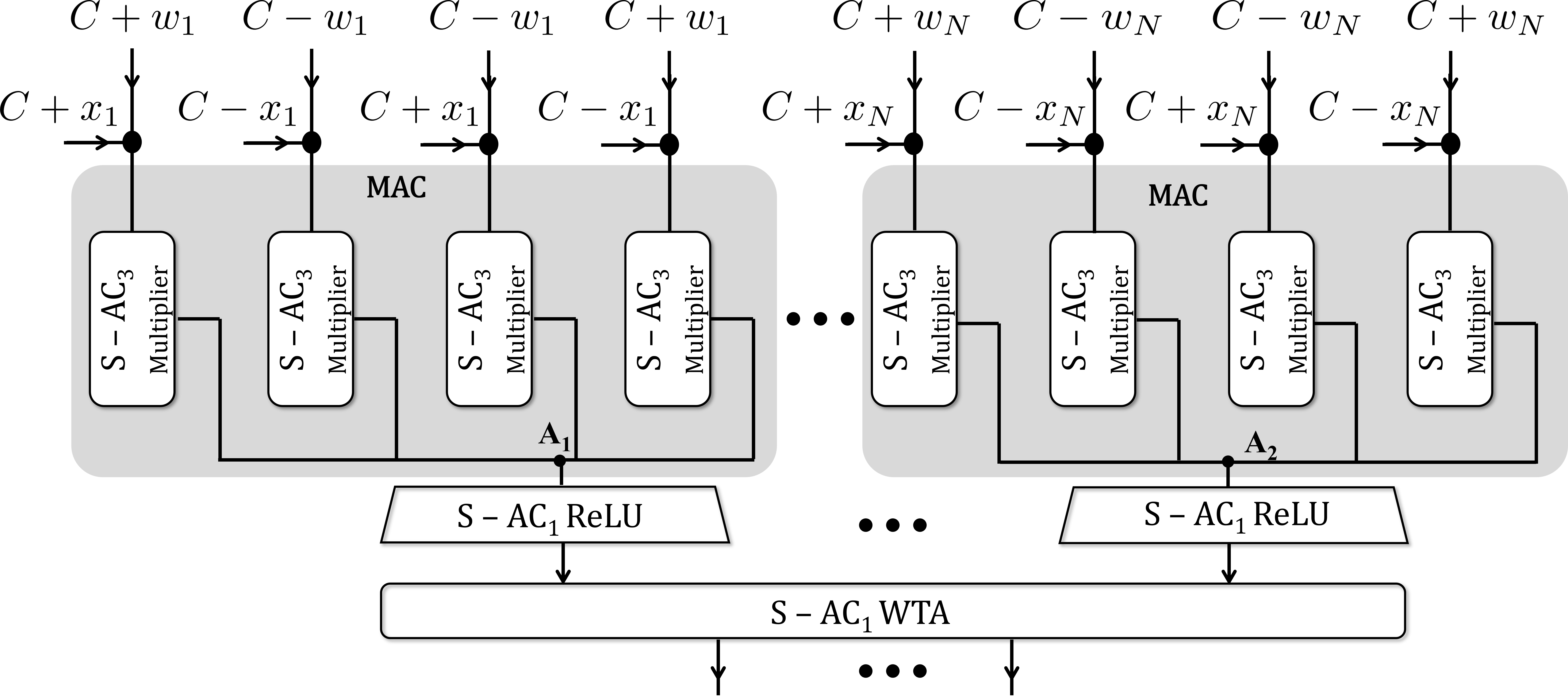}
\hfil
\caption{System-level architecture of a 3-layer neural network using S-AC standard cells.}
	\label{appl}
\end{figure}

\begin{figure}[t]
	\centering
	\subfloat[]{\includegraphics[width=0.5\linewidth]{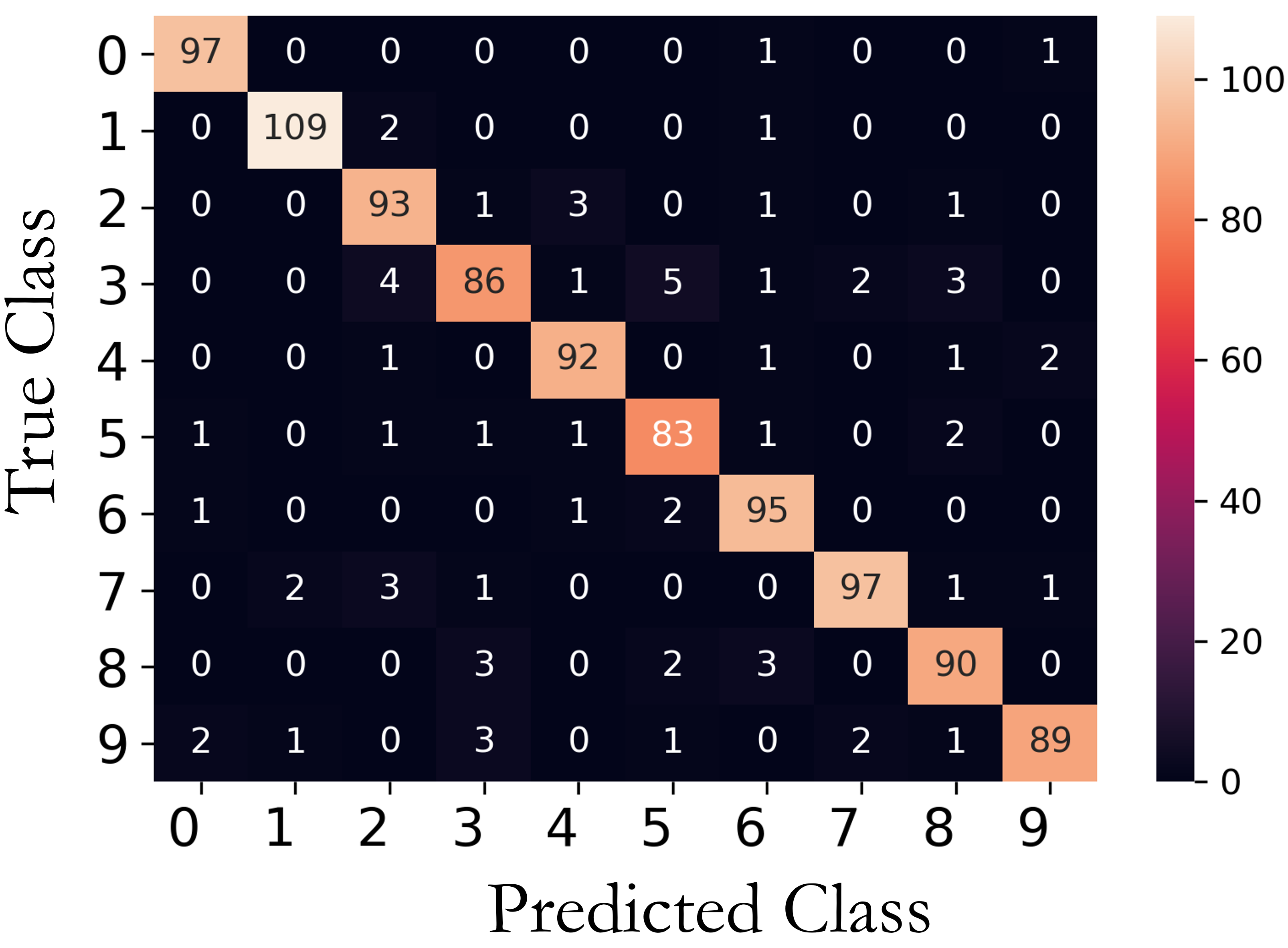}
		\label{confusion}}
	\subfloat[]{\includegraphics[width=0.5\linewidth]{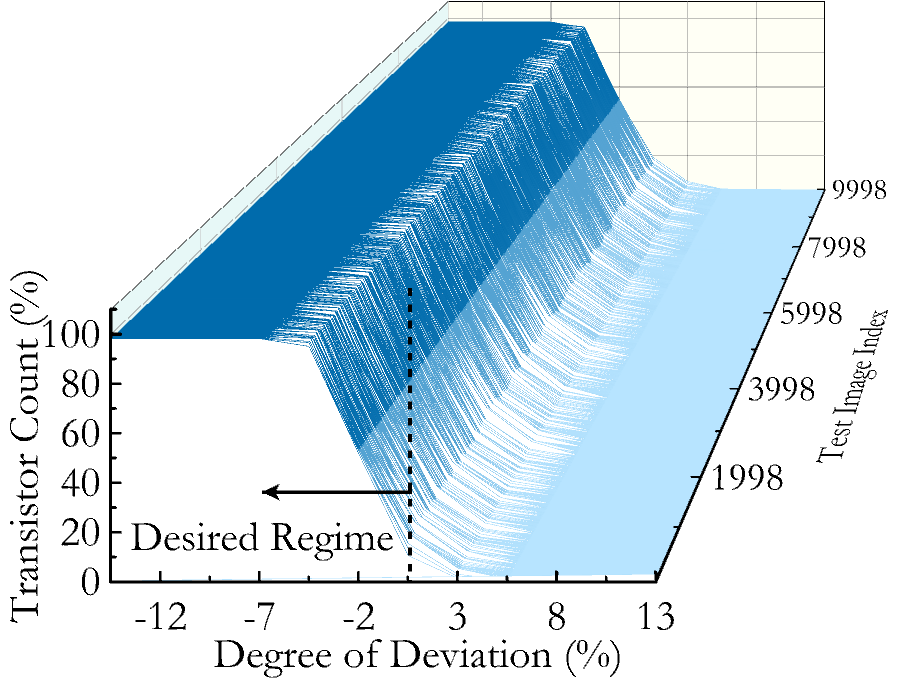}
		\label{error_perc}}
	\caption{\protect\subref{confusion}~{Confusion plot showing the distribution of randomly chosen 1000 test images from MNIST dataset; \protect\subref{error_perc}~Distribution showing the percentage of transistors that have deviated from the desired operating condition for the MNIST test-set.}}
	\label{confP}
\end{figure}

where $\varphi \left(  \cdot  \right)$ is any non-linear function be it tanh($\cdot$), sigmoid, ReLU, etc and ${\eta \left( \textbf{x} \right)}$ be the decision function given by
\begin{align}
    \eta ({\bf{x}}) = {{\bf{w}}^T}{\bf{x}} + b \label{perceptron}
\end{align}
where $\textbf{w} \in \mathbb{R}^N$ is the trained weight vector, $\textbf{x} \in \mathbb{R}^N$ is the input vector, $b \in \mathbb{R}$ is the bias and the function ${\eta }:\mathbb{R}^N \rightarrow \mathbb{R}$ is the decision function. For $\left\{ {x,y} \right\} \in \mathbb{R}$, the decision function $\eta (x)$ can then be rewritten as
\begin{align}
   \eta (x) = w \cdot x{\rm{ + b}}
    \label{perceptron_mapped}
\end{align}
Using equation \eqref{exp0} and \eqref{multipliyeqn} in \eqref{perceptron_mapped} the decision function gets mapped to S-AC based form as
\begin{multline}
\eta (x) = h(2C + w + x) - h(2C + w - x) + \cdots\\
 \cdots h(2C - w - x) - h(2C - w + x) + b
 \label{perceptron_approximated}
\end{multline}
Equation \eqref{perceptron_approximated}  can be viewed as generic decision function $\eta (\cdot)$ mapped into shape domain. This shape equation can then be easily synthesized using only S-AC analog cells. It can be noted that variable $b$ can be assumed as a constant current added to the dot-product $w\cdot x$ and implemented using KCL without additional circuits. Furthermore, to add non-linearity to this output, function $\varphi \left(  \cdot  \right)$ can then be mapped to the shape domain using the formulation demonstrated in Section~\ref{designs}.

\begin{table}[t]
\centering
\caption{Classification Accuracy at Different Operating Regimes \& Different Process Nodes}
\label{classificationaccuracy_table}
\scalebox{0.99}{
\begin{tabular}{|c|c|c|cc|}
\hline
\multirow{2}{*}{Dataset} & \multirow{2}{*}{\begin{tabular}[c]{@{}c@{}}Operating \\ Regime\end{tabular}} & \multirow{2}{*}{\begin{tabular}[c]{@{}c@{}}Classification\\ Accuracy\\ (S/W) \end{tabular}} & \multicolumn{2}{c|}{\begin{tabular}[c]{@{}c@{}}Classification\\ Accuracy\\ (H/W)\end{tabular}} \\ \cline{4-5} 
 &  &  & \multicolumn{1}{c|}{180nm} & 7nm \\ \hline
\multirow{3}{*}{XOR} & SI & \multirow{3}{*}{95} & \multicolumn{1}{c|}{95} & 95 \\ \cline{2-2} \cline{4-5} 
 & MI &  & \multicolumn{1}{c|}{94} & 93 \\ \cline{2-2} \cline{4-5} 
 & WI &  & \multicolumn{1}{c|}{93} & 93 \\ \hline
\multirow{3}{*}{AReM} & SI & \multirow{3}{*}{94} & \multicolumn{1}{c|}{93.5} & 93 \\ \cline{2-2} \cline{4-5} 
 & MI &  & \multicolumn{1}{c|}{93} & 94 \\ \cline{2-2} \cline{4-5} 
 & WI &  & \multicolumn{1}{c|}{93} & 93 \\ \hline
\multirow{3}{*}{MNIST} & SI & \multirow{3}{*}{93} & \multicolumn{1}{c|}{92.5} & 92.1 \\ \cline{2-2} \cline{4-5} 
 & MI &  & \multicolumn{1}{c|}{92} & 92 \\ \cline{2-2} \cline{4-5} 
 & WI &  & \multicolumn{1}{c|}{92.2} & 92.1 \\ \hline
\end{tabular}}
\end{table}

\subsection{S-AC based Neural Network}

Fig.~\ref{appl} shows the system-level implementation of a S-AC based neural network designed using S-AC based analog standard cells. The network was mapped using the algorithm mapping approach described in Section~\ref{designs2} and designed using S-AC based analog cells described in Section~\ref{designs}. {We trained the network using the margin propagation algorithm mentioned in \cite{multiplier_abhishek} with variation aware training~\cite{joshi2020accurate}}. 

Table~\ref{classificationaccuracy_table} summarizes the classification accuracy of synthesized neural networks using S-AC blocks at different process technology nodes and operating regimes. We tested out system classification capabilities on standard Activity Recognition dataset (AReM)~\cite{arem} dataset where we chose two of the activities as positive cases, i.e., bending and lying activities and utilized One versus all approach for binary classification. We further verified the functionality on the benchmark MNIST dataset~\cite{mnist} of handwritten digits. It can be analyzed from Table~\ref{classificationaccuracy_table} that the classification accuracy of implemented hardware nearly matches that of software at both 180nm and 7nm implementation and at different operating regimes. This clearly signifies that the design is both process technology scalable and bias scalable. It can be noted that due to time-complexity in running analog SPICE simulations on the EDA tools(where each full-precision simulation run extended for nearly $6$ hours), we randomly selected $1000$ test images out of the given $10000$ test set.

{Fig.~\ref{confusion} shows the confusion plot showing the distribution for randomly selected $1000$ test images. Furthermore, the implemented MNIST network consisted of 256 input nodes (where each input $28\times28$ image dimension was scaled down to $16\times16$), 15 hidden nodes, and 10 output nodes for which the baseline accuracy from both vanilla network and margin propagation network with variation aware training in PyTorch was obtained as $93\%$. As the hidden node count increases, both vanilla and margin propagation network match the state-of-the-art accuracy.}{Table~\ref{comp_ann} shows a comparison of different analog ANN implementations. It can be observed that the Energy/pixel obtained from S-AC network varies as the operating regime shifts from WI to SI, signifying that as per the application, the user can tune the performance specification. The energy/pixel of the S-AC ANN at 180nm for WI is $3 \times$ lesser than~\cite{sanyal} while in the SI, it is approximately $14 \times$ more.} Fig.~\ref{error_perc} shows the percentage deviation of transistors from the desired biasing regime for the MNIST dataset. It can be found that on average around 8\% of the total transistors in the designed system, shifted their operation from their expected operating regime. For instance, when the designed system was biased to operate in weak inversion regime, near about 8\% of total transistors shifted their operation to moderate inversion regime while for the system biased in moderate inversion regime, a similar percentage of transistors shifted their operation to strong inversion regime. However, despite this change in operating condition, there is no significant drop in accuracy as the classification network leverages the benefits of bias scalability of the S-AC circuit.

\begin{table}[t]
\centering
\caption{{Comparison with State-of-the-art Analog ANNs}}
\label{tab:my-table}
\scalebox{0.89}{
\begin{tabular}{|l|c|c|c|c|c|c|c|} 
\hline
 & {\cite{wang2017low}} & {\cite{sanyal_ref}} & {\cite{sanyal}}  & \multicolumn{4}{c|}{This Work} \\ 
\hline
Process (nm) & 130 & 130 & 65 & \multicolumn{2}{c|}{7} & \multicolumn{2}{c|}{180} \\ 
\hline
Supply (V) & 1.2 & - & 1.2 & \multicolumn{2}{c|}{0.7} & \multicolumn{2}{c|}{1.2} \\ 
\hline
Classifier & binary & binary & ANN & \multicolumn{2}{c|}{ANN} & \multicolumn{2}{c|}{ANN} \\ 
\hline
Speed (MHz) & 1.3 & 50 & 5 & \multicolumn{2}{c|}{6} & \multicolumn{2}{c|}{3.37} \\ 
\hline
Feature Size &  48 & 81 & 25 & \multicolumn{2}{c|}{256} & \multicolumn{2}{c|}{256} \\ 
\hline
\begin{tabular}[c]{@{}l@{}}Biasing \\Regime\end{tabular} & - & - & - & WI & SI & WI & SI \\
\hline
Accuracy (\%) & 90 & 90 & 82 & 92.2 & 92.5 & 92.1 & 92.1 \\ 
\hline
Energy/pixel (pJ)& 11.1$^{*}$ & 7.8$^{*}$ & 6.9 & 0.05$^{\dagger}$& 3.7$^{\dagger}$& 2.3$^{\dagger}$ & 97.6$^{\dagger}$\\
\hline
\end{tabular}}
\begin{tablenotes}
 \item $^{*}$ Only for first layer implemented on-chip;
  \item $^{\dagger}$ Based on SPICE simulations of hidden and output layers.
 \end{tablenotes}
 \label{comp_ann}
\end{table}

\section{Conclusions}\label{conc}
In this work, we proposed a framework for designing analog computing circuits that are process, bias, and temperature scalable. The key differentiator from the previous work~\cite{synth_bias_scale} is the multi-spline approach that allows the framework to trade-off power/area with accuracy (Fig.~\ref{exp}). The framework leads to the design of S-AC standard cells whose responses were found to be robust to variations in biasing and temperature and scalable across process nodes similar to digital standard cells. As a result, a S-AC standard cell designed in a 180nm process could be easily used for design in a 7nm process. Thus, the modules can be potentially used for the automated synthesis of large-scale analog processors. While the focus of this work was to design S-AC circuits for machine learning processors, the approach can be generalized to other analog processors as well. As a proof-of-concept, we demonstrated the approach for a 3-layer neural network whose test accuracy remains minimally affected by changes in biasing and process nodes. Even though the results in the paper show excellent agreement between the accuracy of the software simulation and circuit simulation for small-scale neural networks, future work will leverage this framework to synthesize large-scale analog deep neural networks and reconfigurable machine learning processors.

\begin{appendices}
\section{Proof: Generalized Margin Propagation Formulation using Multi-Spline Approach}
\label{Appendix:A}
Let us consider a log-sum-exp function~\cite{synth_bias_scale} $h_{\log}:\mathbb{R}^N \rightarrow \mathbb{R}$ given by
\begin{align}
{h_{\log }}({\bf{x}}) = C \cdot \log \left( {\sum\nolimits_{i = 1}^N {{e^{\frac{{{x_i}}}{C}}}} } \right)
\label{log-sum-exp2}
\end{align}
where $C$ is a hyper-parameter and $\textbf{x} \in \mathbb{R}^N$ is a vector with elements $x_i \in \mathbb{R}$. 
Equation \eqref{log-sum-exp} can be written as
\begin{align}
\sum\nolimits_{i = 1}^N {{e^{\frac{{{x_i} - {h_{\log }}}}{C}}}}  = 1
\label{log-sum-exp_rewritten2}
\end{align}
which is an equivalent non-linear constraint satisfaction problem. Let us approximate this non-linear function using linear splines. It can be noted that the same methodology can also be extended to other non-linear functions. As a proof-of-concept, we show for log-sum-exponential. In Fig.~\ref{exp}, we have shown the plot of the \textit{exponential} function and its approximation using one-spline $(S=1)$ and three-splines $(S=3)$. Here, $Q_1,Q_2,\cdots,Q_S$ are the tangential points and $T_1,T_2,\cdots,T_{S}$ are the tuning points. The generic line equation for the $j^{th}$ spline where $j \in \left( {1, \cdots ,S} \right)$ when approximated using piece-wise-linear lines can be written using point-slope form as
\begin{align}
{{\theta}_j}\left( x \right) = {e^{{Q_j}}} \cdot x + {e^{{Q_j}}}\left( {1 - {Q_j}} \right)\quad \forall x \ge 0
	\label{line_eqn_jth}
\end{align}
where $e^{Q_j}$ is the slope and $e^{Q_j}(1 - {Q_j})$ is its intercept on the line on the vertical axis.
Similarly, for the $(j+1)^{th}$ spline, we can write its line equation as, 
\begin{align}
{{\theta}_{j+1}}\left( x \right) = {e^{{Q_{j+1}}}} \cdot x + {e^{{Q_{j+1}}}}\left( {1 - {Q_{j+1}}} \right) \forall x \ge 0
	\label{line_eqn_j1th}
\end{align}
The tuning points $T_j$ (intercept between $j^{th}$ and $(j+1)^{th}$ spline) can be obtained by equating the line equations of $j^{th}$ and $(j+1)^{th}$ spline at ${x_i}={T_j}$ and can be written as,
\begin{align}
	{e^{{Q_j}}} \cdot {T_j} + {e^{{Q_j}}}\left( {1 - {Q_j}} \right) =&{e^{{Q_{j + 1}}}} \cdot {T_j} + {e^{{Q_{j + 1}}}}\left( {1 - {Q_{j + 1}}} \right)
	\label{equating_line_eqn}
\end{align} 
Equation \eqref{equating_line_eqn} can be re-written as
\begin{align}
{T_j} = &\frac{{{Q_{j + 1}} \cdot {e^{{Q_{j + 1}}}} - {Q_j} \cdot {e^{{Q_j}}}}}{{{e^{{Q_{j + 1}}}} - {e^{{Q_j}}}}} - 1
\label{Tuning_Point}
\end{align}
Then, the approximation of the function ${e^x}$ using $S$-splines (a case of 3-spline approximation is shown in Fig.~\ref{exp}) can be written using point-slope form and using \eqref{line_eqn_jth} and \eqref{Tuning_Point} as
\begin{align}
{{e^x} \cong }&{{e^{{Q_1}}}{{\left[ {{x} - {T_1}} \right]}_ + } + \left( {{e^{{Q_2}}} - {e^{{Q_1}}}} \right){{\left[ {{x} - {T_2}} \right]}_ + } +  \cdots }\nonumber\\ 
{}&{ \cdot  \cdots  + \left( {{e^{{Q_S}}} -  \cdots  - {e^{{Q_2}}} - {e^{{Q_1}}}} \right){{\left[ {{x} - {T_{S}}} \right]}_ + }}
\label{generic_msmp_expanded}
\end{align}	
Equation \eqref{generic_msmp_expanded} can then be generalized as
\begin{align}
{e^x} \cong \sum\nolimits_{j = 1}^S {\left( {{e^{{Q_j}}} - \sum\nolimits_{k = 1}^{j - 1} {{e^{{Q_k}}}} } \right)} {\left[ {x - {T_j}} \right]_ + }
\label{generic_msmp_app}
\end{align}

The above equation~(\ref{generic_msmp_app}) shows the approximation of $e^x$ using S-splines. For ease of understanding let us choose a specific case of 3-splines, viz. $S=3$. Let the tangential points for this case be ${Q_1} = {\log _e}\left( {0.5} \right)$,${Q_2} = {\log _e}(1)$, ${Q_3} = {\log _e}(2)$. Then by using (\ref{Tuning_Point}) we get,
\begin{align}
	{T_1} &= {\log _e}\left( {0.5} \right) - 1 =  - {\log _e}2 - 1
	\label{tangential_0}
\\
	{T_2}{\rm{ }} &= \frac{{ - {{\log }_e}\left( {0.5} \right) \times 0.5}}{{0.5}} - 1 = {\log _e}2 - 1
	\label{tangential_1}
\\
	{T_3} &= \frac{{2{{\log }_e}2}}{{2 - 1}} - 1 = 2{\log _e}2 - 1
	\label{tangential_2}
\end{align}
Using equation (\ref{tangential_0}) - (\ref{tangential_2}) in (\ref{generic_msmp_app}), we have
\begin{align}
{{e^x} \cong }&{\frac{1}{2}{{\left[ {{x} + {{\log }_e}2 + 1} \right]}_ + } + \frac{1}{2}{{\left[ {{x} - {{\log }_e}2 + 1} \right]}_ + } + {\rm{ }}}\nonumber \\
{}&{ \cdots{} \frac{1}{2}{{\left[ {{x} - 2 \cdot {{\log }_e}2 + 1} \right]}_ + }}
	\label{equation18_3splineLSE_nooffset}
\end{align}
Substituting equation \eqref{equation18_3splineLSE_nooffset} in \eqref{log-sum-exp_rewritten2}, for $i \in \left( {1, \cdots ,N} \right)$, we get
\begin{align}
\sum\nolimits_{i = 1}^N {\left[ {{x_i} + {O_1} - h} \right] + \left[ {{x_i} + {O_2} - h} \right] + \left[ {{x_i} + {O_3} - h} \right]}  = C'
\label{equation18wc_off}
\end{align}
Here, $O_1=C(1+log_e2), O_2=C(1-log_e2)$ and $O_3=C(1-2log_e2)$ are the offsets and $C'=2C$ is a tunable parameter. Equation \eqref{equation18wc_off} shows the approximation of exponential function with 3-splines. The same can then be generalized to $S$-splines, $N$-inputs as
\begin{align}
\sum\nolimits_{i = 1}^N {\sum\nolimits_{j = 1}^S  {\left[ {{x_i} + {O_j} - h} \right]_{+}} } &= C \nonumber \\ 
\sum\nolimits_{i = 1}^N {\sum\nolimits_{j = 1}^S {\left[ {{x_{i,j}} - h} \right]_{+}} }  &= C
	\label{equation18wc1_DUALsUM}
\end{align}
where $O_j$ is the offset due to $j^{th}$ spline, $x_{i,j}$ is the $i^{th}$ input corresponding to $j^{th}$ spline, $C$ is a hyperparameter and $S$ is a design parameter also called splines count. It can be noted from Fig.~\ref{exp} that with the increase in the number of splines ($S$), the computational precision increases, while the input dimension increases along $N$. Equation~\eqref{equation18wc1_DUALsUM} is called Generalized Margin Propagation function (GMP).
\end{appendices}

\subsection{Acknowledgments}
The authors would like to acknowledge the joint IISc-WashU MoU to facilitate the collaboration between the two institutions. This work is also supported by the Department of Science and Technology of India (SERB CRG/2021/005478, DST/IMP/2018/000550).

\bibliographystyle{IEEEtran}
\bibliography{ref}
\end{document}